\documentclass[12pt]{article}
\usepackage{amsmath}
\usepackage{float}
\usepackage{graphicx,psfrag,epsf}
\newcommand{\vtt}[1]{%
  \text{\normalfont\ttfamily\detokenize{#1}}%
}
\usepackage{natbib}
\usepackage{enumerate}
\usepackage{mathtools}
\usepackage{indentfirst}
\usepackage{setspace}
\usepackage{titlesec}
\usepackage{caption}
\usepackage{mathrsfs}
\usepackage{bbm}
\usepackage{tabularx}      
\usepackage[table]{xcolor}
\definecolor{shade}{HTML}{F8F4FF} 
\RequirePackage{amsthm,amsmath,amsfonts,amssymb}
\RequirePackage[colorlinks,citecolor=blue,urlcolor=blue]{hyperref}
\usepackage{url} 
\usepackage{algorithm}
\usepackage[noend]{algorithmic}

\usepackage{caption}

\newlength\myindent
\newcommand\bindent{%
  \begingroup
  \setlength{\itemindent}{\myindent}
  \addtolength{\algorithmicindent}{\myindent}
}
\newcommand\eindent{\endgroup}
\newcommand{\blind}{1}
\usepackage[margin=1in]{geometry}

\newtheorem{theorem}{Theorem}

\newtheorem{corollary}{Corollary}
\newtheorem*{DoobsTheorem*}{Doob's Theorem}

\newtheorem*{restatedtheorem*}{}
\newtheorem{innercustomthm}{Theorem}
\newenvironment{customthm}[1]
  {\renewcommand\theinnercustomthm{#1}\innercustomthm}
  {\endinnercustomthm}
\newtheorem{innercustomcor}{Corollary}
\newenvironment{customcor}[1]
  {\renewcommand\theinnercustomcor{#1}\innercustomcor}
  {\endinnercustomcor}

\def\spacingset#1{\renewcommand{\baselinestretch}%
{#1}\small\normalsize}

\begin{document}
\spacingset{1}

\if1\blind
{
  \title{\bf Nonparametric Copula Models for Multivariate, Mixed, and Missing Data}
  \author{Joseph Feldman\thanks{Corresponding author email: jrf11@rice.edu. An \vtt{R} package implementing the proposed approach is available on the author's github page, found at \url{https://github.com/jfeldman396/GMCImpute} } \ and Daniel R. Kowal\thanks{Research was sponsored by the Army Research Office (W911NF-20-1-0184), the National Institute of Environmental Health Sciences of the National Institutes of Health (R01ES028819), and the National Science Foundation (SES-2214726). The content, views, and conclusions contained in this document are those of the authors and should not be interpreted as representing the official policies, either expressed or implied, of the Army Research Office, the National Institutes of Health, or the U.S. Government. The U.S. Government is authorized to reproduce and distribute reprints for Government purposes notwithstanding any copyright notation herein.} \\
    Department of Statistics, Rice University}
    \date{}
  \maketitle
} \fi

\if0\blind
{
\title{\bf Nonparametric Copula Models for Mixed and missing-at-random Data }
\date{}
\maketitle
} \fi

\bigskip
\begin{abstract}
Modern datasets commonly feature both substantial missingness and many variables of mixed data types, which present significant challenges for estimation and inference. Complete case analysis, which proceeds using only the observations with fully-observed variables, is often severely biased, while model-based imputation of missing values is limited by the ability of the model to capture complex dependencies among (possibly many) variables of mixed data types. To address these challenges, we develop a novel Bayesian mixture copula for joint and nonparametric modeling of multivariate count, continuous, ordinal, and unordered categorical variables, and deploy this model for inference, prediction, and imputation of missing data. Most uniquely, we introduce a new and  computationally efficient  strategy for marginal distribution estimation that eliminates the need to specify any marginal models yet delivers  posterior consistency for each marginal distribution and the copula parameters under missingness-at-random. Extensive simulation studies demonstrate exceptional modeling and imputation capabilities relative to competing methods, especially with mixed data types, complex missingness mechanisms, and nonlinear dependencies. We conclude with a  data analysis that highlights how improper treatment of missing data can distort a statistical analysis, and how the proposed approach offers a resolution.
\end{abstract}

\noindent%
\vfill

\newpage

\spacingset{1.7}

\section{Introduction}
Missing data are ever-present in modern statistics and data analysis. The sources of missingness are vast and varied: participant non-response in surveys \citep{rubin1976inference}, 
participant attrition in longitudinal studies \citep{gustavson2012attrition}, linking multiple data sources \citep{reiter2012bayesian}, or errors in the data collection process all contribute to missingness. Any statistic meant to be computed on a fully-observed sample of data---including frequentist estimators and Bayesian posterior distributions---must be modified carefully in the presence of  missing data. At the broadest level, the goal remains to infer an unknown population quantity $Q$, and specifically to provide accurate point estimates and precise uncertainty quantification for $Q$; here, we focus on the additional challenges and implications of abundant missingness among many variables of mixed data types.

When confronted with missing data, there are two options for analysis. The first is to proceed using only observations for which all variables are observed. However, this \emph{complete case} (CC) analysis, while common in practice, is highly problematic in many settings.  CC analysis often substantially decreases the sample size, leading to imprecise and under-powered analysis. More critically, CC analysis can introduce various and significant forms of bias. Consider a sample of correlated bivariate data  $\{(Y_{i1}, Y_{i2})\}_{i=1}^{n}$, and suppose that the missingness in $Y_1$ is determined by the value of $Y_2$, which is fully observed (missingness-at-random; see below).  Figure~\ref{examp} shows the potential impacts of a CC analysis: the empirical cumulative distribution function (ECDF) of $Y_1$ is severely biased, which implicates inference on $Q(Y_{1})$ as well as popular Bayesian semiparametric copula models discussed subsequently \citep{hoff2007extending,murray2013bayesian, cui2019novel, feldman2022bayesian}.

\begin{figure}[h]
    \centering
    \includegraphics[width=.4\textwidth]{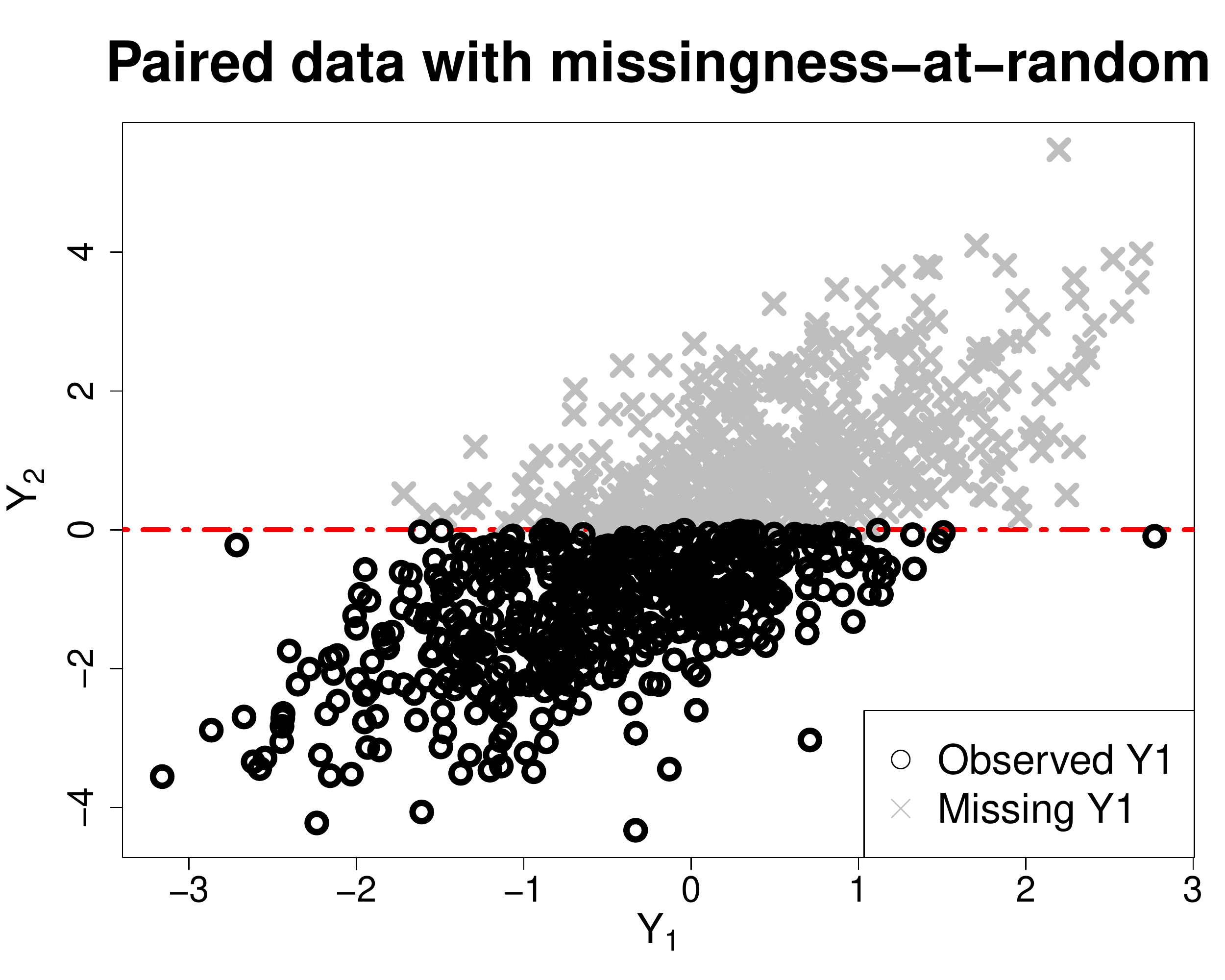}
    \includegraphics[width=.4\textwidth]{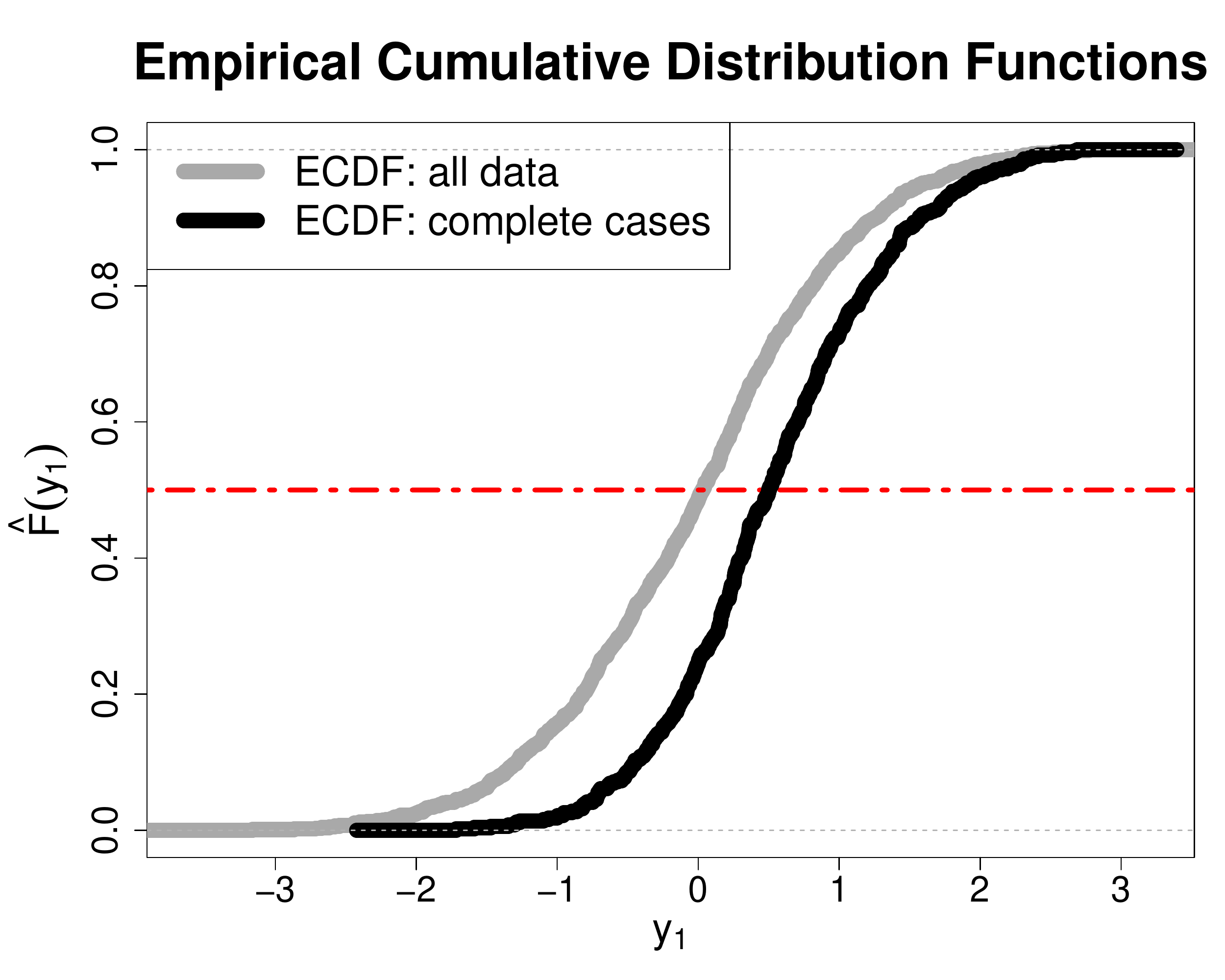}
    \caption{\small Bivariate data $\{(Y_{i1}, Y_{i2})\}_{i=1}^{n}$ with missing-at-random missingness (left) and the corresponding true and empirical cumulative distribution function (ECDF) for $Y_1$ (right). The missing data severely biases the ECDF, which impacts functionals of this term---including traditional statistics as well as  Bayesian semiparametric copula models.}
    \label{examp}
\end{figure}

The second option, which we pursue here, is \emph{imputation} of missing values. Informally, a statistical model is fit to the observed data and then used to repeatedly simulate the missing values, thus forming many completed datasets. Then, estimates $\hat{Q}$ are computed on each completed dataset, and combined to produce point estimates and uncertainty quantification for $Q$. If the model adequately captures the features of the data, we can expect the inference based on an imputation procedure to correct the shortcomings of a CC analysis.

The specification of an imputation model is made precise by considering a joint model for all data $\boldsymbol{Y} = (Y_{ij})$ and binary missingness variables $\boldsymbol{R} = (R_{ij})$, where $R_{ij} = 1$ indicates that $Y_{ij}$ is missing, and $R_{ij} = 0$ means that $Y_{ij}$ is observed. Let $\boldsymbol{Y}^{obs}=(Y_{ij}: R_{ij} = 0)$ denote the observed data and $\boldsymbol{Y}^{mis}=(Y_{ij}:R_{ij} = 1)$ the missing values. We assume that this model is indexed by distinct parameters $\boldsymbol{\theta}$ for $\boldsymbol Y$ and $\boldsymbol \phi$ for $\boldsymbol R$, with joint likelihood 

\begin{equation}\label{jointlike}
p(\boldsymbol R, \boldsymbol Y^{obs} \mid \boldsymbol \theta, \boldsymbol \phi) = \int p(\boldsymbol Y^{obs}, \boldsymbol Y^{mis} \mid \boldsymbol{\theta})p(\boldsymbol R \mid \boldsymbol Y^{obs}, \boldsymbol Y^{mis}, \boldsymbol\phi)d \boldsymbol Y^{mis}
\end{equation}

We focus on \emph{missingness-at-random} (MAR), which allows the missingness mechanism to depend on the observed (but not missing) data: $p(\boldsymbol{R} \mid \boldsymbol Y^{obs}, \boldsymbol Y^{mis}, \boldsymbol{\phi}) = p(\boldsymbol{R} \mid \boldsymbol{Y}^{obs}, \boldsymbol{\phi})$ \citep{rubin1976inference}. In this case the missingness is \emph{ignorable}, and the model specified on the observed data $p(\boldsymbol Y^{obs} \mid \boldsymbol \theta) = \int p(\boldsymbol Y^{obs}, \boldsymbol Y^{mis} \mid \boldsymbol{\theta})d \boldsymbol Y^{mis}$ may be used for imputation. A stronger assumption is \emph{missing-completely-at-random} (MCAR), $p(\boldsymbol{R} \mid \boldsymbol Y^{obs}, \boldsymbol Y^{mis}, \boldsymbol{\phi}) = p(\boldsymbol{R} \mid \boldsymbol{\phi})$, which is a special case of MAR. 

There are several important considerations for MAR. First, CC analysis is strongly inadvisable (see Figure \ref{examp}), and thus imputation is needed in general. Second, MAR is most likely satisfied when $\boldsymbol Y^{obs}$ contains many potentially informative variables \citep{little2021missing}. Thus, MAR demands a model capable of accommodating multiple variables, possibly of mixed types. Finally, the suitability of MAR in practice depends on the adequacy of the assumed model. In aggregate, MAR necessitates a model for multivariate and mixed data that can adapt to complex marginal and joint distributional features.

\begin{figure}[h]
    \centering
        \includegraphics[width=.4\textwidth]{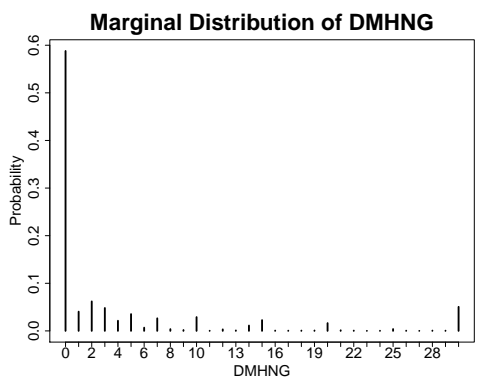}
    \caption{\small The marginal distribution of days of self-reported poor mental health (\vtt{DMHNG}) from the NHANES data, which is the response variable of interest in our real data analysis. Discreteness, boundedness, heaping, and zero-inflation combine to make modeling difficult.}
    \label{margDMHNG}
\end{figure}

Our motivating example comes from a collection of variables (see Table~\ref{tab:data}) in the National Health and Nutrition Examination Survey (NHANES). These variables include count, continuous, ordinal, and unordered categorical variables, with missingness as high as 43\% for some variables and missing values for each data type. Notably, these variables include self-reported mental health---which displays complex and discrete marginal distributional features (Figure~\ref{margDMHNG})---along with demographic and socioeconomic variables, alcohol and drug use variables, and health-related variables with intricate multivariate relationships. Most importantly, CC analysis is unsatisfactory or misleading for these data (see Section~\ref{realdat}). Thus, an imputation model is required---and in particular one capable of accommodating many variables of mixed types with intricate distributional features.

\begin{table}[h]
\centering
\begin{tabular}{m{7cm} m{4cm} m{3cm}} 
\textbf{Variable} &  \textbf{Values} & \textbf{\% Missing} \\
\hline
  \rowcolor{shade}
 \multicolumn{3}{l}{\textbf{Response variable:}} \\
\href{https://wwwn.cdc.gov/Nchs/Nhanes/2011-2012/HSQ_G.htm#HSQ480}{DaysMentHlthNotGood (DMHNG)} &  $\{0, 1, \ldots, 30\}$ & 14\% \\
 \hline
  \rowcolor{shade}
  \multicolumn{3}{l}{\textbf{Demographic and socioeconomic variables:}} \\
 \href{https://wwwn.cdc.gov/nchs/nhanes/2011-2012/demo_g.htm#RIAGENDR}{Gender} &  {Male, Female} & 0 \\
 \rowcolor{shade}
 \href{https://wwwn.cdc.gov/nchs/nhanes/2011-2012/demo_g.htm#RIDAGEYR}{Age (years)}  &   $\{18, \ldots, 80\}$ &0  \\
 \href{https://wwwn.cdc.gov/nchs/nhanes/2011-2012/demo_g.htm#RIDRETH1}{Race$^*$} &  {White, Black, \newline Hispanic, Other} & 0 \\
  \rowcolor{shade}
 \href{https://wwwn.cdc.gov/nchs/nhanes/2011-2012/demo_g.htm#DMDHREDU}{Education Level$^*$} & {$<$ HS, = HS, $>$ HS } & 5\%\\

  \href{https://wwwn.cdc.gov/nchs/nhanes/2011-2012/demo_g.htm#INDFMIN2}{Family Income$^*$ (FI)} & {Low, Middle, High} & 4\%\\
     \rowcolor{shade}
\href{https://wwwn.cdc.gov/Nchs/Nhanes/2011-2012/HIQ_G.htm#HIQ011}{Uninsured$^*$}&{Yes, No} & 0.2\%\\
 \hline
  \rowcolor{shade}
 \multicolumn{3}{l}{\textbf{Alcohol and drug use variables:}} \\
 \href{https://wwwn.cdc.gov/Nchs/Nhanes/2011-2012/ALQ_G.htm#ALQ151}{HeavyDrinker} & {Yes, No} & 29\%\\
  \rowcolor{shade}
 \href{https://wwwn.cdc.gov/Nchs/Nhanes/2011-2012/SMQRTU_G.htm#SMQ680}{UseNicotine} & {Yes, No} & 15\%\\
 \href{https://wwwn.cdc.gov/Nchs/Nhanes/2011-2012/DUQ_G.htm#DUQ200}{UsedMarijuana} & {Yes, No } & 43\%\\
  \rowcolor{shade}
 \href{https://wwwn.cdc.gov/Nchs/Nhanes/2011-2012/DUQ_G.htm#DUQ240}{UsedHardDrug} & {Yes, No} & 30\%\\
\hline
 \rowcolor{shade}
\multicolumn{3}{l}{\textbf{Health-related variables:}} \\

\href{https://wwwn.cdc.gov/nchs/nhanes/2011-2012/BMX_G.htm#BMXBMI}{Body Mass Index (BMI, kg/$m^2$)} & $[13.4, 81.2]$ & 6\%  \\
  \rowcolor{shade}
\href{http://data.nber.org/nhanes/2011-2012/BPQ_G.htm}{HasHighBP (BPQ020 at link) }& {Yes, No } & 0.1\% \\

\href{http://data.nber.org/nhanes/2011-2012/BPQ_G.htm}{HasHighChol (BPQ080 at link) }& {Yes, No} & 6\% \\
  \rowcolor{shade}
\href{https://wwwn.cdc.gov/Nchs/Nhanes/2011-2012/DIQ_G.htm#DIQ010}{HasDiabetes$^*$}&{Yes, No }& 0.08\% \\
 \hline
  \rowcolor{shade}

\end{tabular}
\caption{\small Variables in the analysis dataset with hyperlinks to the online NHANES descriptions. Annotated variables ($*$) include minor modifications (e.g., collapsed categories) from the original NHANES variables. \label{tab:data}}
\end{table} 

The literature on imputation models is extensive, yet limited in its ability to address these critical challenges; see \cite{murray2018multiple} for a thorough review. Broadly, there are two main frameworks for imputation. The first, \emph{fully conditional specification} (FCS), imputes missing values by (i) specifying a univariate regression model for each variable in the dataset conditional on all other variables and (ii) using each regression model to impute (separately) the missing values for each variable \citep{van1999flexible,raghunathan2001multivariate}. This approach offers several advantages: it is amenable to mixed data types, allows customization of each univariate model to increase flexibility \citep{burgette2010multiple, tang2017random}, and is implemented is freely available software  \citep{van2011mice}. However, FCS does not guarantee a valid joint distribution for the data, which is especially problematic for Bayesian inference, and is difficult to tune in high dimensions, since it requires a separate model fit for each variable. Perhaps most important, FCS often cannot capture complex multivariate relationships in the data   \citep{murray2016multiple}, which we confirm in Section~\ref{sim}.

The second main approach constructs a joint distribution for all variables in the dataset and then imputes missing values from the (posterior) predictive distribution.  Bayesian nonparametric models are particularly attractive, including for imputation of multiple categorical \citep{dunson2009nonparametric, manrique2013bayesian, manrique2017bayesian}, ordinal \citep{kottas2005nonparametric, deyoreo2017bayesian}, or  categorical and continuous variables \citep{murray2016multiple,roy2018bayesian}.  Related, \cite{taddy2010bayesian} proposed a multivariate mixture model with kernel components specific to each data type, but did not consider imputation. These existing approaches have several limitations. First, they do not simultaneously accommodate  categorical, continuous, count, and ordinal variables. Second, they often require careful model specification for each variable, which is arduous in moderate to high dimensions. Finally, the accompanying MCMC samplers are typically complex and computationally intensive.  To our knowledge, there is no publicly available software for imputation based on these methods, which limits their practical utility.


Copula models offer a potential avenue to estimate a joint model on mixed data types: they combine arbitrary marginal distributions with a mechanism to model joint dependencies \citep{joe2014dependence}. \cite{zhao2020matrix, zhao2020missing} deployed frequentist Gaussian copula models for imputation of MCAR data with continuous and ordinal variables, but did not consider MAR missingness or other data types. 
\cite{pitt2006efficient} specified parametric families for count and continuous variables within a Bayesian Gaussian copula model which could be extended for imputation. However, parametric  specification of marginal distributions is restrictive and time-consuming, especially when there are complex marginal distributions (Figure~\ref{margDMHNG}) and many variables to consider (Table~\ref{tab:data}). 

\cite{hoff2007extending} partially resolved this issue for count, continuous, and ordinal variables using the extended rank-likelihood (RL) for Gaussian copula estimation, which was extended to higher dimensions using factor models in \cite{murray2013bayesian} and deployed for imputation in \cite{cui2019novel}. The RL uses a rank-based approximation to the likelihood for \emph{semiparametric} inference, whereby the Gaussian copula (correlation) parameters are inferred using only the ranks of the observed data. 
\cite{feldman2022bayesian} introduced the extended rank-probit likelihood (RPL) to include count, continuous, ordinal, and now unordered categorical variables. Most uniquely, the R(P)L delivers inference for the copula parameters \emph{without} requiring any estimation or model specification of the marginal distributions, which is a substantial simplification that facilitates high-dimensional imputation.

Despite these advantages, semiparametric Bayesian copula models  have two glaring shortcomings in the presence of missing data. First, these models do not estimate the marginal distributions and thus do not provide a data generating process for prediction or imputation. Instead, the default approach is to fix each margin at its ECDF and generate posterior predictive variates by repeatedly (i) drawing a latent Gaussian variable under the model and (ii) applying the inverse ECDF.  However, the ECDF is significantly flawed under MAR (see Figure~\ref{examp}), so the resulting posterior predictive imputations will produce inaccurate estimation and uncertainty quantification for $Q$---even if the joint dependencies are well-modeled by the Gaussian copula. \color{black}We demonstrate this limitation in Section~\ref{hybrid}, and conclude that clearly, these models cannot be relied upon for prediction or imputation with MAR data\color{black}.

Second, Gaussian copula models only specify linear associations on the latent scale. As such, they cannot capture complex and nonlinear dependencies and interactions, which we demonstrate empirically in Section~\ref{sim-1}. Gaussian mixture copulas \citep{tewari2011parametric, rajan2016dependency} offer some additional distributional flexibility, but are highly parameterized, less robust than rank-based methods, and limited to certain data types.

To resolve these limitations, we develop a novel  Bayesian mixture copula model for joint and nonparametric modeling and imputation of count, continuous, ordinal, and unordered categorical variables. The model features a rank-based likelihood paired with a latent mixture of factor models that is designed to provide robust, parsimonious, and flexible characterization of complex dependencies among mixed data types. A primary innovation in this work is the introduction and theoretical justification for the \emph{margin adjustment}, which eliminates the reliance on the ECDF in the posterior predictive distribution of rank-based copula models. The margin adjustment features several key properties:
\begin{enumerate}
    \item It requires no specification of any marginal models, no additional assumptions, and no additional parameters;
    \item It delivers posterior consistency and posterior uncertainty quantification for each marginal distribution, even under MAR;
    \item It is computationally scalable and empirically accurate for estimation and imputation.
\end{enumerate}
The importance of these features is highlighted using both simulated and real data, which decisively show that the proposed imputation strategy offers significant improvements over competing methods, especially in the presence of data \color{black} MAR \color{black} and nonlinear dependencies.

This paper is organized as follows. Section~\ref{copmodels} introduces Bayesian copula models for mixed data types. In Section~\ref{margadjust}, we define and study the margin adjustment. Section~\ref{GMC} describes our novel Gaussian mixture copula, with extensions in Section \ref{cat} for unordered categorical variables.  We apply our proposed approach in Section~\ref{sim} with two simulation studies and a real data example in Section~\ref{realdat}. We conclude in Section~\ref{conc}. Supplementary material includes proofs of all results, details on the computations, additional simulation results, and an \texttt{R} package that implements the proposed approach.

\vspace{-5mm}
\section{Bayesian Copula Models for Mixed Data Types}\label{copmodels}

\subsection{The Gaussian Copula}
 Our first objective is to develop a Bayesian model for multivariate and mixed data. This model will be used to generate posterior predictive draws of the missing data, thereby allowing estimation and uncertainty quantification of  arbitrary $Q$ through imputation. Consider the Gaussian copula, which models the $p$-dimensional random vector $\boldsymbol Y = (Y_{1}, \dots, Y_{p})$ using 
\begin{align}
    \label{lat-z}
    \boldsymbol{z}  &\sim N_{p}(\boldsymbol 0, \boldsymbol{C}_{\boldsymbol \theta}), \quad \boldsymbol{z} = (z_1,\ldots,z_p)'\\
    \label{PIT}
    y_{j} &= F_{j}^{-1}\{\Phi(z_{j})\}, \quad j=1,\ldots, p.
\end{align}
The Gaussian copula links the univariate marginal distributions $\{F_{j}\}_{j=1}^{p}$ for each component of $\boldsymbol Y$ with a multivariate model for latent Gaussian data $\boldsymbol Z$ governed by correlation matrix $\boldsymbol C_{\boldsymbol \theta}$, which is indexed by parameters $\boldsymbol \theta$. Thus, each $F_{j}$ describes the marginal features of $Y_j$ while $\boldsymbol C_{\boldsymbol \theta}$ encodes the dependencies among $\boldsymbol Y$. Model \eqref{lat-z}-\eqref{PIT} implies the joint CDF for $\boldsymbol Y$ is $F(y_{1}, \dots, y_{p})= \Phi_{p}[\Phi^{-1}\{F_{1}(y_{1})\}, \dots \Phi^{-1}\{F_{p}(y_{p})\}]$, where $\Phi_p$ is the CDF of a $p$-dimensional Gaussian random vector with mean zero and correlation matrix $\boldsymbol{C}_{\boldsymbol \theta}$ and $\Phi$ is the univariate standard normal CDF.

Bayesian inference for the Gaussian copula requires prior distributions for the unknown $\boldsymbol \theta$ and $\{F_j\}$. Given posterior samples of $\boldsymbol \theta$ and $\{F_j\}$, posterior predictive simulations for the missing data are generated by drawing from \eqref{lat-z}-\eqref{PIT}, i.e., simulating $\tilde{\boldsymbol Z}_i \sim N_p(\boldsymbol 0, \boldsymbol C_{\theta})$ and setting $\tilde{y}^{mis}_{ij} = F_{j}^{-1}\{\Phi(\tilde{z}_{ij})\}$ for each missing component $j$ in observation $i$. This algorithm highlights the mutual importance of the copula correlation $\boldsymbol C_{\boldsymbol \theta}$ and the margins $\{F_{j}\}$, which we explore and generalize in subsequent sections.

The Gaussian copula model has several critical limitations. First, the margins $F_{j}$ must be specified either parametrically, which is restrictive and time-consuming when $p$ is moderate or large, or nonparametrically, which significantly increases the computational burdens. Second, the correlation matrix $\boldsymbol{C}_{\theta}$ captures only latent linear dependencies, and thus may not be suitable for more complex relationships. Lastly, the link \eqref{PIT} is not well-defined for unordered categorical variables, and estimation of the Gaussian copula with discrete $Y_{j}$ is problematic \citep{hoff2007extending}.   Thus, modifications are needed to provide valid joint models for mixed data types, with particular focus on modeling flexibility and computational scalability.

\subsection{Semiparametric Copula Models for Mixed Data}\label{RPLGC}
One approach that bypasses the need to specify individual marginal models is the extended rank-likelihood (RL; \cite{hoff2007extending}). Let $\boldsymbol{Y} = \{\boldsymbol{y}_{i}\}_{i=1}^{n}$ with $\boldsymbol{y}_i = (y_{i1},\ldots, y_{ip})'$ contain $p$ numeric (continuous, count or ordinal) variables; modifications for unordered categorical variables are discussed in Section \ref{cat}. The non-decreasing link in \eqref{PIT} implies a partial ordering on the latent scale for \eqref{lat-z}: for variable $j$ and observations $i$ and $k$ we know that $y_{ij}< y_{kj} \implies z_{ij} <z_{kj}$. This ordering is preserved for each variable in the dataset when  the event $\mathcal{D}(\boldsymbol Y) \coloneqq \{ \boldsymbol{Z} \in \mathbb{R}^{n \times p}: \max\{z_{kj}: y_{kj}< y_{ij}\} <z_{ij}< \min\{z_{kj}: y_{ij}< y_{kj}\}\}$ occurs.

The RL is derived by first expressing the Gaussian copula likelihood in terms of $\mathcal{D}(\boldsymbol Y) $:
    \begin{align}
    p(\boldsymbol{Y} \mid \boldsymbol{\theta}, \{F_{j}\}_{j=1}^p) &= p\{\boldsymbol{Y}, \boldsymbol{Z} \in \mathcal{D}(\boldsymbol{Y}) \mid \boldsymbol{\theta}, \{F_{j}\}_{j=1}^p\}\label{RPL} \\  
    &= p\{\boldsymbol{Z} \in \mathcal{D}(\boldsymbol{Y})\mid \boldsymbol{\theta}\} \  p\{\boldsymbol{Y} \mid  \boldsymbol{Z} \in \mathcal{D}(\boldsymbol{Y}), \boldsymbol{\theta}, \{F_{j}\}_{j=1}^{p}\}. \label{decomp}
    \end{align}
The decomposition~\eqref{RPL}-\eqref{decomp} is made possible since the event $\boldsymbol Z \in \mathcal{D}(\boldsymbol Y)$ does not depend on the marginal distributions $\{F_{j}\}_{j=1}^{p}$ and must occur with observation of $\boldsymbol Y$. \cite{hoff2007extending} argued that the left term in~\eqref{decomp} should contain most of the information about $\boldsymbol \theta$, and \cite{murray2013bayesian} showed that it is indeed sufficient for posterior consistency for $\boldsymbol\theta$.  Thus, the RL enables semiparametric inference on $\boldsymbol \theta$ (and  $\boldsymbol C_{\boldsymbol\theta}$) by targeting the posterior \begin{equation}\label{RL-post}
p\{\boldsymbol \theta \mid \boldsymbol Z \in \mathcal{D}(\boldsymbol Y)\} \propto p\{\boldsymbol Z \in \mathcal{D}(\boldsymbol Y) \mid \boldsymbol{\theta}\} \ p(\boldsymbol \theta).\end{equation}
Notably, \eqref{RL-post} does not include  the marginal CDFs $\{F_{j}\}$. When priority lies in inference for  $\boldsymbol{C}_{\boldsymbol
 \theta}$, which contains substantive information on multivariate relationships in the data, this artifact is particularly convenient for model specification and computation (see Algorithm~\ref{alg1}). However, consideration of $\{F_{j}\}$ is necessary for the posterior predictive distribution---and thus for missing data imputation. We address this challenge in Section~\ref{margadjust}.

\subsection{Estimating Semiparametric Copulas with Missing Data}\label{estim}

 With missing data, we only have access to the observed ranks. Thus, we modify the RL posterior appropriately:
\begin{equation}\label{obsRL}
    p\{\boldsymbol \theta \mid \boldsymbol Z^{obs} \in \mathcal{D}(\boldsymbol Y^{obs})\} \propto   \int p\{\boldsymbol Z^{obs} \in \mathcal{D}(\boldsymbol Y^{obs}), \boldsymbol Z^{mis} \mid \boldsymbol{\theta}\}d \boldsymbol Z^{mis} \ p(\boldsymbol \theta)
\end{equation}
where $\boldsymbol{Z}^{obs} = (Z_{ij}: R_{ij} = 0)$  and $\boldsymbol{Z}^{mis} = (Z_{ij} : R_{ij} = 1)$. Though \eqref{obsRL} non-standard, it is relatively simple to construct a Gibbs sampling algorithm to sample from this distribution. The sampler (Algorithm~\ref{alg1}) alternates between drawing from 
 $[(\boldsymbol Z^{obs}, \boldsymbol Z^{mis}) \mid \boldsymbol Y^{obs}, \boldsymbol \theta]$ and  $[\boldsymbol \theta \mid \boldsymbol Y^{obs}, \boldsymbol Z]$. The first step features univariate truncated normal draws for each $ Z_{ij}^{obs}$ based on~\eqref{lat-z} and the RL, while prediction of $\tilde{Z}_{ij}^{mis}$ is unrestricted by any observed ordering constraints. Next, $[\boldsymbol \theta \mid \boldsymbol Y^{obs}, \boldsymbol Z] = [\boldsymbol \theta \mid\boldsymbol Z]$ is drawn from a posterior which features a multivariate Gaussian likelihood, and thus sampling is straightforward for many choices of priors $p(\boldsymbol \theta)$. 

\begin{algorithm}[h]
\caption{Bayesian RL Gaussian Copula Gibbs Sampler with Missing Data}
\begin{algorithmic}\label{alg1}
     \STATE \textbf{Require:}  prior $p(\boldsymbol \theta)$ 
     \begin{itemize}
         \item  \textbf{Step 1}: Sample $(\boldsymbol Z^{obs}, \boldsymbol Z^{mis}) \mid \boldsymbol \theta $
     \end{itemize}
       \bindent
       
            \FOR{$ Z_{ij} \in \boldsymbol Z^{obs}$}
                \STATE Compute $z_{\ell} = \max\{z^{obs}_{\ell j} : y^{obs}_{\ell j}< y_{ij}\}$ and $z_{u} = \min\{z^{obs}_{\ell j} :y_{ij}< y^{obs}_{\ell j}\},\ell \neq i$
                \STATE Sample $ Z_{ij} \sim \mbox{Normal}(\mu_{ij}, \sigma_{j}^{2}) \mathbbm{1}(z_{\ell},z_{u})$
                \ENDFOR
            \FOR{$ Z_{ij} \in \boldsymbol Z^{mis}$}
                \STATE Sample $\tilde{Z}_{ij} \sim \mbox{Normal}(\mu_{ij}, \sigma_{j}^{2})$
                \ENDFOR
                    
               

                
     \STATE where  $\mu_{ij} = (\boldsymbol C_{\boldsymbol\theta})_{j-j} (\boldsymbol C_{\boldsymbol \theta}^{-1})_{-j-j}\boldsymbol Z_{i- j}$, $\sigma_{j}^{2} = (\boldsymbol C_{\boldsymbol{\theta}})_{jj}- (\boldsymbol C_{\boldsymbol \theta})_{j-j}(\boldsymbol  C_{\boldsymbol \theta}^{-1})_{-j-j}(\boldsymbol C_{\boldsymbol \theta})_{-jj}$
            
     \eindent

          \begin{itemize}
         \item  \textbf{Step 2}: Sample $\boldsymbol \theta \sim p\{\boldsymbol \theta \mid \boldsymbol{Z}^{obs}, \boldsymbol Z^{mis}, \boldsymbol Y^{obs}\} = p(\boldsymbol \theta \mid \boldsymbol{Z}^{obs}, \boldsymbol Z^{mis}) $
              \end{itemize}
              \bindent
         \STATE where $p\{\boldsymbol \theta \mid \boldsymbol{Z}^{obs}, \boldsymbol Z^{mis}\} \propto N_{p}\{(\boldsymbol Z^{obs}, \boldsymbol Z^{mis}); \boldsymbol 0, \boldsymbol C_{\boldsymbol \theta}\} p (\boldsymbol \theta) $
         \eindent
\end{algorithmic}
\end{algorithm}

A remarkable feature of Algorithm \ref{alg1} is the absence of any marginals $\{F_{j}\}$. While this is advantageous for posterior inference on $\boldsymbol \theta$ --- it removes the need to specify or estimate any marginal distributions --- the margins are in fact necessary for prediction and imputation. The default semiparametric procedure is to fix each $\{F_{j}\}$ at the ECDF \citep{hoff2007extending, murray2013bayesian, cui2019novel, feldman2022bayesian}, and the accompanying imputation step would apply Algorithm \ref{alg1} and compute $\hat{F}^{-1}_{j}\{\Phi(\tilde z_{ij}^{mis})\}$. However, the ECDF does not account for the uncertainty about each $F_j$. More critically, the ECDF is  at risk for significant bias under MAR (see Figure \ref{examp} and Sections \ref{sim}-\ref{realdat}), which will lead to inaccurate predictions and imputations --- even if $\boldsymbol \theta$ is inferred correctly.

\section{The Margin Adjustment}\label{margadjust}
To eliminate reliance on the ECDFs for posterior predictive sampling and imputation---while still maintaining the beneficial structure of the Bayesian RL copula model---we propose a new strategy called the \emph{margin adjustment}. The margin adjustment does not require any additional modeling assumptions or parameters, is fully automated (i.e., individual specification of marginal models for each $F_j$ is not needed), and 
provides computationally efficient and consistent posterior inference for the margins $\{F_j\}$,  even in the presence of data MAR. The margin adjustment does not impact posterior inference for $\boldsymbol \theta$, and thus \color{black} Algorithm~\ref{alg1} is unchanged.

\subsection{Derivation and Theory}\label{MA-derive}
 
The key insight of the margin adjustment is that the combination of the RL rank constraints \eqref{RPL}-\eqref{decomp} and the latent data model  \eqref{lat-z} are sufficient to infer the marginal distributions $\{F_j\}_{j=1}^{p}$ with strong theoretical guarantees. 
Under the RL, $Z_j$ is a non-decreasing (and unknown) transformation of $Y_j$ for each $j$. Thus, upon ordering both $\{Z_{ij}\}_{i=1}^{n}$ and $\{Y_{ij}\}_{i=1}^{n}$, the position of $\max \{ Z_{ij}:  Y_{ij} \le x\}$ among $\{Z_{ij}\}_{i=1}^n$ will be identical to the position of $\max\{Y_{ij}:  Y_{ij} \leq x\}$ among $\{ Y_{ij}\}_{i=1}^n$ for any $x$ greater than the minimum of $\{Y_{ij}\}_{i=1}^{n}$. For any $x$ below this value, the set $\{ Y_{ij} \le x\}$ will be empty with probability 1. Thus, we define 
\begin{equation}\label{cutpoint}
    Z_{j}^{n}(x) = \max[\{ Z_{ij}:  Y_{ij} \leq x\} \cup \{ Z_{ij}: Y_{ij} = \min( \{Y_{ij}\}_{i=1}^{n})\}, i \in \{1,\dots,n\}].
\end{equation}
Informally, if $F_{j}(x) = \tau$, then $Z_{j}^{n}(x)$  will approximate the $\tau$th quantile under the marginal latent data model. This motivates the following marginal distribution estimator:
\begin{equation}\label{marginadjust}
    \tilde{F}_{j}(x) =  G_{j}\{Z_{j}^{n}(x)\},
\end{equation}
where $G_{j}$ is the marginal distribution for $Z_{j}$ induced by the latent data model under the copula. Importantly, the margin adjustment is compatible with \emph{any} rank-based copula model. All that is required is the multivariate model for $\boldsymbol Z$, which induces marginals $G_{j}$.  Under the Gaussian copula \eqref{lat-z}-\eqref{PIT}, $G_{j}$ is the standard normal CDF $\Phi$; modifications for the Gaussian mixture copula are available in Section \ref{GMC}.

More formally, Theorem~\ref{ma-cdf2} provides a general setting in which a continuous random variable $Z$ may be used to infer the distribution function of $Y = h(Z)$ with almost sure convergence, where $h$ is any monotone increasing function. Our primary example is the RL posterior, where $h$ ensures the necessary ordering across realizations of $(Z,Y)$. 
\begin{theorem}\label{ma-cdf2}
    Suppose $\{Z_i\}_{i=1}^n \overset{i.i.d}{\sim} F_Z$  and $\{Y_i\}_{i=1}^n  = \{h(Z_{i})\}_{i=1}^{n} \sim F_{Y}$ , where $F_Z$ is continuous and $h$ is a monotone increasing function. Defining $Z^{n}(x)$ as \eqref{cutpoint}, the margin adjustment satisfies 
$\tilde F(x) \coloneqq F_Z\{Z^{n}(x) \} \stackrel{a.s.}{\to}F_Y(x)$ for all $x \in \mathbb{R}$.
\end{theorem}
The more challenging setting occurs when data are MAR. In the presence of missing data, we modify \eqref{cutpoint} appropriately:
\begin{align}\label{cutpointobs}
   Z_{j}^{n}(x) =\max[\{Z^{obs}_{ij}: Y^{obs}_i \le x\} &\cup \{Z^{obs}_{ij} :Y^{obs}_{ij} = \min(Y^{obs}_{ij})\}, i \in \{1,\dots, n\}].
\end{align}
Crucially, with this modification, the margin adjustment remains consistent under MAR. For simplicity, we demonstrate our result for $p=2$ variables, one of which is MAR.  


\begin{theorem}\label{consistMA-misv2}
Suppose $\{\boldsymbol{Z}_{i}\}_{i=1}^{n}=\{(Z_{i1},Z_{i2})\}_{i=1}^{n}\overset{i.i.d}{\sim} G$ where $G$ is continuous with marginal distributions $G_{1},G_{2}$, and $\{\boldsymbol{Y}_{i}\}_{i=1}^{n}=\{(Y_{i1},Y_{i2})\}_{i=1}^{n} =[(F_1^{-1}\{G_1(Z_{i1})\}, F_2^{-1}\{G_2(Z_{i2})\})]_{i=1}^{n}$ has joint distribution function $F$ with marginal distributions $F_1,F_2$. Suppose that $Y_2$ is completely observed and $Y_{1}$ is MAR.  Define $Z_{1}^{n}(x) \coloneqq \max[\{Z_{i1}: Y^{obs}_i \le x\} \cup \{Z_{i1} :Y^{obs}_{i1} = \min(Y^{obs}_{i1})\}], i \in \{1,\dots, n\}$. Then the margin adjustment satisfies
$\tilde{F}_{1}(x) \coloneqq G_{1}\{Z_{1}^{n}(x)\} \overset{a.s.}{\to}F_{1}(x)$
for all $x \in \mathbb{R}$. 
\end{theorem}
Generalizations beyond the bivariate case are straightforward. The theorem also applies for discrete $Y_j$, where $F_j^{-1}$ maps quantile intervals defined by the left and right limits of the step function $F_{j}$ to elements in the support of $Y_{j}$. Notably, the consistency in Theorem~\ref{consistMA-misv2} is \emph{not} valid for the ECDF under MAR
 (see Figure~\ref{examp}), which undermines any methods that rely on the ECDF for prediction and imputation \citep{hoff2007extending, murray2013bayesian, cui2019novel, feldman2022bayesian}.

\subsection{Bayesian Estimation and Imputation}
The margin adjustment \eqref{marginadjust} is a function of the latent data $\boldsymbol Z$, and thus inherits a posterior distribution under the Bayesian RL copula model. Under Algorithm \ref{alg1}, $\boldsymbol Z^{obs}$ is sampled from its joint posterior, and so the margin adjustment may be seamlessly integrated in any Bayesian RL copula model to provide margin estimation and uncertainty quantification with minimal computational cost.  For clarity, we outline the procedure to obtain posterior samples of the margin adjustment for each $j \in \{1,\dots, p\}$ under the RL Gaussian copula in Algorithm~\ref{alg2}; modifications for the Gaussian mixture copula are available in Section \ref{GMC}. 

\begin{algorithm}[h]
\caption{The Margin Adjustment Sampling under the Bayesian RL Gaussian Copula}
\begin{algorithmic}\label{alg2}
     \STATE \textbf{Require:}  One posterior sample of $\boldsymbol Z^{obs}$ from Algorithm~\ref{alg1}
     \STATE \textbf{Return:} One posterior sample of $\tilde{F}_{j}(x)$, $j \in \{1,\dots, p\}$
      \FOR{$j \in \{1,\dots, p\}$ and any $x$}
            \STATE Compute  $Z_{j}^{n}(x)$ as \eqref{cutpointobs}
            \STATE Compute $\tilde{F}_{j}(x) = \Phi\{Z_{j}^{n}(x)\}$

                \ENDFOR

\end{algorithmic}
\end{algorithm}
In practice, we compute the margin adjustment for each unique $x \in \{Y_{ij}^{obs}\}_{i=1}^{n}$. Since the resulting $\tilde F_j$ is a step function with jumps determined by these observed values, we then fit a monotone interpolating spline to $\{x, \tilde F_j(x)\}$ with pre-specified upper and lower bounds. The interpolating spline preserves $\tilde F_j$ at the observed data values but expands the support of the data-generating process beyond only those observed values, which is important for imputation of count and continuous variables. 

Most important, we apply the margin adjustment to deliver model-based imputation. We demonstrate this using the RL Gaussian copula in Algorithm \ref{alg3}, but once again this algorithm may be generalized to any RL copula model. 

\begin{algorithm}[H]
\caption{Bayesian RL Gaussian Copula Imputation using the Margin Adjustment }
\begin{algorithmic}\label{alg3}
     \STATE \textbf{Require:}  One posterior sample of $\boldsymbol{\tilde{Z}}^{mis}$ from Algorithm~\ref{alg1} and $\{\tilde{F}_{j}\}_{j=1}^{p}$ from Algorithm~\ref{alg2}\vspace{-1.2em}
     \STATE \textbf{Return:} One completed data set $\boldsymbol Y = (\boldsymbol Y^{obs}, \boldsymbol Y^{mis})$
      \FOR{$ \tilde{z}_{ij}^{mis} \in \boldsymbol{\tilde{Z}}^{mis}$}
            \STATE Set $ \tilde{y}_{ij}^{mis} = \tilde{F}_{j}^{-1}\{\Phi( \tilde{z}_{ij}^{mis})\}$ 

                \ENDFOR

\end{algorithmic}
\end{algorithm}
 \noindent Thus, the margin adjustment replaces the default ECDF for imputation with an estimator that remains consistent in the presence of MAR. This benefit is explored empirically in Sections \ref{sim}-\ref{realdat}, and yields substantial improvements in prediction and imputation inference.

 \vspace{-1em}
\subsection{Strong Posterior Consistency under MAR}\label{post consist}
We now establish the asymptotic properties of the posterior distribution of the Gaussian copula correlation parameter $\boldsymbol C_{\boldsymbol\theta}$ under the RL with ignorable missing data, and demonstrate how this posterior consistency extends to the margin adjustment~\eqref{marginadjust}. First, we adapt the result in \cite{murray2013bayesian}, which established posterior consistency of $\boldsymbol C_{\boldsymbol \theta}$ under the RL without missingness. However, that proof relied upon the almost sure convergence of the ECDF $\hat F_j$ to $F_j$, which is not maintained under MAR. 


\begin{theorem}\label{consist_complete}
Suppose $\{\boldsymbol Y_i\}_{i=1}^n \stackrel{i.i.d}{\sim} G^{\infty}_{\boldsymbol C_{0}, F_{1},\dots,F_{p}}$ for true copula parameters $\boldsymbol C_0$ and true marginal CDFs $F_1,\ldots,F_p$. Let  $\Pi$ be a prior distribution on the space of all $p\times p$ positive semi-definite correlation matrices $\boldsymbol C_{\boldsymbol \theta}$ with corresponding density $\pi(\boldsymbol C_{\boldsymbol\theta})$ with respect to a measure $\nu$. Suppose $\pi(\boldsymbol C_{\boldsymbol{\theta}})>0$ almost everywhere with respect to $\nu$ and assume that the missigness is ignorable. Then, for $\boldsymbol C_{0}$ a.e.\! $[\nu]$ and any neighborhood $\mathcal{A}$ of $\boldsymbol C_{0}$, we have that 
     $\lim_{n\rightarrow \infty}\Pi\{\boldsymbol C_{\boldsymbol{\theta}} \in \mathcal{A} \mid \boldsymbol{Z}_{n}^{obs} \in \mathcal{D}(\boldsymbol{Y}_{n}^{obs})\} =  1 \ a.s \  [G^{\infty}_{\boldsymbol C_{0}, F_{1},\dots,F_{p}}]$.
\end{theorem}

In conjunction with Theorems~\ref{ma-cdf2}--\ref{consist_complete}, the strong posterior consistency of $\boldsymbol C_{\boldsymbol\theta}$ also yields posterior consistency for the margin adjustment.

\begin{corollary}\label{MA-postconsist}
Under the conditions of Theorem~\ref{consist_complete}, define $\tilde{F}_j$ as in \eqref{marginadjust} with $Z_{j}^{n}(x)$ as \eqref{cutpointobs} and $G_j = \Phi$ for each $j\in \{1,\ldots,p\}$. Then for any $x \in \mathbb{R}$ and any neighborhood $\mathcal{A}$ of $F_{j}(x)$ 
     $\lim_{n\rightarrow \infty}\Pi\{\tilde{F}_{j}(x) \in \mathcal{A} \mid \boldsymbol{Z}_{n}^{obs} \in \mathcal{D}(\boldsymbol{Y}_{n}^{obs})\} =  1 \ a.s \  [G^{\infty}_{\boldsymbol{C}_{0}, F_{1},\dots,F_{p}}]$.
\end{corollary}

\noindent These results are powerful: the RL Gaussian copula with the margin adjustment delivers fully Bayesian inference with strong posterior consistency for both the marginal distributions and the copula parameters. Notably, these results apply for mixed and MAR data.

\section{Gaussian Mixture Copulas via Latent Factors}\label{GMC}
Although we have established theoretical guarantees for the RL and the margin adjustment under a Gaussian copula model---including for mixed (count, continuous, ordinal) variables and missing data---the Gaussian copula  only captures linear associations on the latent scale through latent correlation matrix $\boldsymbol C_{\boldsymbol\theta}$. As such, it may not be sufficiently powerful to capture nonlinearities and interactions on the observed scale (see Section~\ref{sim}), which is vital to justify MAR and for imputation under complex dependencies. However, generalizations of the Gaussian copula must carefully consider computational scalability, model parsimony, and suitable adaptations of the margin adjustment.

 To build an imputation model capable of adapting to unanticipated features in data, we develop a novel Gaussian mixture copula (GMC). When paired with the RL and MA, the resulting copula model is fully nonparametric, and may be used to impute variables of arbitrary type. The GMC extends the Gaussian copula by replacing the latent data model \eqref{lat-z} with a finite mixture: 
\begin{equation}\label{finitemultmix}
    \boldsymbol{z}\sim \sum_{h=1}^{H}\pi_{h}N_p(\boldsymbol{\alpha}_{h},\boldsymbol{C}_{h})
\end{equation}
where the marginal distribution of the $j$th component is $z_{j} \sim \sum_{h=1}^{H}\pi_{h}N(\{\boldsymbol{\alpha}_{h}\}_{j},\{\boldsymbol{C}_{h}\}_{jj})$. The multivariate mixture on the latent data can be combined with the observed data marginals to define the GMC. For completeness, we show that is indeed a valid copula. 
\begin{theorem}\label{copdefine}
Let $\mathbb{C}_{GMC}(\boldsymbol u)= \Psi(\psi^{-1}_{1}\{F_{1}(y_{1})\}, \dots,\psi^{-1}_{p}\{F_{p}(y_{p})\})$, where $\Psi = \sum_{h=1}^{H} \pi_{h}\Phi_{p}(\boldsymbol{\alpha}_{h},\boldsymbol{C}_{h})$, $\psi_{j} = \sum_{h=1}^{H} \pi_{h}\Phi(\{\boldsymbol{\alpha}_{h}\}_{j},\{\boldsymbol{C}_{h}\}_{jj})$, and $\{F_{j}\}_{j=1}^{p}$ are the marginals of $\{Y_{j}\}_{j=1}^{p}$. Then
 $\mathbb{C}_{GMC}$ defines a valid copula.
\end{theorem}
\noindent The data generating representation of $\mathbb{C}_{GMC}$ simulates $\boldsymbol Z$ from~\eqref{finitemultmix} and links the realization to the observed scale via $y_{j} = F_{j}^{-1}\{\psi_{j}(z_{j})\}$\color{black}.

Although the GMC latent data model \eqref{finitemultmix} provides greater representational ability than the Gaussian copula \eqref{lat-z}, especially for nonlinearities and interactions, the GMC modeling and computational capabilities are limited in higher dimensions. In particular, the GMC is parameterized by $\boldsymbol \theta = \{\boldsymbol \pi_{h}, \boldsymbol \alpha_h, \boldsymbol C_h\}_{h=1}^H$, which contains many parameters when $p$ is moderate or large. Further, Gaussian mixture models tend to over-cluster when $p$ is large, which results in more clusters---and thus more parameters---than necessary.

Instead, we apply our mixture on lower-dimensional latent factors $\boldsymbol \eta \in \mathbb{R}^k$
 with $k \ll p$: 
 \begin{equation}\label{factormix}
     \boldsymbol{\eta}_{i} \sim \sum_{h =1}^{H}\pi_{h}N_k(\boldsymbol \mu_{h}, \boldsymbol \Delta_{h}), \quad \boldsymbol z_i \mid \boldsymbol \eta_{i} \sim N_p( \boldsymbol \Lambda \boldsymbol \eta_{i}, \boldsymbol\Sigma)
 \end{equation}
where $\boldsymbol{\Sigma} = \mbox{diag}(\sigma_{1}^{2}, \dots, \sigma_{p}^{2})$,   $\boldsymbol{\Lambda}$ is a $p\times k$ dimensional matrix of factor loadings, and $\boldsymbol{\eta}_{i}$ is a $k$-dimensional vector of latent factors. The latent factor mixture model \eqref{factormix} induces a mixture model \eqref{finitemultmix}  for $\boldsymbol Z$ through marginalization over $\boldsymbol \eta$, and specifically with    $\boldsymbol \alpha_h = \boldsymbol \Lambda \boldsymbol \mu_h$ and $ \boldsymbol C_h = \boldsymbol \Lambda \boldsymbol \Delta_{h} \boldsymbol\Lambda' + \boldsymbol\Sigma$ for $h=1,\ldots,H$. Thus, the latent data $\boldsymbol Z$ are still endowed with a flexible mixture model, but the clustering is directed to a lower-dimensional space. This feature offers important benefits relative to existing GMCs \citep{tewari2011parametric, rajan2016dependency}, namely, that it affords parsimonious estimation of the dependence structure among moderate to high-dimensional data.  \cite{chandra2020escaping} recently applied this strategy for clustering of continuous data and demonstrated how it alleviates the curse of dimensionality in model-based clustering---i.e., as $p$ grows, the number of nonempty clusters trivially tends toward $n$---but this approach has not been deployed for copula models or mixed data types. 

The remaining challenge lies in Bayesian modeling of the finite mixture on $\boldsymbol{\eta}$. In practice, the number of latent clusters $H$ will be unknown and should be determined based on the data. Thus, we propose a Dirichlet process (DP) to allow $H \to \infty$, and use a stick-breaking process for the mixing weights $\{\pi_h\}$ \citep{ishwaran2001gibbs}: $\pi_{h} = V_{h}\prod_{l<h}(1-V_{l})$ with $V_{\ell}\stackrel{i.i.d}{\sim} \mbox{Beta}(1,\alpha_{\pi})$
and $\alpha_{\pi} \sim \mbox{Gamma}(a_{\alpha}, b_{\alpha})$. For computational convenience, we implement a truncated DP \citep{ishwaran2002approximate} that resembles \eqref{factormix}, where now $H$ is a conservative upper bound. The component-wise mean and covariance are assigned a Normal-Inverse Wishart prior, $(\boldsymbol \mu_{h}, \boldsymbol \Delta_{h}) \sim \mbox{NIW}(\boldsymbol \mu_{0}, \delta^{2}\boldsymbol I_{k},\kappa_{0}, \nu_{0})$, while the diagonal elements of $\boldsymbol \Sigma$ are assigned $\sigma_j^{-2}\sim \mbox{Gamma}(a_\sigma, b_\sigma)$. Lastly, we apply a global-local shrinkage prior for the loadings matrix $\boldsymbol{\Lambda} = \{\lambda_{jt}\}$ that encourages columnwise shrinkage for rank selection, which reduces sensitivity to the choice of $k$ \citep{bhattacharya2011sparse}. 

 Our approach for Bayesian inference, prediction and imputation combines the GMC with the RL and margin adjustment (GMC-MA). Despite the complexity of the GMC-MA, only minor modifications are needed for Algorithms \ref{alg1}-\ref{alg3}. A particular convenience comes from~\eqref{factormix}: conditional on $\boldsymbol \theta$ under the mixture of factor models, $Z_{ij} \sim N(\sum_{t=1}^{k} \lambda_{jt} \eta_{it},  \sigma_{j}^{2})$, which implies the components of $\boldsymbol Z_i$ are independent univariate Gaussian. Therefore, even in the presence of missing data, the core elements of  Algorithm \ref{alg1} are the same: each $ Z_{ij}^{obs}$ is sampled from a truncated Gaussian, the posterior of $ \tilde{Z}_{ij}^{mis}$ is once again unrestricted, and sampling of $\boldsymbol \theta$ involves standard steps for Gaussian factor and mixture models; details are provided in the supplementary material. For the margin adjustment, we simply modify Algorithm \ref{alg2} to use  $G_{j} = \psi_{j}$. Finally, imputations are similarly generated by modifying Algorithm \ref{alg3} to use $\psi_{j}$ in place of $\Phi$. 
 
\section{Extensions for Unordered Categorical Variables}\label{cat}

We incorporate unordered categorical variables via the extended rank-probit likelihood (RPL) \citep{feldman2022bayesian}, which generalizes the RL. Suppose $\boldsymbol Y= \boldsymbol{Y}^{r} \cup \boldsymbol{Y}^{q} $, where $r$ indexes the numeric variables, $q$ indexes the unordered categorical variables and $p = r + q$.
For each categorical variable $\boldsymbol Y_c$ with $k_c$ levels, the RPL encodes a vector of $k_c$ binary variables  $\boldsymbol
\gamma_{c}$ with the corresponding latent data restriction $\{\gamma_{ic_{m}} = 1 \cap \gamma_{ic_{l}} = 0, l \neq m\} \implies  \{z_{ic_{m}} >0 \cap z_{ic_{l}} < 0, l \neq m\}$, i.e., $y_{ic} = m$ implies that only the $m$th component is positive and the others are all negative. This representation avoids the need to select reference groups. Aggregating this representation across all $q$ unordered categorical variables, the observed categorical memberships must satisfy the event $\mathcal{D}'(\boldsymbol Y^{q}) \coloneqq \cup_{c=1}^{q}
     \{\boldsymbol{Z}^{n \times k_{c}}:  \gamma_{ij} = 1 \implies z_{ij}> 0 \cap \{z_{i\ell}<0\}_{\ell \ne j}\}$.
 This representation is also recommended for ordinal variables with few levels \citep{feldman2022bayesian}.

 The RPL joins the rank event on the $r$ numeric variables with the probit-style representation of the $q$ categorical variables to define $\mathcal{E}(\boldsymbol Y) = \mathcal{D}(\boldsymbol Y^{r}) \cup \mathcal{D}'(\boldsymbol Y^{q})$,  which substitutes for $\mathcal{D}(\boldsymbol Y)$ in \eqref{RPL}-\eqref{decomp} for joint modeling of continuous, count, ordinal and now unordered categorical variables.
 Imputation for unordered categorical variables is carried out by estimating the multinomial probabilities of each level. These quantities are easily computed with posterior samples of $\boldsymbol \theta$ and the latent data model, as the probability that an observation assumes a particular level is given by the probability that the corresponding latent component is positive and all others are negative; the supplement provides the Gibbs sampling and imputation algorithm for the GMC-MA under the RPL.
\section{Simulation Studies}\label{sim}

\subsection{Mixed Data, Nonlinearity, and MAR}\label{sim-1}

In the first simulation study, we evaluate (i) the impact of MAR on marginal distribution estimation via the ECDF---and show how the margin adjustment corrects the resulting biases---and (ii) assess whether the proposed GMC-MA is capable of accurate imputation under nonlinear dependencies. We generate mixed datasets with nonlinear dependencies by simulating $Y_1 \sim N(0,1)$ , $Y_{2} \mid Y_{1} = y_{1} \sim \mbox{Poisson}(5 \vert y_{1}\vert)$ , and $Y_3  \mid Y_{2} = y_{2}, Y_{1} = y_{1} \sim \mbox{Bernoulli}\{\Phi(-0.5+ y_{2_{scale}})\}$ 
for  $n \in \{500,1000,2000\}$, where  $y_{2_{scale}}$ is the centered and scaled version of $y_2$. Next, we introduce missingness using the MAR mechanism $    R_j \mid Y_1 = y_1 \sim \mbox{Bernoulli}\{\Phi(-0.5 + \beta \lvert y_1 \rvert)\}$ for $j = 2,3,$ which links the missingness in both $Y_2$ and $Y_3$ with the observed value of $Y_1$.
The parameter $\beta$ determines both the amount of missingness and the impact of $Y_1$ on the missingness for each of $Y_2$ and $Y_3$. We consider $\beta \in \{0.5,1\}$, with the lower value resulting in approximately 30\% complete cases and 50\% of each variable missing, and the higher value yielding approximately 20\% complete cases with 60\% marginal missingness. 

\begin{figure}[h]
    \centering
    \includegraphics[width = .24\textwidth, keepaspectratio]{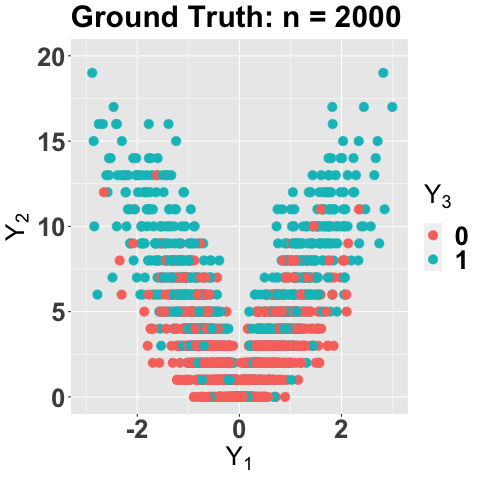}
    \includegraphics[width = .24\textwidth, keepaspectratio]{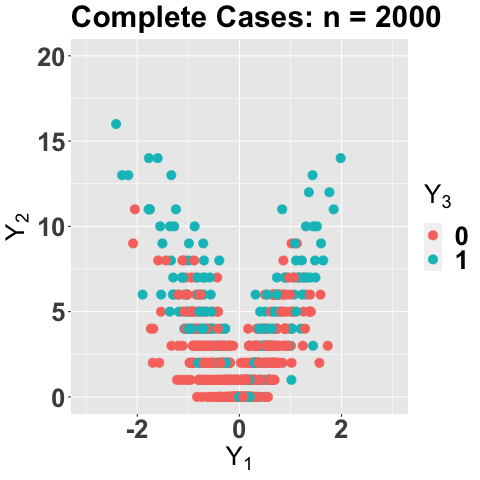}
    \includegraphics[width = .24\textwidth, keepaspectratio]{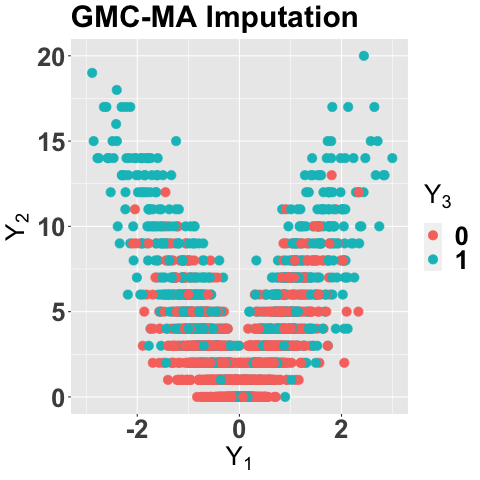}
    \includegraphics[width = .24\textwidth, keepaspectratio]{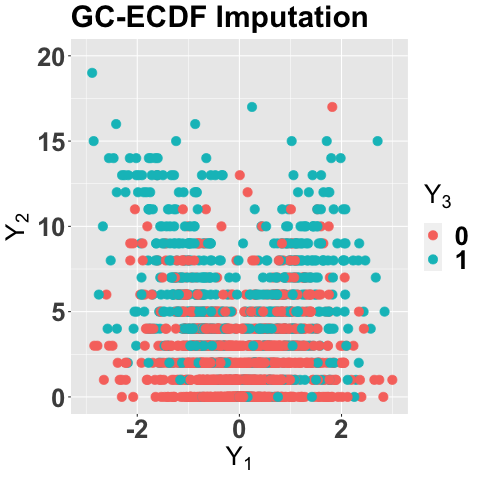}
    \caption{\small A simulated dataset without missingness (left) and the complete cases after applying the MAR mechanism (left-middle) with $\beta = 0.5$. The proposed approach (right-middle) is significantly better than the Gaussian copula (right) at capturing the challenging nonlinear relationship between $Y_1$ and $Y_2$ and correctly imputing additional $Y_3 = 1$ values (blue) when $\vert Y_1\vert$ is large.}
    \label{imp}
\end{figure}

We highlight the challenging nonlinearities and missingness under this data-generating mechanism ($n=2000, \beta = 0.5$) with a single simulated dataset in Figure~\ref{imp}. Compared to the full dataset, the complete cases  omit larger values of $Y_2$ and many instances of $Y_3 = 1$. To visualize the comparative imputation methods, we provide a single imputed dataset from the proposed GMC-MA and compare it to \cite{hoff2007extending} using the \verb|sbgcop| package in \verb|R|, which uses a single component Gaussian copula  with the ECDF for posterior predictive simulations (similar results are obtained for additional realizations in the supplement). Clearly, the GMC-MA offers significant improvements in detecting non-linearity: it is capable of capturing the complex relationship between $Y_1$ and $Y_2$ and correctly imputing additional $Y_3=1$ values when $\vert Y_1 \vert$ is large. 
We emphasize that GMC-MA does not leverage any aspect of the true data-generating process beyond the variable data type.

\begin{figure}[h]
    \centering
\includegraphics[width =.32\textwidth, keepaspectratio]{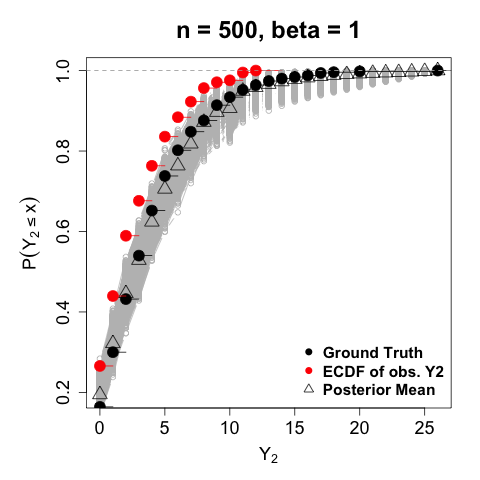}
\includegraphics[width =.32\textwidth, keepaspectratio]{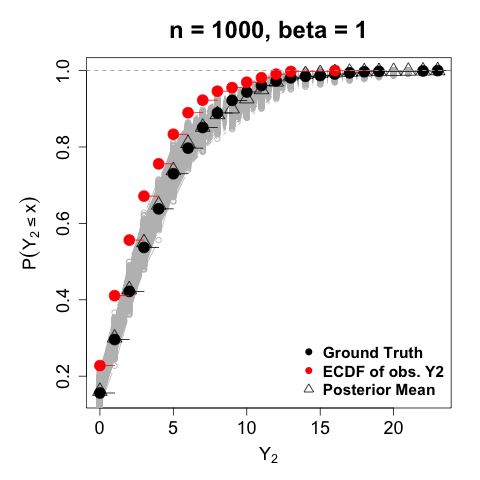}
\includegraphics[width =.32\textwidth, keepaspectratio]{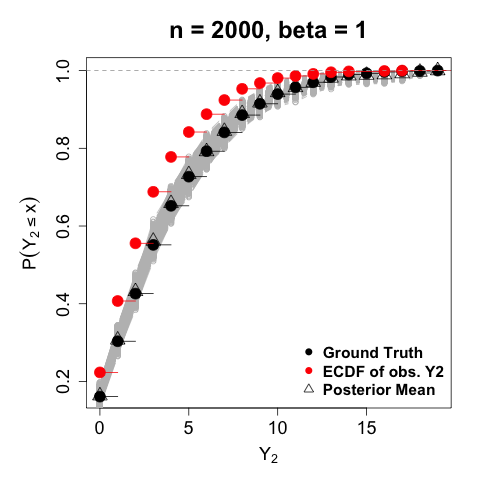}
    \caption{\small Estimation and inference for the marginal distribution of $Y_2$ under MAR ($\beta = 1$) with varying $n$. The ECDF of $Y_2^{obs}$ (red points) deviates significantly from the ECDF of $Y_2$ prior to removing missingness (black points). The posterior draws (gray lines) and posterior mean (triangles) from the margin adjustment show that the proposed approach is highly accurate, even under severe MAR. Similar results are observed for $\beta = 0.5$ in the supplement.}
    \label{SS1}
\end{figure}

Next, we evaluate the margin adjustment, and specifically seek to assess whether it corrects the biases of the ECDF in the presence of data MAR. In Figure~\ref{SS1}, we focus on the marginal distribution for $Y_2$, which is a count variable subject to MAR. We compute the ECDF of $Y_2$ prior to removing missingness, which we treat as the ground truth (black points);  the ECDF computed on the observed data $ Y_2^{obs}$ (red points); and  posterior draws (gray lines) and the posterior expectation (triangles) based on the  GMC-MA. Posterior inference uses the estimators described in Section~\ref{GMC} and the Gibbs sampler from the supplement, which we run for 5,000 iterations. Trace plots of the draws of the marginal distribution functions for $Y_{1}$ and $Y_{2}$ indicate that the MCMC algorithm converges after about 1,500 samples, which we discard as a burn-in. We display the results for the more challenging missingness setting ($\beta = 1$); similar results are observed for $\beta = 0.5$ in the supplement.

 Most notably, the ECDF on the observed data is badly biased, and this flawed estimator would be used for imputation under default semiparametric (rank-based) copula models \citep{hoff2007extending, murray2013bayesian,cui2019novel, feldman2022bayesian}. By comparison, the posterior distribution for the margin adjustment concentrates quickly around the ground truth as $n$ grows, while the point estimates of the marginal distribution are highly accurate. These results suggest that Corollary~\ref{MA-postconsist} may be applicable more broadly, including for GMCs. Finally, we note that the interpolation strategy for the marginal distribution is effective: several values in the support of $Y_{2}$ are unobserved in $Y_2^{obs}$, yet the margin adjustment remains accurate for these cumulative probabilities. Analogous results for binary $Y_3$ are presented in the supplement, and demonstrate the exceptional performance of the margin adjustment

\subsection{Imputation for Regression Analysis}\label{hybrid}

In the second simulation study, we study the impacts of imputation within the broader context of a regression analysis, and include comparisons with Bayesian nonparametric models and popular non-Bayesian alternatives for multiple imputation. To incorporate the challenges of real-world data analysis while maintaining partial control over the data-generating process, we use hybrid synthetic data. First, we select three variables from the 2011-2012 NHANES data (Table~\ref{tab:data}): a categorical variable (\verb|Family Income| \ (\verb|FI|)), a count variable (\verb|Age|), and a continuous variable (\verb|BMI|). Both \verb|Age| and \verb|BMI| are centered and scaled, while \verb|FI| has three levels: \verb|Low|, \verb|Middle|, and \verb|High|. Next, for each of the $n = 2434$ complete NHANES observations, we generate a continuous response variable $\vtt{New}$ using a Gaussian linear model  with an \verb|FI:BMI| interaction, $\vtt{New}_{i} \mid -   \sim N(\boldsymbol{x}_i'\boldsymbol{\beta
}_{true}, \sigma^{2})$, where $\boldsymbol{\beta}_{true} = (\alpha,\beta_{\vtt{Middle}},\beta_{\vtt{High}}, \beta_{\vtt{Age}}, \beta_{\vtt{BMI}},\beta_{\vtt{Middle:BMI}}, \beta_{\vtt{High:BMI}}) = (1,1,2,0.5,-2,2,4)$, and
set $\sigma^2$ via the signal-to-noise-ratio  $\mbox{SNR} = \mbox{var}(\boldsymbol{X\boldsymbol\beta}_{true})/\sigma^{2} \in \{1,3\}$. 
Finally, we introduce MAR for each variable $j$ with the exception of \verb|BMI|, $ R_{ij} \mid -  \sim \mbox{Bernoulli}\{\Phi(-0.7 + \vtt{BMI}_i + \omega_{ij})\}$, 
where $\omega_{ij}$ are Gaussian  with $\mbox{Corr}(\omega_{ij}, \omega_{ij'}) = 0.3$ and mean $-0.2$.  
Thus, $\boldsymbol{R}$ introduces correlated and data-dependent (via \verb|BMI|) patterns of missingness across both the response variable and the covariates. The missingness mechanism $\boldsymbol{R}$ is applied to all but 300 observations, and yields on average 49\% complete cases. We repeat this data-generating process to create 100 hybrid synthetic datasets.

Since the missingness in \vtt{New}, \vtt{Age}, and \vtt{FI} is linked to \vtt{BMI}, CC analysis is at risk of significant bias. We illustrate this point in Table~\ref{tab1}, where we compute ordinary least squares estimators ($\boldsymbol{\hat{\beta}}_{CC}$) and standard confidence bounds ($1.96\hat \sigma_{CC}$) for the regression coefficients using only the completely-observed data.  For all variables besides \vtt{Age}, CC analysis yields estimates and inference that depart significantly from the ground truth. Thus, alternative estimation and inference techniques are required, and specifically ones that can properly account for the MAR missingness.


\begin{table}[h]
    \centering
    \begin{tabular}{c|c|c|c|c|c|c|c}
    \rowcolor{shade}   &\vtt{Intercept}  & \vtt{Middle} &
      \vtt{High} & 
      \vtt{BMI}&
      \vtt{Age}&
      \vtt{Middle:BMI}&
      \vtt{High:BMI}\\    \hline
        $\boldsymbol \beta_{true}$ & 1 & 1 &2 & -2, &0.5 & 2 &  4\\\hline
        $\boldsymbol{\hat{\beta}}_{CC}$ & 1.77 & 0.19 & 0.39 &-1.50 & 0.51 & 1.50 & 3.01 \\ 
        $(1.96\hat{\sigma}_{CC})$& (0.10) & (0.12) & (0.16) & (0.09) & (0.05) & (0.11) & (0.17)
    \end{tabular}
    \caption{\small Complete case coefficient estimates and standard confidence bounds averaged across simulations. The CC analysis is severely biased for all variables except \vtt{Age}.}
    \label{tab1}
\end{table}

For evaluations and comparisons among imputation methods, we generate $m=20$ multiple imputations for each hybrid synthetic dataset using several distinct approaches.  
First, we use the proposed GMC-MA approach to generate posterior samples and imputations. Next, we use the same model and posterior draws, but replace the margin adjustment with the ECDF (GMC-ECDF). This comparison isolates the downstream impact of biased margin estimates for prediction under copula models.  Specifically, improvements in imputation accuracy under the proposed approach demonstrate the benefits of the margin adjustment extend more broadly to multivariate inference, and also the accumulating risk of using the ECDF for imputation with MAR data. These imputations are based on the Gibbs sampler (see the supplementary material) run for 10,000 iterations, with the first 5,000 discarded as a burn-in and the imputations computed every 50th sample to achieve $m=20$. 

To compare our approach to a Bayesian nonparametric alternative, we estimate the Gaussian mixture of factor models~\eqref{factormix} on $\boldsymbol Y^{obs}$ (instead of $\boldsymbol Z^{obs}$), using the same priors and hyperpriors discussed in Section~\ref{GMC}. This model is closely related to \cite{chandra2020escaping}, and provides a flexible model for multivariate data. Notably, this method treats each variable as continuous, and offers an opportunity to evaluate the gains of employing a rank-based copula model for mixed variable types. To convert \vtt{FamIncome} to a numeric variable, its levels are relabeled $(\vtt{low}, \vtt{middle}, \vtt{high}) = (1,2,3)$ which captures the ordinal properties of the variable, and we center and scale the observed values of \vtt{BMI}, \vtt{Age}, and \vtt{New}. Imputations of \verb|Age| and \verb|FamIncome| under the Gaussian mixture model are rounded (GM-RND) to preserve these variables' discreteness in the completed datasets. 

Among FCS (and non-Bayesian) methods, we create multiple imputations from the popular algorithm MICE (multiple imputation using chained equations; \citealp{van1999flexible}) under default settings in the \verb|R| package \verb|mice|. In addition to the default MICE algorithm, which employs linear models with main effects for each variable,  we include a modified version that features classification and regression trees for each variable (MICE-CART), which is better suited to capture interactions  \citep{burgette2010multiple}.

\begin{figure}[h]
    \centering
        \includegraphics[width = .42\textwidth, keepaspectratio]{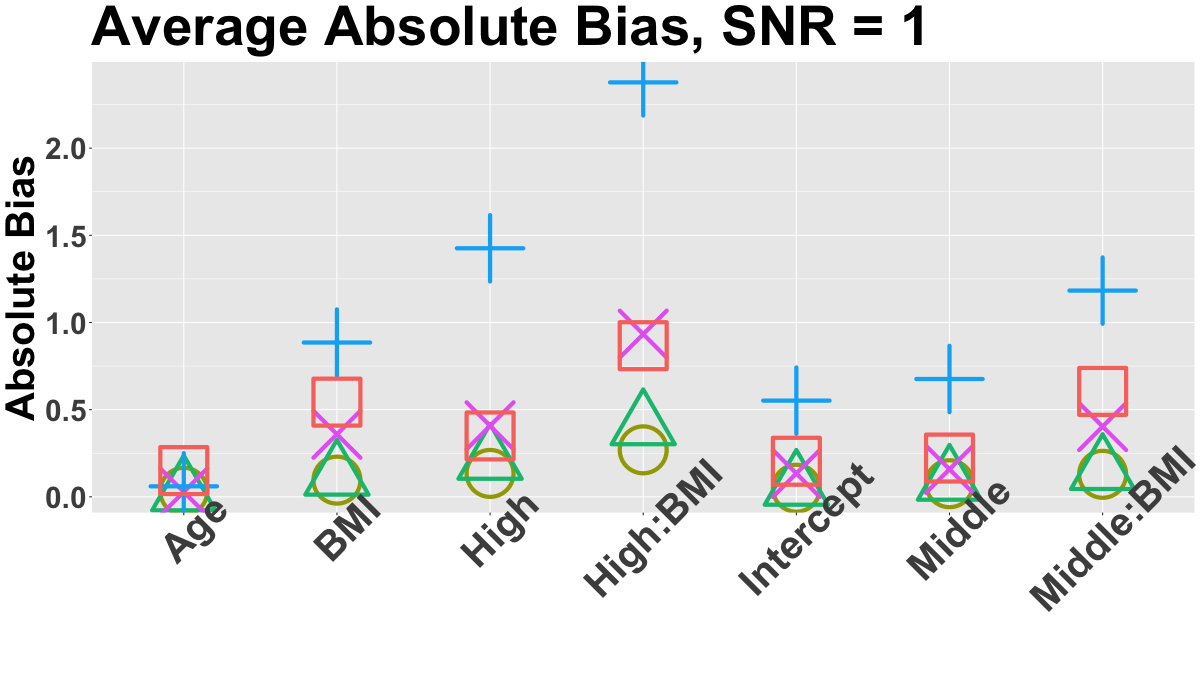}
        \includegraphics[width = .42\textwidth, keepaspectratio]{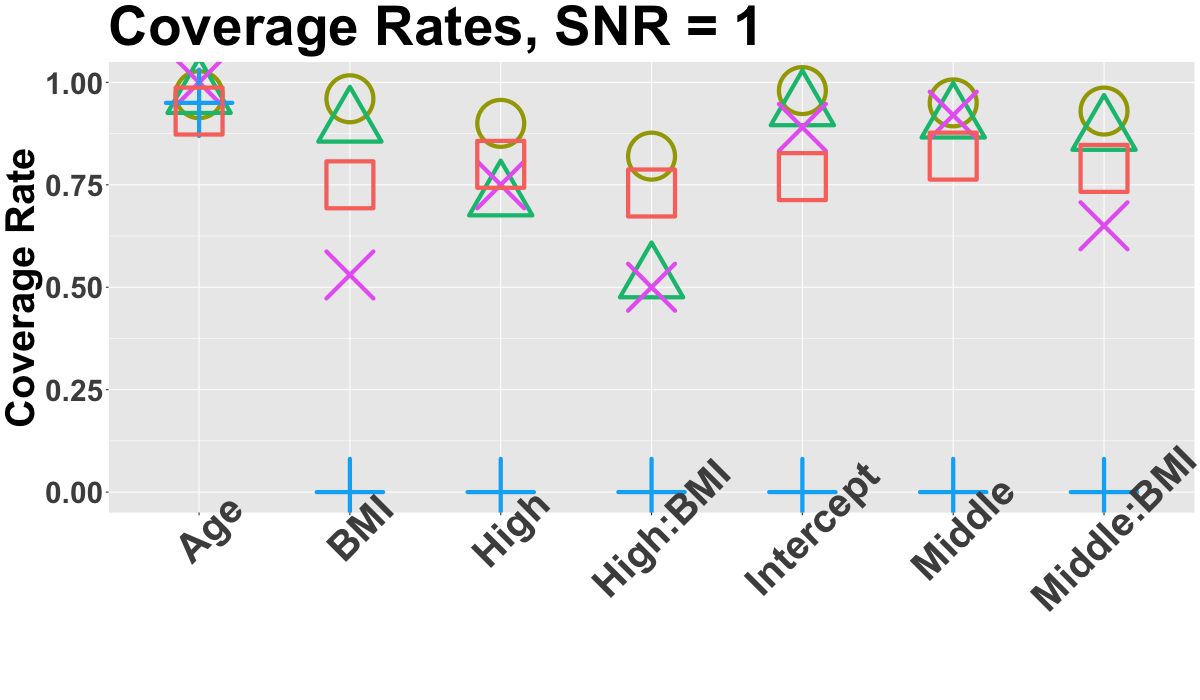}
        \includegraphics[width = .51\textwidth, keepaspectratio]{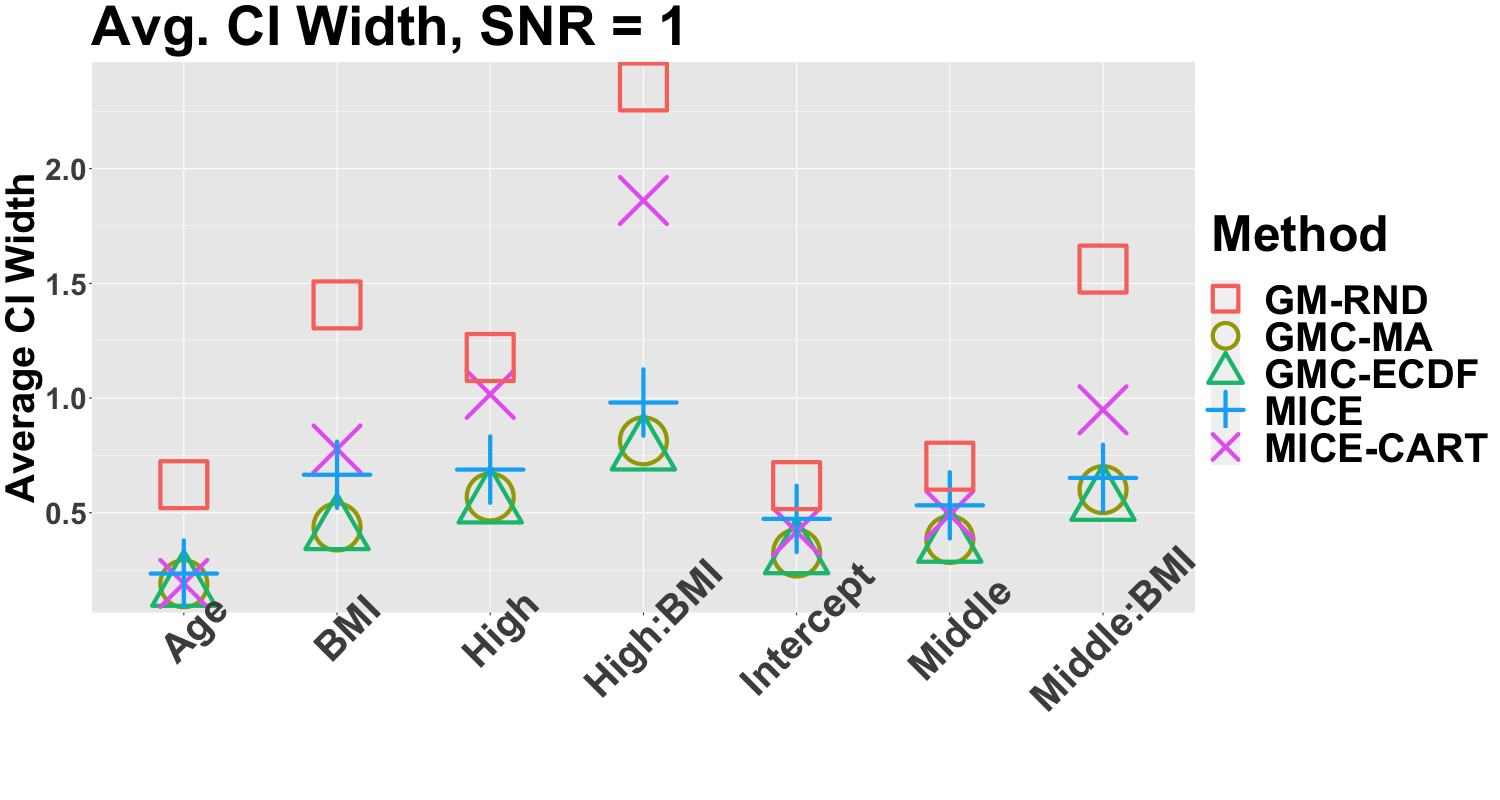}

    \caption{\small Absolute bias (left), interval coverage rates (center), and interval widths (right) for point and 99\% interval estimates computed under each imputation method. The GMC-MA approach consistently provides the most accurate point estimates (small absolute bias), the most well-calibrated intervals (large coverage rates), and highly precise inference (small interval widths). Similar  results for $\mbox{SNR}=3$ are presented in the supplement.} 
    \label{coverage}
\end{figure}

For each completed dataset, we fit a linear regression model for $\vtt{New}$ (using the correct covariates and interactions) and use the combining rules from~\cite{rubin2004multiple} to create point estimates and 99\% confidence intervals. The results  are summarized in Figure~\ref{coverage} for $\mbox{SNR = 1}$ via the absolute bias for each point estimate, and the coverage rates and widths for each interval estimate, averaged across 100 simulations. In this highly challenging scenario, the proposed GMC-MA imputations --- marked by the yellow circles in each plot --- consistently provide the most accurate point estimates (smallest absolute bias), the most well-calibrated intervals (largest coverage rates), and among the most precise inference (smallest interval widths). The results are similar for $\mbox{SNR} =3$, and available in the supplement.

Clearly, the margin adjustment is crucial: the GMC-ECDF intervals (green triangles) do not provide close to the nominal coverage for several coefficients---despite using the same underlying model as the GMC-MA---due to the bias in the ECDF under MAR. 
As expected, the GMC-MA intervals are slightly wider than the GMC-ECDF intervals: the former account for the uncertainty in the marginal distributions via the posterior distribution, while the latter treat the marginal distributions as fixed (at the ECDFs). 

GM-RND provides mostly unsatisfactory results, evidenced by high average interval widths and poor coverage rates. The Gaussian mixture  fit to the observed data clearly does not have the distributional flexibility to model mixed data types, even with a helpful rounding step. The RPL is specifically designed for this purpose, which yields significant improvement in modeling and imputation.

The default MICE approach also performs quite poorly: the point estimates are the least accurate and the interval estimates provide less than 5\% coverage for all variables except \vtt{Age}. MICE-CART offers some improvements, but still lags in estimation accuracy and the intervals are not close to the nominal coverage while substantially wider.  Further, significant coverage gaps remain in both SNR settings for the interaction terms. As \cite{burgette2010multiple} note, one potential disadvantage of MICE-CART is the decreased efficiency when a parametric imputation model is suitable. In the supplement, we highlight instances of this inefficiency across multiple imputations;  the CART imputations often misclassify \vtt{FI}, which creates significant problems for estimating the interaction effects.
\section{Real Data Application}\label{realdat}

\subsection{Setting and Goals}
The 2011-2012 National Health and Nutrition Experimentation Survey (NHANES) asks the question, ``For how many days during the past 30 days was your mental health not good?" The responses can be linked to other demographic and behavioral variables included in the questionnaire, enabling important insights into self-reported mental health. Of particular importance is the identification of key associative behaviors for at-risk individuals, as more broadly, mental health indicators are proxies for quality of life, depression, and risk for self-harm \citep{horwitz1999handbook}.

We study the association between self-reported marijuana use (\verb|UsedMarijuana|), gender, race, and high levels of self-reported poor mental health (\verb|DMHNG|). However, there are several significant challenges for this analysis. First, the data are subject to substantial missingness (Table~\ref{tab:data}), especially for the variables of interest. In particular, \verb|UsedMarijuana| is not asked of any individual older than 59, and is over 40\% missing. Thus, CC analysis of the association between \verb|UsedMarijuana| and \verb|DMHNG|---among other variables---would omit all individuals older than 59, and potentially  bias the results. Fortunately, \verb|Age| is recorded, which suggests that MAR may be  reasonable for these missing values. To strengthen this assumption, we include all variables in Table~\ref{tab:data} in our GMC-MA model. 

Second, the dataset contains many variables of mixed data types, with $n = 5856$ observations of $p = 15$ variables  (Table~\ref{tab:data}). The marginal distributions are nontrivial:  \verb|DMHNG| exhibits   discreteness, zero-inflation, boundedness, and heaping (Figure~\ref{margDMHNG}), which are challenging for regression analysis  \citep{kowal2021semiparametric}. To focus on at-risk individuals with high \verb|DMHNG| values, we study the upper quantiles of \verb|DMHNG| and the association with key variables of interest. However, the sample quantiles of this discrete variable  do not satisfy asymptotic normality, and thus are ill-suited for traditional multiple imputation \citep{rubin2004multiple}. 

Instead, we propose an alternative and general strategy for uncertainty quantification based on the posterior predictive distribution.  Specifically, we generate 500 posterior predictive datasets $\{\tilde Y_{ij}\}$ of size $n \times p$ and compute our summary statistics $\hat Q( \boldsymbol{\tilde Y})$ on each of these predictive datasets, which delivers posterior predictive inference for $Q$. This observation-driven inference leverages the GMC-MA model to capture challenging marginal and joint distributions across mixed data types in the presence of missingness, and simultaneously accounts for the joint uncertainty arising from (i) model parameters, (ii) missing data, and (iii) the replicability for a new dataset of the same size. By comparison, statistics computed on the imputed data $(\boldsymbol Y^{obs}, \boldsymbol{Y}^{mis})$ only incorporate uncertainty from $\boldsymbol \theta$ and $\boldsymbol{Y}^{mis}$, which limits the generalizability of the inference and conclusions. 

Lastly, we consider the sampling design in the NHANES survey. The NHANES sampling design includes certain over-sampled subgroups, most notably stratified by race. We stratify our analysis by race (and gender) and compute these quantities for each subgroups, which avoids the need to re-weight for population-level inference that aggregates across all strata.

\subsection{Checking Calibration}\label{calib}

Our aim is to quantify the extent to which model-based inferences change between CC analysis and a full data analysis that accounts for the missing values. To this end, we fit the GMC-MA model on both the CC data and the full data. First, by fitting the model on the CC data, we can compute posterior predictive diagnostics to assess whether the model is adequate for these data. Posterior predictive diagnostics compare the posterior predictive distribution of a statistic $\hat Q(\boldsymbol{\tilde Y})$ to the observed value $\hat Q(\boldsymbol Y)$. However, $\hat Q(\boldsymbol Y)$ is unavailable for the full dataset due to abundant missingness. Thus, the  CC diagnostics are the best available option. Second, by comparing the GMC-MA model output from the CC and full datasets, we can assess the impact of missingness on the analysis, and in particular whether the CC analysis is biased or misleading.

The CC data is created by dropping any observation with missing values, yielding a data set of size $n_{CC} = 2434$. We fit the GMC-MA on both the CC and the full datasets and generate predictive datasets of size $n_{CC} \times p $ and $n \times p$, respectively. 
For our statistics $\hat Q$, we compute three quantities stratified by race, gender, and marijuana use: the empirical distribution of \verb|DMHNG|, which is useful for posterior predictive diagnostics, and the 75th and 90th sample quantiles of \verb|DMHNG|, which target the at-risk individuals within each stratum. 
The GMC-MA fit to the full dataset must account for the additional uncertainty due to the missing observations, but also benefits from a much larger sample size. Differences in location for these posterior (predictive) distributions, however, would suggest bias due to missing data.

\begin{figure}[h]
    \centering
    \includegraphics[width = .24\textwidth,keepaspectratio]{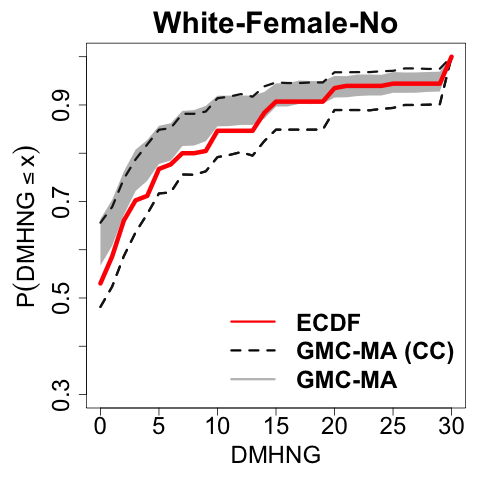}
    \includegraphics[width = .24\textwidth,keepaspectratio]{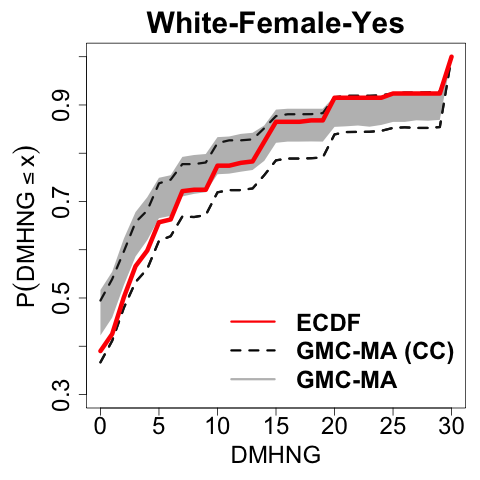}
    \includegraphics[width = .24\textwidth,keepaspectratio]{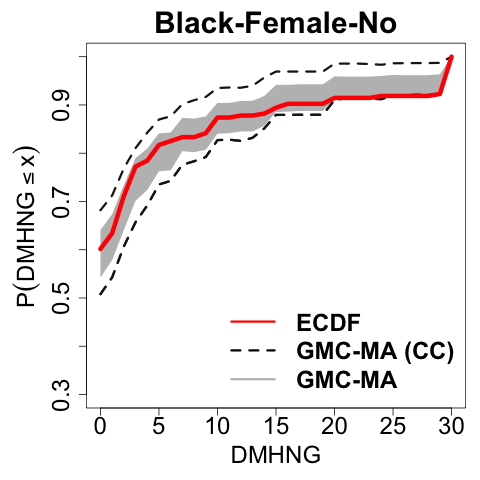}
    \includegraphics[width = .24\textwidth,keepaspectratio]{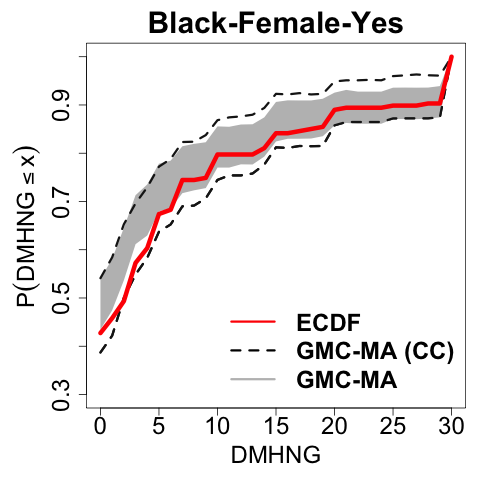}

    \caption{\small Posterior predictive summaries for models fit to the CC dataset (GMC-MA (CC)) and the full dataset (GMC-MA). For each race-gender-marijuana use stratum, we compare the 95\% HPD intervals for the posterior predictive ECDFs, and include the ECDF on the CC data for reference. The GMC-MA (CC) output is well-calibrated to the observed data. By comparison, the GMC-MA fit to the full dataset produces intervals that are narrower and shifted, which provides evidence of MAR---and that CC analysis is unreliable.}
    \label{compecdf}
\end{figure}

We compare the posterior predictive samples of the empirical distribution of \verb|DMHNG| from the CC and full dataset fits in Figure~\ref{compecdf}, and include the ECDF computed on the CC data. Specifically, we compute the ECDF on each posterior predictive dataset $\{\tilde Y_{ij}\}$---stratified by race, gender, and marijuana use---for the CC and full datasets, and report the pointwise 95\%  highest posterior density (HPD) intervals. %

The ECDF on the CC data falls within the 95\% HPD intervals from the GMC-MA (CC) fit. Thus, the model accurately describes
the challenging features of \verb|DMHNG|: zero-inflation, heaping (the large jumps around \verb|DMHNG| $\in \{7, 10, 14, 15, 20\}$), and boundedness at 30 (the lower interval converges to one at \verb|DMHNG| $= 30$). These results are stratified by race, gender, and marijuana use, and thus evaluate the \emph{joint} distribution. Similar results for the remaining race-gender-marijuana use strata 
are presented in the supplementary material.

Next, we compare the fitted GMC-MA models on the CC and full datasets. Most notably, the GMC-MA fit to the full dataset has substantially narrower 95\% HPD intervals, and often \emph{shifts} the predictive intervals.  For some strata, the predictive ECDF from the GMC-MA actually excludes the ECDF fit to the CC data (i.e. for white females) while the GMC-MA and the GMC-MA (CC) intervals do not fully overlap. Because the GMC-MA (CC) output broadly agrees with the empirical version, we argue that these discrepancies are not due to model misspecification, but rather due to the significant impacts of missing data. 
These results confirm our expectations based on the simulation results (Section~\ref{sim}) and suggest that a CC analysis of these data is unreliable.

\vspace{-1em}
\subsection{Associating Marijuana Use with Self-Reported Mental Health}
To investigate the associations between at-risk self-reported mental health and race, gender, and marijuana use, we compute the posterior predictive statistics $\hat Q(\boldsymbol{\tilde Y})$ for the 75th (see the supplement) and 90th quantile of \verb|DMHNG|, stratified by gender-race-marijuana use. These quantities are computed for both GMC-MA on the full dataset and GMC-MA (CC), and provide posterior predictive uncertainty quantification, which is summarized using the posterior median and 95\% HPD intervals. 

\begin{figure}[h]
    \centering
    \includegraphics[width = .8\textwidth, keepaspectratio]{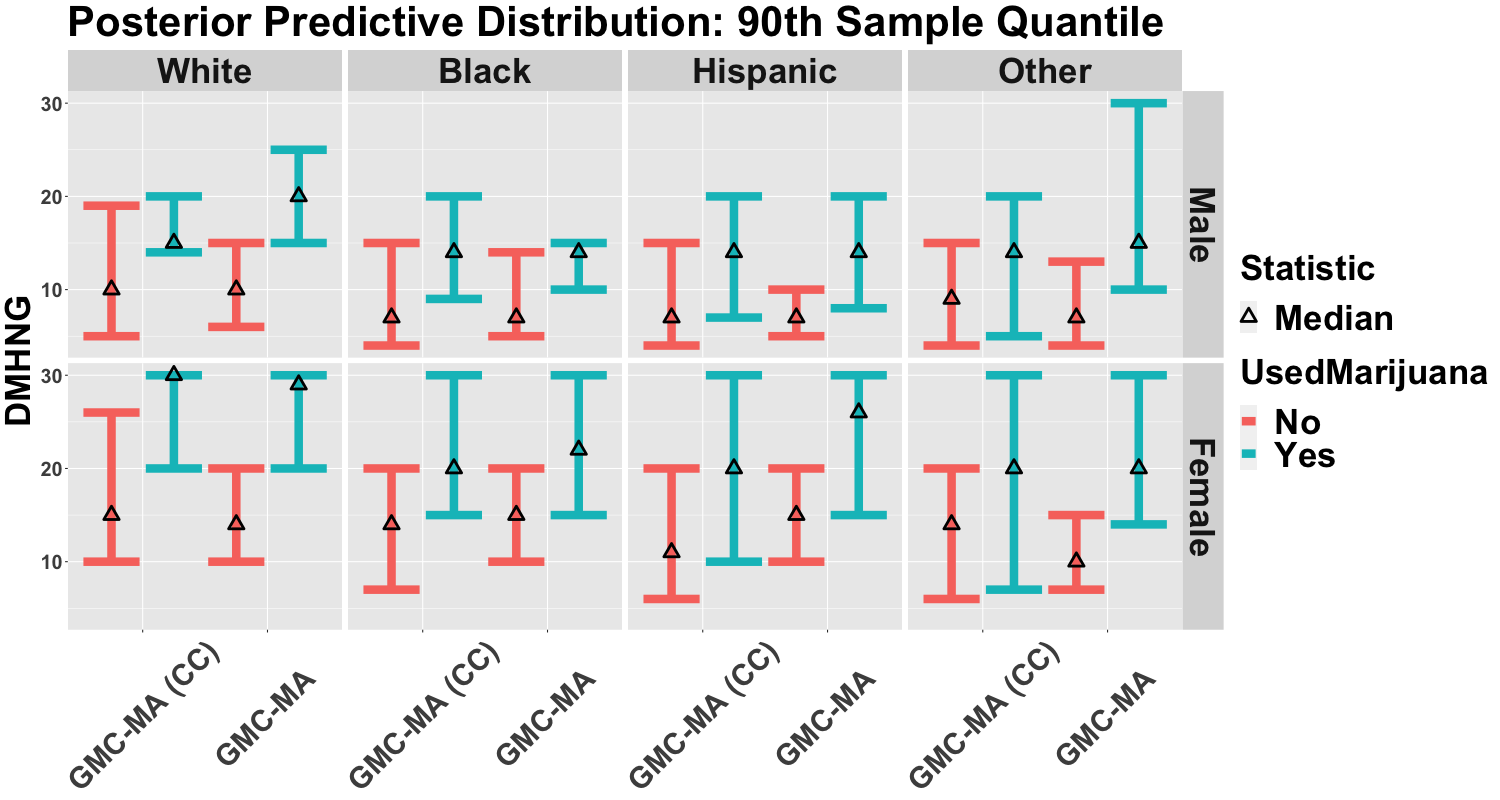}
\caption{\small 
Posterior predictive medians and 95\% HPD intervals for the predictive 90th quantiles of \vtt{DMHNG} by race-gender-marijuana use and comparing models fit to the complete case (CC) dataset (GMC-MA (CC)) and the full dataset (GMC-MA).  The CC analysis produces wider intervals with more overlap between  marijuana users and non-users across all strata, which dilutes the strong, significant, and adverse effects detected by GMC-MA fit to the full dataset.}

    \label{qcomp-int}
\end{figure}

Figure~\ref{qcomp-int} summarizes these point and interval estimates for the 90th quantile of \verb|DMHNG| for each stratum. Across all strata, there are several intervals from the GMC-MA (CC) fit with substantial overlap between marijuana users and non-users; yet many of these intervals become well-separated under the full dataset analysis with GMC-MA. The point estimates (posterior predictive medians) are similarly attenuated for the CC analysis. Specifically, consider the difference in predictive medians between marijuana users and non-users (the supplement provides a direct graphical comparisoin of these point estimates). The estimated differences are positive for all strata: the 90th quantile of \verb|DMHNG| is greater for marijuana users than non-users. However, these estimated differences are consistently larger for the GMC-MA on the full dataset compared to GMC-MA (CC).  For example, the estimated difference  in the 90th quantile of \verb|DMHNG| between marijuana users and non-users for white males is 10 for the GMC-MA on the full dataset, but only 5 for GMC-MA (CC). Similar trends are observed for the 75th quantile, although the discrepancies are less pronounced (see the supplementary material). 
Thus, CC analysis fails to identify certain strong, significant, and adverse associations between marijuana use and larger values of \verb|DMHNG|, which are detected clearly under the full dataset analysis.

These results emphasize the serious risks posed by CC analysis, which can produce biased or misleading conclusions. The implications are important for mental health studies: accurate estimation of the relationship between specific behaviors or attributes  and proxies for at-risk individuals is vital. By using the GMC-MA, we were able to perform full dataset analysis---despite abundant missingness, mixed data types, and complex marginal and joint distributions---and highlight the specific limitations of CC analysis. 

\section{Conclusion}\label{conc}
We proposed a nonparametric copula model for mixed (count, continuous, ordinal, and unordered categorical) data types subject to values missing-at-random. The model features a latent mixture of factor models to induce a nonlinear and scalable Gaussian mixture copula model. We employed the rank-probit likelihood for posterior inference, which circumvents the need to specify marginal distributions yet maintains strong posterior consistency for the parameters of the underlying copula model. A central innovation was the introduction and  theoretical analysis of the \emph{margin adjustment}, which delivers consistent inference for each marginal distribution under rank-based copula models with no further modeling assumptions and minimal additional computing cost. The margin adjustment eliminates any reliance on the ECDF for prediction and imputation, which is the default approach in rank-based copula models yet can be severely biased under MAR. Carefully-designed simulation studies showed significant improvements in imputation and marginal distribution estimation for the proposed approach relative to state-of-the-art alternatives, especially in the presence of nonlinear dependencies, mixed data types, and MAR missingness. We applied our model and imputation strategies to self-reported mental health data and demonstrated the pitfalls of complete case analysis---and showed how the proposed approach may resolve these issues.

There are numerous interesting directions for future work. First, the proposed Gaussian mixture copula model and the margin adjustment apply not only to imputation, but also to prediction. Our strong theoretical results suggest that the proposed framework may prove useful for posterior prediction of multivariate and mixed data, especially in the presence of missingness. Similarly, our theoretical analyses suggest that rank-likelihoods and the margin adjustment may apply more broadly to non-Gaussian copula models. Such developments would broaden the applicability of Bayesian inference for copula models, while simultaneously eliminating the need to specify models for each marginal distribution and potentially delivering consistent inference for these margins. Finally, an important and challenging extension is to adapt the proposed framework for missingness-not-at-random. The latent factor mixture \eqref{factormix} is an attractive option for parsimonious joint modeling of the missingness mechanism and the observed data in a low-rank, shared parameter model (e.g., \citealp{creemers2010sensitivity}).

\newpage

\bibliographystyle{apalike}
\bibliography{Bibliography-MM-MC}
\newpage
\begin{center}
    \bf{\large Supplement to \\ ``Nonparametric Copula Models for Multivariate, Mixed, and Missing Data"}
\end{center}
\appendix
\section{Proofs}

\begin{customthm}{1}
    Suppose $\{Z_i\}_{i=1}^n \overset{i.i.d}{\sim} F_Z$  and $\{Y_i\}_{i=1}^n  = \{h(Z_{i})\}_{i=1}^{n} \sim F_{Y}$ , where $F_Z$ is continuous and $h$ is a monotone increasing function. Defining $Z^{n}(x)$ as \eqref{cutpoint}, the margin adjustment satisfies 
$\tilde F(x) \coloneqq F_Z\{Z^{n}(x) \} \stackrel{a.s.}{\to}F_Y(x)$ for all $x \in \mathbb{R}$.

\end{customthm}

\begin{proof}
Fix $\tau = F_{Y}(x) \in (0,1]$ for a given $x$
and let $S$ denote the position of $Z^n(x)$, which is also the position of $\max\{Y_{i}:Y_{i}\leq x\}$ due to the consistent orderings of $\{Z_i\}_{i=1}^{n}$ and $\{Y_i\}_{i=1}^{n}$ since $ Y_{i}$ is a monotone transformation of $Z_{i}$. By the Glivenko-Cantelli Theorem,  $S/n = n^{-1}\sum_{i = 1}^{n}\mathbb{I}\{Y_{i} \leq x\} \overset{a.s.}{\to} \tau$. Now, consider the random variable $U_{i} = F_{Z}(Z_{i})$, which is uniform on $(0,1)$ with $\{U_{i}\}_{i=1}^{n}$. Therefore, the $S$th order statistic of $\{U_{i}\}_{i=1}^{n}$ satisfies $U^{n(S)}= F_{Z}\{Z^{n}(x)\}$. It is well known that $U^{n(S)} \sim \mbox{Beta}(S,n-S+1)$, with $E[U^{n(S)}] = S/(n+1)$ and $V[U^{n(S)}] = Sn/\{(S+n)^{2}(S + n + 1)\} < n^{-1}$. As such, $V[U^{n(S)}] \to 0$ as $n\to \infty$, so $U^{n(S)}$ converges in distribution to a degenerate random variable with point mass at the limit of its expectation:  $S/(n+1) \overset{a.s.}{\to} \tau$. Since $\tau$ is fixed, this also implies that $U^{n(S)} \overset{p}{\to} \tau$. Finally, observe that 
the sequence $Z^{n}(x) $ is monotone in $n$, 
which implies that the sequence $U^{n(S)}$ is also monotone in $n$. Coupling monotonicity and convergence in probability, we have that $U^{n(S)} = F_Z\{Z^{n}(x)\}  \overset{a.s.}{\to} \tau = F_Y(x).$ 

For any $x$ such that $\tau = F_Y(x) = 0$, $S = 1$ since the position of $Z^{n}(x)$ is now the same as $\min(Y_{i})$. Applying the same argument above, the sequence of first order statistics of a uniform random variable will converge in probability to a degenerate random variable with point mass at $0$. In addition, the sequence is monotone decreasing, which maintains the almost sure convergence. Thus, the almost sure convergence holds for any $x \in \mathbb{R}$
\end{proof}

\begin{customthm}{2}
Suppose $\{\boldsymbol{Z}_{i}\}_{i=1}^{n}=\{(Z_{i1},Z_{i2})\}_{i=1}^{n}\overset{i.i.d}{\sim} G$ where $G$ is continuous with marginal distributions $G_{1},G_{2}$, and $\{\boldsymbol{Y}_{i}\}_{i=1}^{n}=\{(Y_{i1},Y_{i2})\}_{i=1}^{n} =[(F_1^{-1}\{G_1(Z_{i1})\}, F_2^{-1}\{G_2(Z_{i2})\})]_{i=1}^{n}$ has joint distribution function $F$ with marginal distributions $F_1,F_2$. Suppose that $Y_2$ is completely observed and $Y_{1}$ is MAR.  Define $Z_{1}^{n}(x) \coloneqq \max[\{Z_{i1}: Y^{obs}_i \le x\} \cup \{Z_{i1} :Y^{obs}_{i1} = \min(Y^{obs}_{i1})\}], i \in \{1,\dots, n\}$. Then the margin adjustment satisfies
$\tilde{F}_{1}(x) \coloneqq G_{1}\{Z_{1}^{n}(x)\} \overset{a.s.}{\to}F_{1}(x)$
for all $x \in \mathbb{R}$. 
\end{customthm}
\begin{proof}
For this proof, we will use upper case letters to denote random variables, lower case letters for observed data, and bold face for vectors. Probabilities are given by $P_\cdot(x)$, where the subscript refers to the respective (marginal, conditional, or joint) distribution. The proof will show that  $Z_{1}^{n}(x) \overset{a.s.}{\to}G_{1}^{-1}\{F_{1}(x)\}$ as $n\to \infty$, which yields the stated result via the continuous mapping theorem.

Note that because $G_{1}, G_{2}, F_{1}, F_{2}$ are non-decreasing, we have that  $y^{obs}_{ij} < y^{obs}_{lj} \implies z^{obs}_{ij} < z^{obs}_{lj},\ j =1,2, \ \forall l\neq i$ -- i.e. the orderings between $Y^{obs}_{j}$ and $Z^{obs}_{j}$ are consistent. Next, suppose that $Y_1$ and $Y_2$ are continuous; the discrete case is addressed subsequently. Given that $\boldsymbol{Y}$ is a component-wise monotone transformation of $\boldsymbol{Z}$, the joint distribution $F$ may be expressed in terms of $G$: 
 \begin{equation}\label{joint}
     F(y_{1},y_{2}) = G[G_{1}^{-1}\{F_{1}(y_{1})\}, G_{2}^{-1}\{F_{2}(y_{2})\}].
 \end{equation}
Similarly, the conditional probability $P(Y_{1} \leq x \mid Y_{2} = y)$ can be expressed in terms of  $\boldsymbol Z$:
\begin{align}
    P(Y_{1} \leq x \mid Y_{2} = y) &= P[Z_{1} \leq G_{1}^{-1}\{F_{1}(x)\} \mid Z_{2}= G_{2}^{-1}\{F_2(y)\}]\nonumber\\
    &=
    \int_{-\infty}^{G_{1}^{-1}\{F_{1}(x)\}} \frac{g[z, G_{2}^{-1}\{F_{2}(y)\}]}{g_{2}[G_{2}^{-1}\{F_{2}(y)\}]}\ d z\label{partial1}
\end{align}
where $g$ and $g_{2}$ are the density functions of $G$ and $G_{2}$, respectively.  


Now, consider the sequence of  probabilities $P_{Z_{1} \mid R_1 = 0}\{Z_{1} \leq Z_{1}^{n}(x)\}$, i.e., the marginal probability that $Z_{1}$ is less than $Z_{1}^{n}(x)$ given that $Y_{1}$ is observed  ($R_1=0$). This sequence is monotone increasing because $Z_{1}^{n}(x)$ is monotone increasing, and clearly bounded above by one. Thus, it converges almost surely to its limit by the monotone convergence theorem. Therefore, $Z_{1}^{n}(x)$ must also converge almost surely to a limit, which we will label $Z^{\infty}_{1}(x)$. 

The conditional probability $P_{Z_{1} \mid R_1 = 0}\{Z_{1} \leq Z_{1}^{n}(x)\}$ is equivalently
\begin{equation}\label{tower}
      P_{Z_{1} \mid R_1 = 0}\{Z_{1} \leq Z_{1}^{n}(x)\} = E_{Z_{2} \mid R_1 = 0}[    P_{Z_{1} \mid Z_{2}, R_1 = 0}\{Z_{1} \leq Z_{1}^{n}(x)\}]
\end{equation}
where the expectation is taken with respect to the distribution of $Z_{2}$ given that $Y_{1}$ is observed (i.e., not missing). Because $\boldsymbol{R}$ is missing-at-random, $Z_{1}$ is conditionally independent of $R_{1}$ given $Z_{2}$, so 
 \begin{equation}\label{G-terms}
     P_{Z_{1} \mid Z_{2}, R_1 = 0}\{Z_{1} \leq Z_{1}^{n}(x)\} = P_{Z_{1} \mid Z_{2}}\{Z_{1} \leq Z_{1}^{n}(x)\} = \int_{-\infty}^{Z_{1}^{n}(x)} \frac{g(z, Z_{2})}{g_{2}(Z_{2})}\ dz
 \end{equation}
 and  thus
\begin{equation}\label{ignorability}
   P_{Z_{1} \mid R_1 = 0}\{Z_{1} \leq Z_{1}^{n}(x)\} = E_{Z_{2} \mid R_1=0}[P_{Z_{1} \mid Z_{2}}\{Z_{1} \leq Z_{1}^{n}(x) \}].
\end{equation}
Denoting \eqref{ignorability} by $h\{Z_1^n(x)\}$, note that \eqref{G-terms}--\eqref{ignorability} imply that  $h$ is a continuous function.  Consequently, we can write the limit of $P_{Z_{1} \mid R_1 = 1}\{Z_{1} \leq Z_{1}^{n}(x)\}$ explicitly, yielding that 
\begin{equation}\label{Z-limit}
   P_{Z_{1} \mid R_1 = 0}\{Z_{1} \leq Z_{1}^{n}(x)\} = h\{Z_{1}^{n}(x)\}  \overset{a.s.}{\to}h\{Z_{1}^{\infty}(x)\}.
\end{equation}

Next, consider an application of Theorem~\ref{ma-cdf2} to the \emph{observed} data. By construction, $Z_{1}^{n}(x)$ and the maximum position of $Y_1 \leq x$ will have the same position because $Y_{1}$ is a monotone transformation of $Z_{1}$. By Theorem~\ref{ma-cdf2}, this implies that if $x$ is the $\tau$th quantile under the distribution of observed $[Y_{1} \mid R_1=0]$, then $Z_{1}^{n}(x)$ will converge to the $\tau$th quantile under the distribution of $Z_{1}$ corresponding to observed $[Y_1 \mid R_1=0]$:
\begin{equation}
        P_{Z_{1} \mid R_1 = 0}\{Z_{1} \leq Z_{1}^{n}(x)\} \overset{a.s.}{\to} P_{Y_{1} \mid R_1 = 1}(Y_{1} \leq x ).
\end{equation}
We can also re-write $P_{Y_{1} \mid R_1 = 0}(Y_{1} \leq x )$ in terms of $Z_{1}$ and $Z_{2}$ by~\eqref{partial1} and \eqref{ignorability}:
\begin{align}
     P_{Y_{1} \mid R_1 = 0}(Y_{1} \leq x ) &= P_{Z_{1} \mid R_1 = 0}[Z_{1} \leq G_{1}^{-1}\{F_{1}(x)\}] \\
    &= E_{Z_{2} \mid R_1 = 0}(P_{Z_{1} \mid Z_{2}, R_1 =  0}[Z_{1} \leq G_{1}^{-1}\{F_{1}(x)\}]) \\
    &= E_{Z_{2} \mid R_1 = 0}(P_{Z_{1} \mid Z_{2}}[Z_{1} \leq G_{1}^{-1}\{F_{1}(x)\}])\\
    &= h[G_{1}^{-1}\{F_{1}(x)\}]. \label{Y-limit}
\end{align}
Finally, we see the equivalence between~\eqref{Z-limit} and~\eqref{Y-limit} holds if and only if $Z_{1}^{\infty}(x) = G_{1}^{-1}\{F_{1}(x)\}$, which implies that $Z_{1}^{n}(x) \overset{a.s.}{\to} G_{1}^{-1}\{F_{1}(x)\}$. Once again, an application of the continuous mapping theorem demonstrates the consistency of the MA estimator at $x$.

When $Y_1$ or $Y_2$ is discrete, the proof requires only minor modifications. First, the conditional probability $P(Y_{1} \leq x \mid Y_{2} = y)$ can still be written in terms of $\boldsymbol Z$. Specifically, $G_{2}^{-1}\{F_{2}(y)\}$ maps $y$ to the interval $(G_{2}^{-1}\{F_{2}^-(y)\}, G_{2}^{-1}\{F_{2}(y)\}]$ where $F_{2}^-(y)$ is the left limit of $F_{2}$ at $y$ \citep{zhao2020missing}. Therefore, for discrete $Y_{2}$, $P(Y_{1} \leq x \mid Y_{2} = y)$ is now
\begin{align*}\label{partial-discrete}
    P(Y_{1} \leq x \mid Y_{2} = y) &= P[Z_{1} \leq G_{1}^{-1}\{F_{1}(x)\} \mid Z_{2} \in (G_{2}^{-1}\{F_2^-(y)\},G_{2}^{-1}\{F_2(y)\}]]\\
    &=\int_{-\infty}^{G_{1}^{-1}\{F_{1}(x)\}} \int_{G_{2}^{-1}\{F_{2}^-(y)\}}^{G_{2}^{-1}\{F_{2}(y)\}}
    \frac{g(z_1, z_2)}{g_{2}(z_{2})}\ dz_{2}\ dz_{1}
\end{align*}
 If $Y_{1}$ is discrete, the event $Y_{1} \leq x$ is equivalent to $Z_{1} \leq G^{-1}\{F_{1}(x)\}$, where $F_{1}(x)$ is the right limit of $F_{1}$ at $x$. Therefore, the argument does not change, and the rest of the proof follows identically.

 To extend this result to $p$-dimensions, we first extend the structure of the joint distributions for $\boldsymbol Z$ and   $\boldsymbol Y$ into $p$ dimensions, where $G, F$ are $p$-dimensional distribution functions with marginals $\{G_{j}\}_{j=1}^{p}$ and $\{Y_{j}\}_{j=1}^{p}$. As in \eqref{joint}-\eqref{partial1}, joint and conditional distributions for $\boldsymbol Y$ can similarly be expressed in terms of $\boldsymbol Z$.
 
 We then partition $\boldsymbol Y$ into ($\boldsymbol Y^{comp}, \boldsymbol Y^{part}$), where $Y_{j} \in \boldsymbol Y^{part} \implies \sum_{i=1}^{n}R_{ij} >0$ i.e. a variable is partially observed if it has at least one missing value in the sample. $Y^{comp}$ is comprised of all variables which are completely observed, i.e. $Y_{j} \in \boldsymbol Y^{comp} \implies\sum_{i=1}^{n}R_{ij} = 0$. We assume that all variables in $\boldsymbol Y^{part}$ are MAR, which implies that $\boldsymbol Y^{comp}$ is non-empty. Then, for any $Y_{j} \in \boldsymbol Y^{part}$ and $x$, we define $Z_{j}^{n}(x)$ as in \eqref{cutpointobs} and consider the sequence of probabilities $P_{Z_{j} \mid R_{j} = 0}\{Z_{j}\leq Z_{j}^{n}(x)\}$. Defining $\boldsymbol Z_{-j}$ to be $\boldsymbol Z \setminus Z_{j}$, the rest of the proof follows identically by noticing that $P_{Z_{j} \mid R_{j} = 0}\{Z_{j}\leq Z_{j}^{n}(x)\} = E_{\boldsymbol Z_{-j} \mid R_j = 0}[P_{Z_j \mid \boldsymbol Z_{-j}, R_j =0}\{Z_{j}\leq Z_{j}^{n}(x)\}]$ converges almost surely to a limit since $Z_{j}^{n}(x)$ is monotone and bounded, while Theorem \ref{ma-cdf2} implies that $P_{Z_{j} \mid R_{j} = 0}\{Z_{j}\leq Z_{j}^{n}(x)\} \overset{a.s.}{\to} P_{Y_{j} \mid R_{j}=0}(Y_{j} \leq x) = h[G^{-1}_{j}\{F_{j}(x)\}]$.  
\end{proof}

\begin{customthm}{3}
Suppose $\{\boldsymbol Y_i\}_{i=1}^n \stackrel{i.i.d}{\sim} G^{\infty}_{\boldsymbol C_{0}, F_{1},\dots,F_{p}}$ for true copula parameters $\boldsymbol C_0$ and true marginal CDFs $F_1,\ldots,F_p$. Let  $\Pi$ be a prior distribution on the space of all $p\times p$ positive semi-definite correlation matrices $\boldsymbol C_{\boldsymbol \theta}$ with corresponding density $\pi(\boldsymbol C_{\boldsymbol\theta})$ with respect to a measure $\nu$. Suppose $\pi(\boldsymbol C_{\boldsymbol{\theta}})>0$ almost everywhere with respect to $\nu$ and assume that the missigness is ignorable. Then, for $\boldsymbol C_{0}$ a.e.\! $[\nu]$ and any neighborhood $\mathcal{A}$ of $\boldsymbol C_{0}$, we have that 
     $\lim_{n\rightarrow \infty}\Pi\{\boldsymbol C_{\boldsymbol{\theta}} \in \mathcal{A} \mid \boldsymbol{Z}_{n}^{obs} \in \mathcal{D}(\boldsymbol{Y}_{n}^{obs})\} =  1 \ a.s \  [G^{\infty}_{\boldsymbol C_{0}, F_{1},\dots,F_{p}}]$.
\end{customthm}
\begin{proof}
Because the missingness mechanism is ignorable, the priors for the parameters governing the data generating process for $\boldsymbol{Y}$ and the missingness mechanism $\boldsymbol{R}$ are independent \citep{rubin1976inference}. Therefore,  we can utilize the following variant of Doob's theorem to prove the result, as was done in \cite{murray2013bayesian}.

\begin{DoobsTheorem*} \citep{gu2009bayesian} Let $X_{i}$ be observations whose distributions depend on a parameter $\boldsymbol \theta$, both taking values in Polish spaces. Assume $\boldsymbol \theta \sim \Pi$ and $X_{i}\mid \boldsymbol \theta \sim P_{\boldsymbol \theta}$. Let $\mathcal{X}_{N}$ be the $\sigma$-field generated by $X_{1},\dots,X_{N}$, and $\mathcal{X}_{\infty} = \sigma(\bigcup_{i}^{\infty}\mathcal{X}_{i})$. If there exists a $\mathcal{X}_{\infty}$ measureable function $f$ such that for $(\boldsymbol \omega, \boldsymbol \theta) \in \Omega^{\infty} \times \Theta, \ \boldsymbol \theta = f(\boldsymbol \omega) \ a.e. \ [P_{\boldsymbol \theta}^{\infty} \times \Pi]$ then the posterior is strongly consistent at $\boldsymbol \theta$ for almost every $\boldsymbol \theta \ [\Pi]$.
\end{DoobsTheorem*}
To adopt Doobs Theorem to this setting, note the data generating process for $\boldsymbol Y_i$
\begin{align}
    \boldsymbol Z_i &\overset{i.i.d}{\sim} N(0, \boldsymbol C_{0})\\
    y_{ij} &= F_{j}^{-1}\{\Phi(z_{ij})\}
\end{align}
implies each $\boldsymbol Z_{n}^{obs}$, generated i.i.d by a probability distribution indexed by $\boldsymbol C_{0}$, must satisfy the event $\mathcal{D}(\boldsymbol Y_{n}^{obs})$. Therefore, it suffices to establish the existence of a consistent estimator $\boldsymbol C_{\boldsymbol \theta}$ of $\boldsymbol C_{0}$, the data generating Gaussian copula correlation matrix, that is measureable with respect to the sigma-field generated by the sequence $\{\boldsymbol Z_{n}^{obs} \in \mathcal{D}(\boldsymbol{Y}_{n}^{obs})\}_{n=1}^{\infty}$, where $n$ indexes the sample size.

 Suppose $\boldsymbol Y_n$ is comprised of $n_{1}>1$ complete cases without any missing values $(\boldsymbol Y^{CC})$ and $n_{2}$ cases with missing values for at least one variable ($\boldsymbol Y^{inc}$) such that $n_{1} + n_{2} = n$.  For each observation $i$ in $\boldsymbol Y^{CC}$, for each variable $j \in \{1,\dots,p\}$ with $y_{ij} =x$, consider the margin adjustment~\eqref{marginadjust} with $Z_{j}^{n}(x)$ as in \eqref{cutpointobs}. Let $T_{nij} = \sum_{i =1}^{n} \mathbbm{1}\{z_{ij}\leq Z_{j}^{n}(x)\}$, with $\boldsymbol{T}_{ni}(\boldsymbol{Y}_{i}^{obs}) = (T_{ni1},\ldots, T_{nip})$ and $\boldsymbol{T}_{n}(\boldsymbol{Y}_{n}^{obs}) = \{\boldsymbol{T}_{ni}(\boldsymbol{Y}_{i}^{obs})\}_{i=1}^{n}$. 
As noted in \cite{murray2013bayesian}, the information contained $\boldsymbol{T}_{n}(\boldsymbol{Y}_{n}^{obs})$ is also contained in $\boldsymbol Z_{n}^{obs}$. Namely, $\boldsymbol{T}_{n}(\boldsymbol{Y}_{n}^{obs})$ may be extracted from the boundary conditions  of $\boldsymbol{Z}_{n}^{obs} \in \mathcal{D}(\boldsymbol{Y}_{n}^{obs})$. Therefore, any function measureable with respect to $\mathscr{T}_{n}$, the sigma-algebra generated by the sequence $\{\boldsymbol{T}_{n}(\boldsymbol{Y}_{m}^{obs})\}_{m=1}^{n}$ is also measureable with respect to the sigma algebra generated by the corresponding sequence $\{\boldsymbol Z_{m}^{obs} \in \mathcal{D}( \boldsymbol{Y}_{m}^{obs})\}_{m=1}^{n}$. Consequently, as in \cite{murray2013bayesian}, we exclusively work with $\mathscr{T}_{n}$.
\color{black}

Now, $Z_{j}^{n}(x)$ is a random variable, and hence $T_{nij}$ is a random variable.  We work with its expectation under the true data generating model. Define $\hat{U}_{nij} = E[T_{nij}]/(n+1)$ and $\boldsymbol{ \hat{U}}_{ni} = (\hat{U}_{ni1},\dots, \hat{U}_{nip^{*}})$. Then,

\begin{align}
   \hat{U}_{nij} &= \frac{1}{n+1}E     \sum_{l=1}^{n} \mathbbm{1}\{z_{lj}\leq Z_{j}^{n}(x)\}\\
   &=\frac{1}{n+1}    \sum_{l=1}^{n} P\{z_{lj}\leq Z_{j}^{n}(x)\} \label{re-write}
\end{align}

\noindent By the strong law of large numbers and Theorem~\ref{consistMA-misv2}, ~\eqref{re-write} converges almost surely to
\begin{align}
    P\left[z_{j} \leq \Phi^{-1}\{F_{j}(x)\}\right] = F_{j}(x)
\end{align}

\noindent since $\Phi\{Z_{j}^{n}(x)\} \overset{a.s.}{\to}F_{j}(x)$. Therefore, we have that $\hat{U}_{nij} \overset{a.s.}{\rightarrow} U_{ij}$, where $U_{ij} = F_{j}(x)$, the cumulative marginal probability of $x$ for variable $j$ under the true distribution $F_{j}$. Consequently, $\hat{\boldsymbol{U}}_{ni} \overset{a.s.}{\rightarrow} \boldsymbol{U}_{i} = (U_{1},\dots,U_{p})$, and $\boldsymbol{U}_{i}$ is $\mathscr{T}_{\infty}$ measureable.


  Therefore, $\boldsymbol{U}_{i}$ is a sample from a Gaussian copula with correlation matrix $\boldsymbol C_{0}$ where the continuous margins are $\mbox{Uniform}[0,1]$ while the discrete margins are merely re-labeled with their ground-truth cumulative probabilities. We then may apply the argument of \cite{murray2013bayesian} which establishes the existence of a consistent estimator of $\boldsymbol C_{0}$ which is a function of $\boldsymbol{U}_{i}$ and thus $\mathscr{T}_{\infty}$ measureable. Specifically, the problem reduces to estimating polychoric/polyserial correlations with fixed margins, where $\boldsymbol{U}_{i}$ is a regular parametric family admitting a sequence of consistent estimators of $\boldsymbol C_{0}$, for instance using the estimators of \cite{olsson1979maximum} and \citep{olsson1982polyserial}.
  \end{proof}
\begin{customcor}{1}
Under the conditions of Theorem~\ref{consist_complete}, define $\tilde{F}_j$ as in \eqref{marginadjust} with $Z_{j}^{n}(x)$ as \eqref{cutpointobs} and $G_j = \Phi$ for each $j\in \{1,\ldots,p\}$. Then for any $x \in \mathbb{R}$ and any neighborhood $\mathcal{A}$ of $F_{j}(x)$ 
     $\lim_{n\rightarrow \infty}\Pi\{\tilde{F}_{j}(x) \in \mathcal{A} \mid \boldsymbol{Z}_{n}^{obs} \in \mathcal{D}(\boldsymbol{Y}_{n}^{obs})\} =  1 \ a.s \  [G^{\infty}_{\boldsymbol{C}_{0}, F_{1},\dots,F_{p}}]$.
    
\end{customcor}
\begin{proof}


This result follows from an application of Doob's Theorem presented above and Theorem~\ref{consistMA-misv2}. To apply Doob's theorem, consider the marginal distributions of $\{Z_{j}\}_{j=1}^{p}$. Specifically, for any $j \in \{1,\dots,p\}$, each $Z_{ij}^{obs}$ is standard normal, but restricted to fall in the subset of the real line determined by the right and left limits of the true marginal distribution $F_{j}$ evaluated at the realized value of $Y^{obs}_{ij}$. Then, defining the marginal event $\mathcal{D}(Y_{j_{n}}^{obs}) \coloneqq \{Z^{n\times 1}: y^{obs}_{lj} <y^{obs}_{kj} \implies z^{obs}_{lj}>z^{obs}_{kj}, k \neq l\}$ and $Z_{j_{n}}^{obs}$ the latent vector corresponding to $Y_{j_{n}}^{obs}$, it must be the case that $Z_{j_{n}}^{obs} \in \mathcal{D}(Y_{j_{n}}^{obs})$. In addition, the sigma-field generated by the sequence $\{Z_{j_{m}} \in \mathcal{D}(Y_{j_{m}}^{obs})\}_{m=1}^{n}$ is a sub sigma-field of the sigma-field generated by $\{\boldsymbol Z_{m}^{obs} \in \mathcal{D}(\boldsymbol Y_{m}^{obs})\}_{m=1}^{n}$, and hence any function measureable with respect to the former is also measureable with respect to the latter.

Therefore, to apply Doob's theorerm, it suffices to demonstrate the existence of a strongly consistent estimator of $F_{j}(x)$ that is measureable with respect to the sigma-field generated by the sequence of $\{Z_{j_{n}} \in \mathcal{D}(Y_{j_{n}}^{obs})\}_{n=1}^{\infty}$. For any $n$, $Z_{j}^{n}(x)$ is clearly an element of $Z^{obs}_{j_{n}} \in \mathcal{D}(Y_{j_{n}}^{obs})$. Therefore, $\tilde{F}_{j}(x) = \Phi\{Z_{j}^{n}(x)\}$ is measureable with respect to sigma-field generated by the sequence $\{Z_{j_{m}} \in \mathcal{D}(Y_{j_{m}}^{obs})\}_{m=1}^{n}$. By Theorem~\ref{consistMA-misv2}, it follows that $F_{j}(x)$ is a strongly consistent estimator of $\tilde{F}_{j}(x)$ and hence $F_{j}(x)$ is measureable with respect to the sigma-field generated by the sequence of $\{Z_{j_{n}} \in \mathcal{D}(Y_{j_{n}}^{obs})\}_{n=1}^{\infty}$.  Thus, the posterior of $\tilde{F}_{j}(x)$  under the RL is strongly consistent at $F_{j}(x)$. This applies for each continuous and count variable $j  = 1, \dots, p$.  
\end{proof}
\begin{customthm}{4}
Let $\mathbb{C}_{GMC}(\boldsymbol u)= \Psi(\psi^{-1}_{1}\{F_{1}(y_{1})\}, \dots,\psi^{-1}_{p}\{F_{p}(y_{p})\})$, where $\Psi = \sum_{h=1}^{H} \pi_{h}\Phi_{p}(\boldsymbol{\alpha}_{h},\boldsymbol{C}_{h})$, $\psi_{j} = \sum_{h=1}^{H} \pi_{h}\Phi(\{\boldsymbol{\alpha}_{h}\}_{j},\{\boldsymbol{C}_{h}\}_{jj})$, and $\{F_{j}\}_{j=1}^{p}$ are the marginals of $\{Y_{j}\}_{j=1}^{p}$. Then,
 $\mathbb{C}_{GMC}$ defines a valid copula.
\end{customthm}

 \begin{proof}
To prove that $\mathbb{C}_{GMC}$ defines a valid copula, we verify that it satisfies the following three properties:

\begin{enumerate}
    \item $\mathbb{C}_{GMC}(u_{1},\dots, u_{p})$ is non-decreasing in each component $j \in \{1,\dots,p\}$
    
    Let $j \in \{1,\dots,p\}$ be arbitrary and consider $u_{j1}<u_{j2}$.  Define $z_{j1} = \psi_{j}^{-1}(u_{j1})$ and $z_{j2} = \psi_{j}^{-1}(u_{j2})$. Because $\psi_{j}$ is a valid continuous distribution function, it  is monotone, and therefore $z_{j1}<z_{j2}$. \\~\\
    Consider the ratio
    \begin{align*}
        \frac{\mathbb{C}_{GMC}(u_{1},\dots,u_{j1}\dots,u_{p})}{\mathbb{C}_{GMC}(u_{1},\dots,u_{j2}\dots,u_{p})} &= \frac{\Psi\{\psi^{-1}_{1}(u_{1}),\dots, z_{i1}\dots,\psi^{-1}_{p}(u_{p})\}}
      {\Psi\{\psi^{-1}_{1}(u_{1}),\dots, z_{j2}\dots,\psi^{-1}_{p}(u_{p})\}}\\
      &=\sum_{h=1}^{H} \pi_{h} \frac{\Phi\left[\{\psi^{-1}_{1}(u_{1}),\dots, z_{j1}, \dots,\psi^{-1}_{p}(u_{p})\};\boldsymbol{\alpha}_{h}, \boldsymbol{C}_{h}\right]}{\Phi\left[\{\psi^{-1}_{1}(u_{1}),\dots, z_{j2}, \dots,\psi^{-1}_{p}(u_{p})\};\boldsymbol{\alpha}_{h}, \boldsymbol{C}_{h}\right]}
    \end{align*}
By the properties of multivariate Gaussian random vectors, the sum simplifies to
  \begin{align}
      &\sum_{h=1}^{H} \pi_{h} \frac{\Phi\left[z_{j1};\{\psi_{l}^{-1}(u_{l})\}_{l\neq j},\alpha_{h}^{*},\sigma_{h}^{2^{*}}\right]}{\Phi\left[z_{j2};\{\psi_{l}^{-1}(u_{l})\}_{l\neq ij},\alpha_{h}^{*},\sigma_{h}^{2^{*}}\right]}< \sum_{h=1}^{H} \pi_{h} = 1\label{ineq} \\
      &\implies \mathbb{C}_{GMC}(u_{1},\dots,u_{i1}\dots,u_{p})< \mathbb{C}_{GMC}(u_{1},\dots,u_{i2}\dots,u_{p})
  \end{align}
   
  where $\alpha_{h}^{*}, \sigma_{h}^{2^{*}}$ are the conditional mean and variance of the Gaussian random variable obtained by conditioning on $\{\psi^{-1}_{l}(u_{;})\}_{l\neq j}$ for cluster $h$. The inequality is due to the fact univariate Gaussian distribution functions are strictly monotone, implying that the ratio inside the sum in~\eqref{ineq} is strictly less than 1 for each component $h$. 

    \item  For any $j \in \{1,\dots,p\}, \ \mathbb{C}_{GMC}(u_{1} = 1,\dots,u_{j} = u,\dots,u_{p} = 1) = u$

    Note that $\forall j \in \{1,\dots,p\}$, $\psi_j^{-1}(1) =\sum_{h=1}^{H} \pi_{h}\Phi^{-1}\{1;{(\boldsymbol{\alpha}_{h})_{j}, (\boldsymbol{C}_{h})_{jj}}\} = \infty $.
    
     Using the above result, it is simple to see that
    \begin{align*}
    \mathbb{C}_{GMC}(u_{1} = 1,\dots,u_{j} = u,\dots,u_{p} = 1) &= \Psi\{\infty,\dots, \psi_{j}^{-1}(u),\dots,\infty\}\\
    &= \psi_{j}\{\psi_{j}^{-1}(u)\}\\ &= u
    \end{align*}
    \item For $a_{j}<b_{j},\  a_{j},b_{j} \in[0,1] ,\  j = 1,\dots,p,\
    \mathbb{C}_{GMC}(u_{1} \in [a_{1},b_{1}], \dots, u_{p}\in[a_{p},b_{p}])\geq 0$
    \begin{align*}
        \mathbb{C}_{GMC}(u_{1} \in [a_{1},b_{1}], \dots, u_{p}\in[a_{p},b_{p}]) &= \mathbb{C}_{GMC}(u_{1}\leq b_{1}, \dots, u_{p}\leq b_{p})- \mathbb{C}_{GMC}(u_{1}\leq a_{1}, \dots, u_{p}\leq a_{p})\\
        &= \Psi\{\psi^{-1}_{1}(b_{1}),\dots, \psi^{-1}_{p}(b_{p})\} - \Psi\{\psi^{-1}_{1}(a_{1}),\dots, \psi^{-1}_{p}(a_{p})\} \\
        &\geq 0 \ \text{(By 1.)}
    \end{align*}
\end{enumerate}
\end{proof}


\section{Bayesian RPL Gaussian Copula Sampling with Unordered Categorical}

Section \ref{estim} outlines the sampling algorithm for the Bayesian RL Gaussian copula with missing data for $\boldsymbol Y$ comprised of numeric variables. To incorporate unordered categorical in to model, one simple modification is required to Step 1 of Algorithm~\ref{alg1}.  To see this, the probit event $\mathcal{D}'(\boldsymbol Y^{q})$ dictates that for any categorical variable $Y_{c}$ with $k_{c}$ levels, if $y_{ic} = m$, the $k_{c}$ dimensional latent data vector must satisfy the event $z_{i_{k_{m}}} >0 \cap z_{i_{k_{\ell}}} <0, \ell \neq m $. Therefore, the upper bounds for each $\boldsymbol Z^{obs}_{ij}$ are pre-specified at either $0$  or $\infty$, while the lower bounds are similarly pre-specified at $-\infty$ or $0$ for any $j$ corresponding to a categorical level. If the indicator $\gamma_{ij} = 1$, then $z_u = \infty, z_\ell = 0$. On the other hand, for $\gamma_{ij} = 0 $ then $z_u = 0, z_\ell = -\infty$.
The sampling step for $\boldsymbol Z_{ij}^{mis}$ first calculates the predictive probability that $Y^{mis}_{ic} = m$, and then samples the corresponding latent vector with identical upper and lower truncation bounds as the observed case depending on the predictive level. Step 3 in Section \ref{Gibbs} contains details on this computation under the GMC-MA.
\section{Model Specification and Gibbs Sampling Algorithm}\label{MCMC}
\subsection{Global-local shrinkage priors for $\boldsymbol{\Lambda}$}
In the main paper, we mention the use of global-local shrinkage priors for the parameters of the factor-loading matrix $\boldsymbol \Lambda = \{\lambda_{jk}\}$. The following prior encourages columnwise shrinkage for rank selection \citep{bhattacharya2011sparse}: $\lambda_{jt} \sim N(0,\phi_{jh}^{-1}\tau^{-1}_{h})$ with local scale parameters $\phi_{jh} \sim \mbox{Gamma}(\nu_{\phi}/2,\nu_{\phi}/2)$ and global scale parameters $\tau_{h} = \prod_{l=1}^{h}\delta^{\tau}_{l}$, with $\delta^{\tau}_{1} \sim \mbox{Gamma}(a_{1},1) \ \text{and} \ \delta^{\tau}_{l} \sim \mbox{Gamma}(a_{2},1), \ l\geq 2, \ a_{2} \geq 1$. By design, this ordered shrinkage prior reduces sensitivity to the choice of $k$, provided $k$ is sufficiently large. Throughout our simulation studies and real data analysis, we set $a_1 = 2, a_2 = 3, \nu_\phi = 3$.
\subsection{Gibbs Sampling for the RPL GMC-MA}\label{Gibbs}
Bayesian estimation of the GMC-MA under the RPL in the presence of missing data alternates sampling model parameters from their marginal posteriors conditional on complete latent data, and then sampling latent data corresponding to $\boldsymbol{Y}^{mis}$ given $\boldsymbol{Z}^{obs}$ and model parameters. Two aspects of our model simplify this task. First, the margin adjustments $\{\tilde{F}_{j}\}_{j=1}^{r}$ are functionals of posterior samples of GMC parameters, and second, GMC parameters depend only latent $\boldsymbol{Z} = (\boldsymbol{Z}^{mis},\boldsymbol{Z}^{obs})$ through the RPL. Conjugate priors for GMC parameters allow for simple Gibb's sampling steps, while the factor model~\ref{factormix} allows for independence among the components of $\boldsymbol{Z}_{i}$ conditional on $\boldsymbol{\eta}_{i}$. Consequently, the sampling of  $\boldsymbol{Z}^{mis}$ is quite efficient, as the predictive distribution for each component is conditionally univariate normal. 

The algorithm is broken down into five blocks for simplicity. In each, $\boldsymbol{z_{i}}$ is assumed complete, meaning that components corresponding to missing values in $\boldsymbol{y}_{i}$ ($\boldsymbol{z}^{mis}_{i}$) have been sampled.

\begin{enumerate}
    \item \textbf{Sample Cluster Specific Parameters}
    For each cluster $1, \dots, H$
    \begin{itemize}
        \item $c_{i}\mid - \sim \mbox{Multinomial}(\boldsymbol{p})$, where $\boldsymbol{p} = (p_{1},\dots, p_{H})$ and $p_{h} \propto \pi_{h}  \psi_{k}(\boldsymbol{\eta}_{i};\boldsymbol{\mu}_{h}, \boldsymbol{\Delta}_{h})$
        \item $V_{h} \mid - \sim \mbox{Beta
        }(1+n_{h}, \alpha_{\pi} + \sum_{v = h+1}^{H}n_{v})$, $h = 1,\dots, H-1$, $n_{h} = \sum_{i=1}^{N}\mathbb{I}(c_{i} = h)$ 
        \item $\boldsymbol{\Delta}_{h}\mid - \sim \mbox{IW}(\nu_{post}, \Psi_{post})$, $\nu_{post} = \nu_{0} + n_{h}$, $\Psi_{post} = \boldsymbol{I}_{k} + \boldsymbol{S}_{h} + \frac{\kappa_{0}n_{h}}{\kappa_{0} + n_{h}}\boldsymbol{T}_{h}$,\newline $\boldsymbol{S}_{h} = \sum_{i:c_{i} = h} (\boldsymbol{\eta}_{i} - \bar{\boldsymbol{\eta}}_{h})(\boldsymbol{\eta}_{i} - \bar{\boldsymbol{\eta}}_{h})^{T}, \boldsymbol{T}_{h} = (\boldsymbol{\mu}_{0} - \bar{\boldsymbol{\eta}}_{h})(\boldsymbol{\mu}_{0} - \bar{\boldsymbol{\eta}}_{h})^{T}$, $\bar{\boldsymbol{\eta}}_{h} = n_{h}^{-1}\sum_{i: c_{i} = h}\boldsymbol{\eta}_{i}$
        \item $\boldsymbol{\mu}_{h} \mid - \sim N(\boldsymbol{\mu}_{post}, \kappa_{post}^{-1}\boldsymbol{\Delta}_{h})$, $\boldsymbol{\mu}_{post} = \frac{\kappa_{0}\boldsymbol{\mu}_{0} + n_{h} \bar{\boldsymbol{\eta}}_{h}}{\kappa_{0}+n_{h}}$, $\kappa_{post} = n_{h} + \kappa_{0}$
        \item $\alpha_{\pi}\mid - \sim \mbox{Gamma}(a_{\alpha} + H - 1, b_{\alpha} - \sum_{h=1}^{H-1}\log(V_{h}))$
    \end{itemize}
    \item \textbf{Sample Factor Model Parameters}
    \begin{itemize}
        \item $\boldsymbol{\eta}_{i} \mid c_{i} = h, - \sim N_{k}((\boldsymbol{\Delta}_{h}^{-1} + (\boldsymbol{\Lambda}^{T}\boldsymbol{\Sigma}^{-1}\boldsymbol{\Lambda})^{-1})^{-1}(\boldsymbol{\Lambda}^{T}\boldsymbol{\Sigma}^{-1}\boldsymbol{z}_{i} + \boldsymbol{\Delta}_{h}^{-1}\boldsymbol{\mu}_{h}), (\boldsymbol{\Delta}_{h}^{-1} + (\boldsymbol{\Lambda}^{T}\boldsymbol{\Sigma}^{-1}\boldsymbol{\Lambda})^{-1})^{-1})$
        \item $\lambda_{j,-} \mid - \sim N((\boldsymbol{D}_{j}^{-1} + \sigma_{j}^{-2}\boldsymbol{\eta}^{T}\boldsymbol{\eta})^{-1}\boldsymbol{\eta}^{T}\sigma_{j}^{-2}\boldsymbol{z}_{j})$, 
        where $\boldsymbol{D}_{j}^{-1} = diag(\phi_{j1}\tau_{1}, \dots, \phi_{jk}\tau_{jk})$, $\boldsymbol{z_{j}} = (z_{1j}, \dots z_{nj})^{T}$, and $\boldsymbol{\eta} = (\eta_{1j}, \dots \eta_{nj})^{T}$, for $j = 1, \dots, p$
        \item $\sigma_{j}^{-2} \mid - \sim \mbox{Gamma}(a_{\sigma} + \frac{n}{2}, b_{\sigma} + \frac{1}{2}\sum_{i = 1}^{n}\{z_{ij} - ( \lambda_{j}^{T}\eta_{ij})\}^{2}$, for $j = 1, \dots, p$
         \item $\phi_{jh} \mid - \sim \mbox{Gamma}( \frac{\nu + 1}{2}, \frac{\nu + \tau_{h}\lambda_{jt}^{2}}{2})$, for $j = 1, \dots, p, \ h = 1, \dots, k$ 
        \item $\delta_{1} \mid - \sim \mbox{Gamma}(a_{1} + \frac{pk}{2}, 1 + \frac{1}{2}\sum_{l = 1}^{k}\tau_{l}^{(1)}\sum_{j =1}^{p}\phi_{jl}\lambda_{jl}^{2})$, and for $h \geq 2$ \\
        $\delta_{h} \mid - \sim \mbox{Gamma}(a_{1} + \frac{p(k -h + 1)}{2}, 1 + \frac{1}{2}\sum_{l = 1}^{k}\tau_{l}^{(h)}\sum_{j =1}^{p}\phi_{jl}\lambda_{jl}^{2})$, where $\tau_{l}^{h} = \prod_{t = 1, t \neq h}^{l} \delta_{t}$, for $h = 1, \dots, k$
    \end{itemize}
    \item \textbf{Re-sample $Z^{obs}_{j},Z^{mis}_{j}$} \\
    Given the conditional independence among the components of $\boldsymbol{Z}_{i}$ given $\boldsymbol{\eta}_{i}$, components of $\boldsymbol{Z}$ corresponding to observed data points are sampled column-by-column, consistent with the ordering induced by the RPL. For components of $\boldsymbol{Z}_{i}$ associated with missing values, no ordering is imposed, and only the diagonal orthant restriction for categorical variables is enforced.
    
    \begin{itemize}
        \item \textit{Missing categorical/binary data}: If $j$ corresponds to one of the levels of categorical variable $q$ with $k_{q}$ levels, $Z^{mis}_{ij}$ and the other associated levels of $q$ must be sampled consistently with the diagonal orthant set restriction of the RPL. That is, one component of the vector $\boldsymbol{Z}_{k_{q}}$ must be positive while the others negative. To ensure this condition is met, we first calculate the predictive probability that $Y^{mis}_{iq}$ assumes the $j$th level for each $j$ among the $k_{q}$ levels, which is equivalent to 
        \begin{align}\label{imputecat}
            &P(Z^{mis}_{ij} >0 \cap Z^{mis}_{i\ell_{-j}}<0 \mid -) \propto\\& 1- \Phi(0; \sum_{t=1}^{k}\lambda_{jt}\eta_{it}, \sigma_{j}^{2})\prod_{\ell  \in \{c_{1},\dots,c_{k_{q}}\}, \ell \neq j} \Phi(0; \sum_{t=1}^{k}\lambda_{\ell h}\eta_{it}, \sigma_{\ell}^{2}),  j \in \{c_{1},\dots,c_{k_{q}}\} \nonumber
        \end{align}
        Then, we sample the level of $Y^{mis}_{ij}$ using these probabilities, with the resulting classification used in the re-sampling of $Z^{mis}_{ij}$ under the RPL. Let $\mbox{TN}(\mu, \sigma^{2},a,b)$ denote a truncated univariate normal with  mean $ \mu$, variance $\sigma^{2}$, lower truncation $a$, and upper truncation $b$. The re-sampling step for $Z^{mis}_{ij}$ is given by

    \begin{equation}\label{trunc}
        z^{mis}_{ij}\sim \begin{cases} \mbox{TN}(\sum_{t=1}^{k}\lambda_{jt}\eta_{it},\sigma_{j}^{2},0, \infty), & y^{mis}_{ij} = 1\\ \mbox{TN}(\sum_{t=1}^{k}\lambda_{jt}\eta_{it},\sigma_{j}^{2},-\infty,0), & y^{mis}_{ij} = 0 \end{cases}
    \end{equation}
    If $j$ is binary, the probability of one level versus the other is instead given by $P(z^{mis}_{ij} >0\mid c_{i} = h, -) = 1- \Phi_{h}(0; \sum_{t=1}^{k}\lambda_{jt}\eta_{it}, \sigma_{j}^{2})$, but the re-sampling step~\ref{trunc} remains the same
    \item \textit{Missing numeric data:}
    In this case, latent $Z^{mis}_{ij}$ is sampled from the unrestricted univariate Gaussian 
    \begin{equation}
        Z^{mis}_{ij} \mid - \sim N(\sum_{t=1}^{k}\lambda_{jt}\eta_{it},\sigma_{j}^{2})
    \end{equation}
    \item \textit{Observed data:} for each column, sample $Z^{obs}_{ij}$ from a truncated normal, with lower and upper bounds for each observation specified by the RPL:
    \begin{equation}
    Z^{obs}_{ij} \mid - \sim \mbox{TN}(\sum_{t=1}^{k}\lambda_{jt}\eta_{it},\sigma_{j},z^{obs^{\ell}}_{ij}, z^{obs^{u}}_{ij})
    \end{equation}
For ordinal, count, and continuous variables, the truncation limits are $z^{obs^{\ell}}_{ij} = \max\{z^{obs}_{-ij}: y_{obs_{-ij}}<y^{obs}_{ij}\}$, and $z^{obs^{u}}_{ij} = \min\{z^{obs}_{-ij}: y^{obs}_{-ij}>y^{obs}_{ij}\}$, where $z^{obs}_{-ij} = z^{obs}_{j}\setminus z^{obs}_{ij}$. For columns corresponding to categorical levels, the upper and lower truncation limits are
    \begin{equation}\label{trunc2}
        z^{obs^{\ell}}_{ij} = \begin{cases} 0, & Y^{obs}_{ij} = 1\\ -\infty, &Y^{obs}_{ij} = 0 \end{cases}, \quad    \quad          
        z^{obs^{u}}_{ij} = \begin{cases} \infty, & Y^{obs}_{ij} = 1\\ 0, & Y^{obs}_{ij} = 0\end{cases}
    \end{equation}
    \end{itemize}

    \item \textbf{Sample} $\tilde{F}_{j}$\\
    For each unique $x \in \boldsymbol{Y}^{obs}_{j}$, we first 
    find $Z^{n}_{j}(x) = \max\{Z_{ij}^{obs}: Y^{obs}_{ij} \leq x\}$, and compute 
    \begin{equation*}
        \tilde{F}_{j}(x) = \psi_{j}\{Z^{n}_{j}(x)\}
    \end{equation*}
    Where $\psi_{j}$ is a function of the current draw of GMC parameters.  To estimate $\tilde{F}_{j}$ across unobserved values, we then fit a monotone interpolating spline to $\{x,\tilde{F}_{j}(x)\}_{x \in \boldsymbol{Y}_j^{obs}}$ as described in Section~\ref{GMC}, and use this estimate to approximate $\tilde{F}_{j}(x')$ for $x' \notin \boldsymbol{Y}^{obs}_{j}$.
\end{enumerate}
The smoothing step in the sampling of $\tilde{F}_{j}$ is crucial in multiple imputation, as the transformation $Y^{mis}_{ij} = \tilde{F}^{-1}_{j}(Z^{mis}_{ij})$ provides realizations that may assume values across the entire support of variable $j$, instead of only values that were observed.

\section{Hyperparameter Tuning}\label{hyper}
Throughout our simulated and real data studies, we find that the default model specification given in Section~\ref{GMC} requires little hyperparameter tuning. 

Specifically, there are three parameters that we vary to bolster model performance. The first is the dimension of latent $\boldsymbol{\eta}$. For all of our studies, we use a default value of $\lceil 0.7p^{*}\rceil$, where $p^{*}$ is the dimension of the augmented data matrix under the RPL. 
Next, we modify the scaling constant, $\delta$, from the normal-inverse Wishart prior specified for the cluster specific components from the mixture model on $\boldsymbol{\eta}_{i}$. Recall the following hierarchical model structure
\begin{align*}\label{eq3}
    \boldsymbol{z}_{i}  &\sim N(\boldsymbol{\Lambda}\boldsymbol{\eta}_{i}, \boldsymbol{\Sigma})\\
    \boldsymbol{\eta}_{i} &\sim \sum_{h=1}^{H}\pi_{h}N_k(\boldsymbol{\mu}_{h}, \boldsymbol{\Delta}_{h})\\
    (\boldsymbol{\mu}_{h}, \boldsymbol{\Delta}_{h}) &\sim \mbox{NIW}(\boldsymbol{\mu}_{0}, \delta^{2}\boldsymbol{I}_{k},\kappa_{0} = 0.001, \nu_{0} = k+2)
\end{align*}

In the simulation studies and real data analysis, we found that  $\delta$ impacts model fit through the number of clusters that are discovered.  Though we recommend a default value of $\delta = 10$, we find that generally, decreasing $\delta$ has the effect of increasing the number of clusters discovered.  As such, we use $\delta = 5$ in the second simulation study as a lack of separability in the hybrid data reduces the stability of the repeated model fits with $\delta = 10$. Posterior predictive diagnostics---for instance checking marginal and multivariate properties of posterior predictive datasets created via the sampling algorithm in Section \ref{postpredsamp}---may be used to tune this parameter .

The other parameter tuned is the number of unique values present among each numeric variable for re-sampling to occur under the RPL data augmentation mentioned in Section~\ref{MCMC}. For continuous variables, the number of unique levels observed in the data will be $n$.  As such, re-sampling columns of $\boldsymbol{Z}$ associated with such variables above could be computationally intensive, as it would necessitate looping through all unique values of the continuous variable at each iteration of the MCMC.

Instead of enduring this expense, we instead choose an upper bound for the number of unique levels that a particular variable may have to engage in the re-sampling in step 3 of our MCMC algorithm for copula estimation. Any latent columns corresponding to observed variables having more than this number of unique levels are not re-sampled within the MCMC. Instead, they are initialized through a ``pseudo-data" transformation, which results from scaling the column to have mean zero and unit variance. This is done to maintain the orderings and marginal properties of each continuous variable, like skew and multi-modality, which effectively improves the model and margin estimation while placing latent variables on a common scale. In our applications, we choose this threshold to be 350.

\section{Posterior Predictive Sampling Algorithm}\label{postpredsamp}
Section~\ref{realdat} utilizes posterior predictive inference to highlight discrepancies between a complete case analysis and one that accounts for potentially MAR missing data.  We include here the algorithm developed to produce the posterior predictive data sets used for this analysis. 

The procedure is facilitated by the conditional independence implied by the factor model developed in Section~\ref{GMC}. The algorithm begins with ordinary Gaussian mixture sampling steps for sampling of predictive $\tilde{\boldsymbol{\eta}}_{i}$, which in turn enables sampling of predictive $\tilde{\boldsymbol{z}}_{i}$. We then link each component of $\tilde{\boldsymbol{z}}_{i}$ with $\tilde{F}^{-1}_{j}\{\psi_{j}(\tilde{z}_{ij})\}$ for numeric variables, or with the categorical link mentioned in Section~\ref{MCMC}. 
\begin{algorithm}[H]
\caption{Simulation of a posterior predictive data set of size $n$}\label{predsamp}
\includegraphics[width = \textwidth, keepaspectratio]{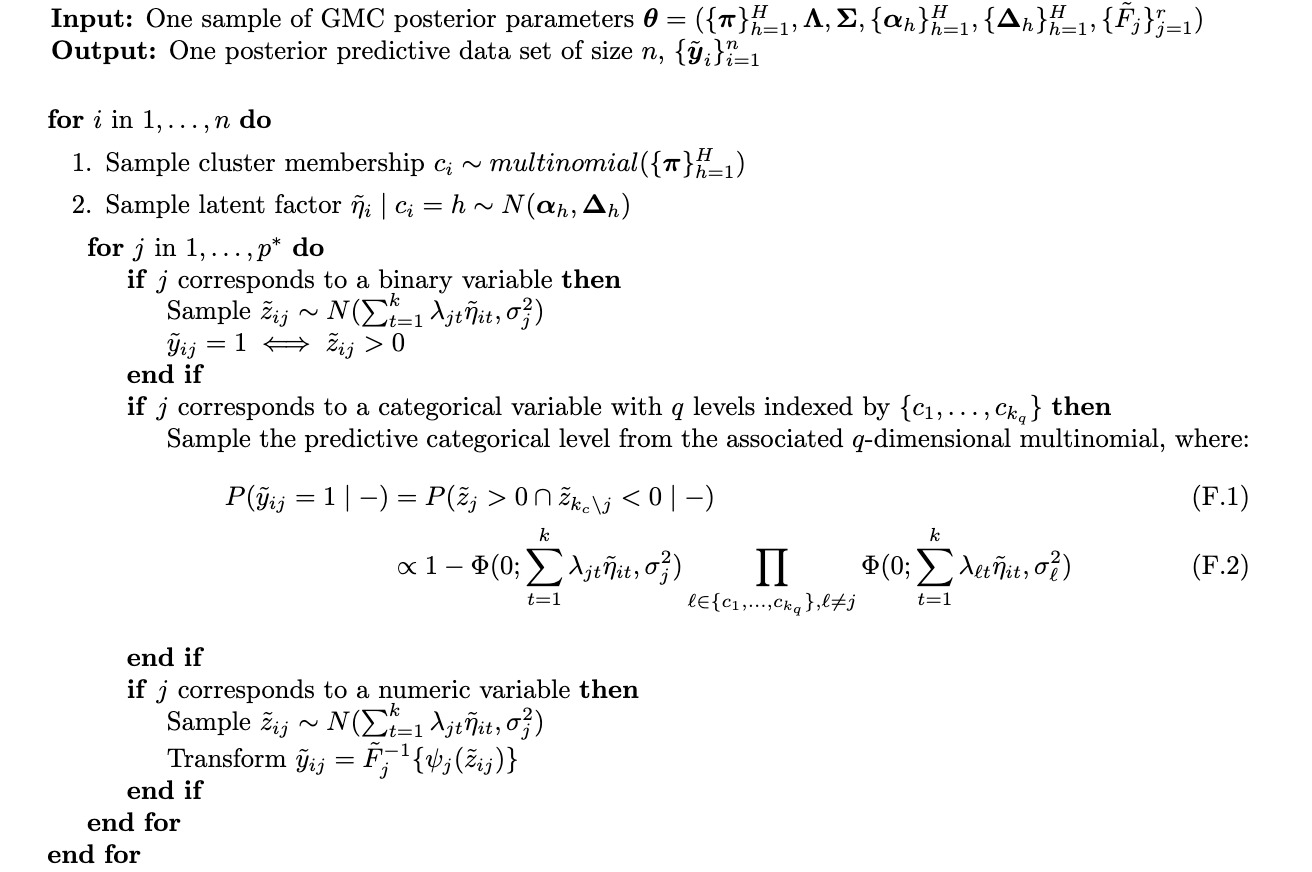}
\end{algorithm}

\section{Further Simulation Results}
\subsection{Mixed Data Types, Nonlinearity, and MAR}
\begin{figure}[h]
    \centering
\includegraphics[width =.27\textwidth, keepaspectratio]{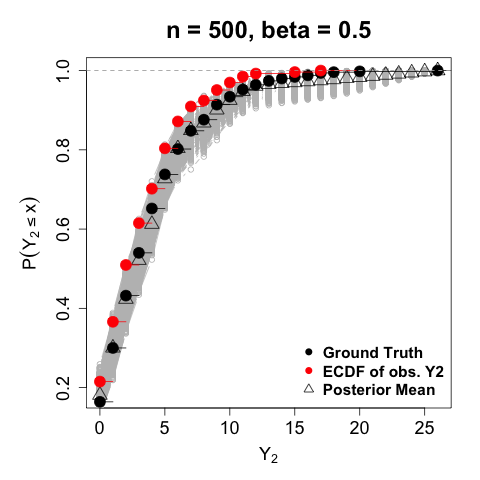}
\includegraphics[width =.27\textwidth, keepaspectratio]{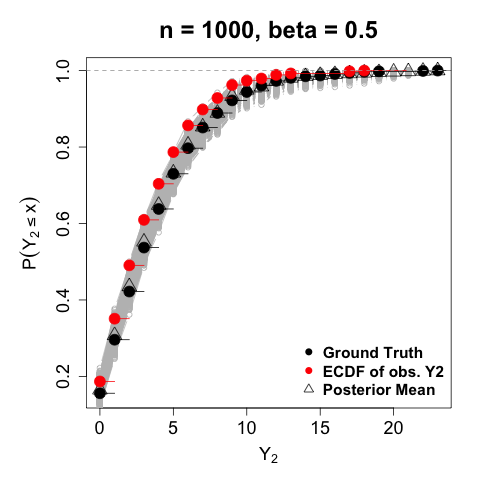}
\includegraphics[width =.27\textwidth, keepaspectratio]{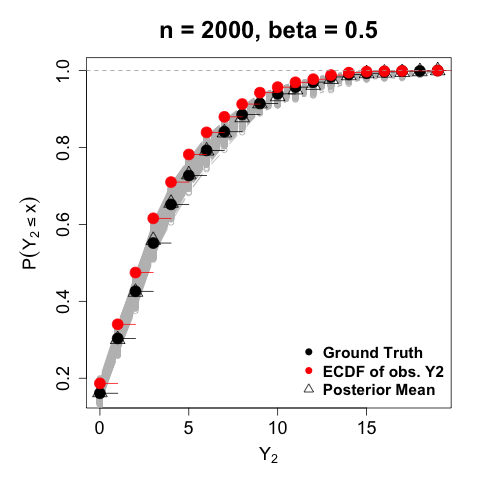}
    \caption{\small Estimation and inference for the marginal distribution of $Y_2$ under MAR ($\beta = 0.5$) with varying $n$. The ECDF of $Y_2^{obs}$ (red points) deviates significantly from the ECDF of $Y_2$ prior to removing missingness (black points). The posterior draws (gray lines) and posterior mean (triangles) from the margin adjustment show that the proposed approach is highly accurate, even under severe MAR.}
    \label{SS1-.5}
\end{figure}

In Figure~\ref{SS1-.5}, we include the evaluation of the margin adjustment for $\beta = 0.5$, analogous to what was presented in Figure~\ref{SS1} in the main paper. With less severe missingness and bias, the margin adjustment more efficiently recovers the ground truth marginals.

 As mentioned in Section~\ref{sim-1}, we also estimate the posterior distribution of the probability of a positive indicator for binary variable $X_{3}$. Under Model~\eqref{factormix} and the RPL, this is simply the probability that latent $Z_{3}$ is greater than zero. This quantity is computed as $1- \sum_{h=1}^{H}\pi_{h}^{s}\Phi\{0;(\boldsymbol{\Lambda}^{s}\boldsymbol{\mu}^{s}_{h})_{3},(\boldsymbol{\Lambda}^{s}\boldsymbol{\Delta}_{h}^{s}\boldsymbol{\Lambda}^{s} + \boldsymbol{\Sigma}^{s})_{33}\}$, where the superscript $s$ denotes the $s$th posterior sample of model parameters.
 
\begin{figure}[h]
    \centering
    \includegraphics[width =.46\textwidth, keepaspectratio]{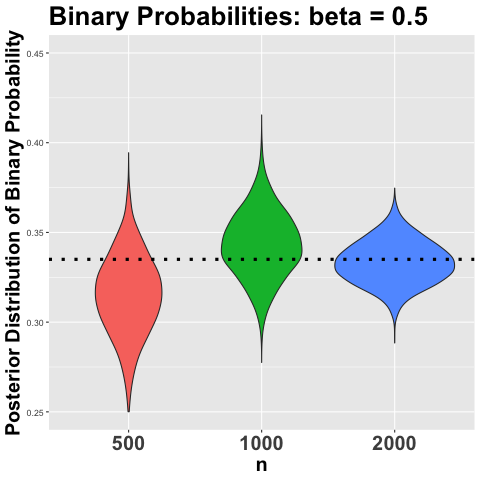}
    \includegraphics[width =.46\textwidth, keepaspectratio]{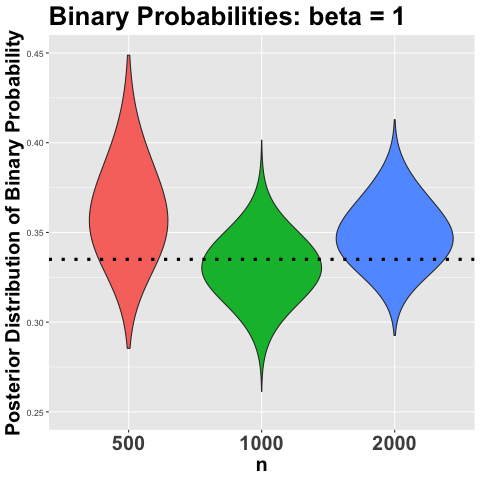}
    \caption{Posterior Probability of a positive indicator for $X_{3}$: As $n$ increases the uncertainty decreases and the distribution concentrates around the ground truth (dotted line)}
    \label{binary}
\end{figure}

 To evaluate the proposed model, we compare the posterior probability of $X_{3}$ to a ``ground truth" value of 0.335, which is the empirical probability of a positive indicator upon simulating 10,000,000 observations under the data generating model.  In Figure~\ref{binary}, we plot the posterior probability of a positive indicator for each $(n,\beta)$ combination. In both plots, we use 10,000 posterior samples for inference, and see that the distributions contract around the ground truth value of 0.335 as the sample size increases, with expected precision loss due the amount of missing data caused by varying $\beta$. This result further develops the potential for the categorical margin adjustment to be extended to more general copula models. For both $\beta$ settings, the missingness mechanism badly biases the empirical estimate of the probability of a positive $X_{3}$; for $\beta = 0.5$, this probability is on average 0.26, while for $\beta = 1$, this probability is $0.23$. Like numeric margins, we see posterior inference for binary proportions under the proposed approach correcting the bias caused by missing data.

\begin{figure}[h]
    \centering
    \includegraphics[width = .33\textwidth, keepaspectratio]{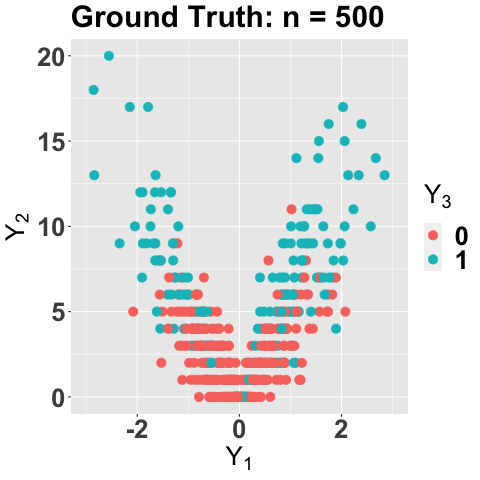}
    \includegraphics[width = .33\textwidth, keepaspectratio]{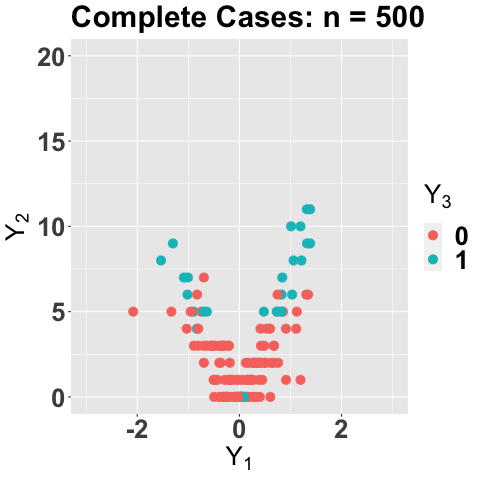}
    \includegraphics[width = .33\textwidth, keepaspectratio]{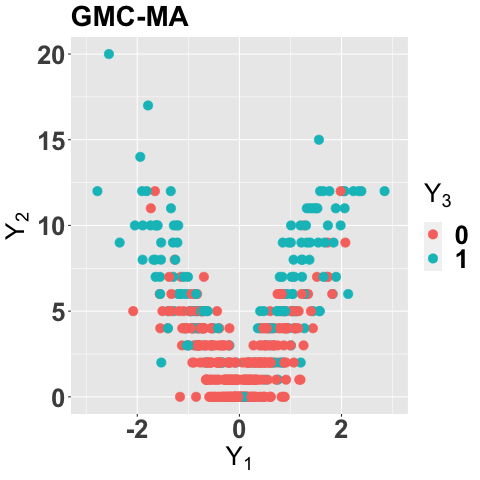}
    \includegraphics[width = .33\textwidth, keepaspectratio]{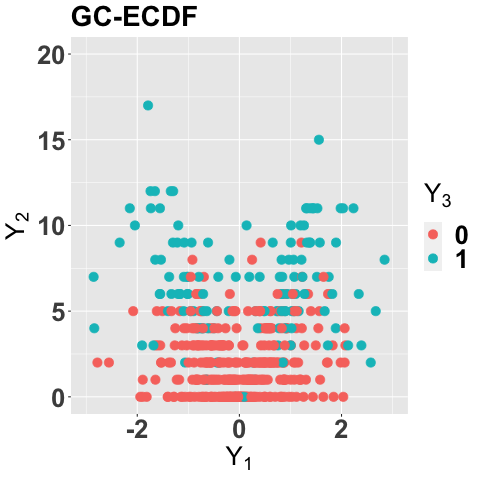}
    \caption{$n = 500, \beta = 0.5$}
    \label{500_5}
\end{figure}
\begin{figure}[h]
    \centering
    \includegraphics[width = .33\textwidth, keepaspectratio]{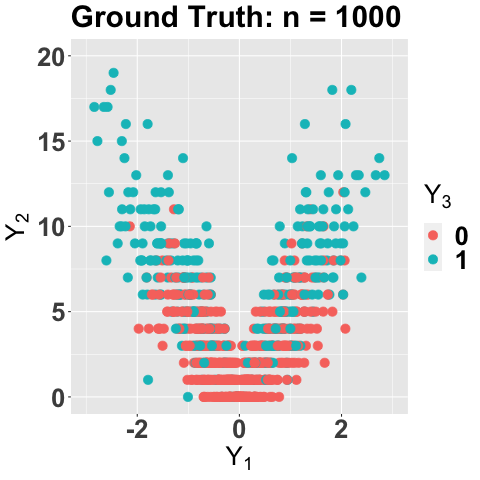}
    \includegraphics[width = .33\textwidth, keepaspectratio]{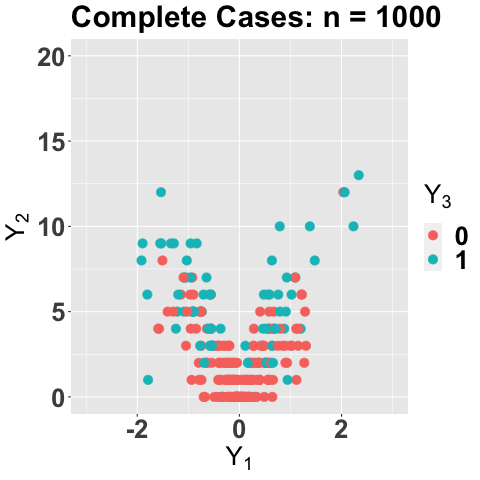}
    \includegraphics[width = .33\textwidth, keepaspectratio]{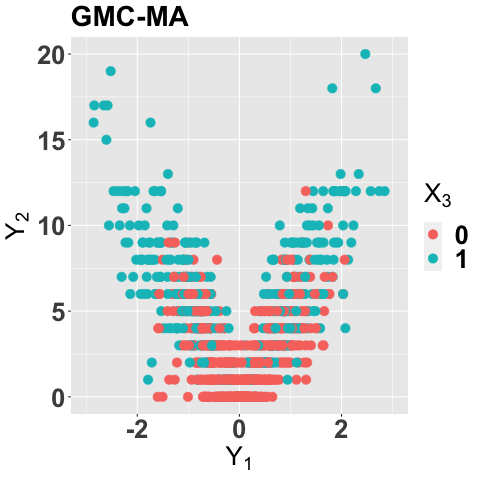}
    \includegraphics[width = .33\textwidth, keepaspectratio]{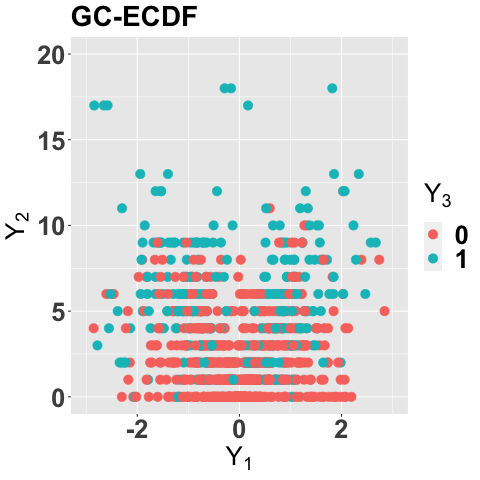}
    \caption{$n = 1000, \beta = 0.5$}
    \label{1000_5}
\end{figure}
\begin{figure}[h]
    \centering
    \includegraphics[width = .33\textwidth, keepaspectratio]{Images/GTn500.png}
    \includegraphics[width = .33\textwidth, keepaspectratio]{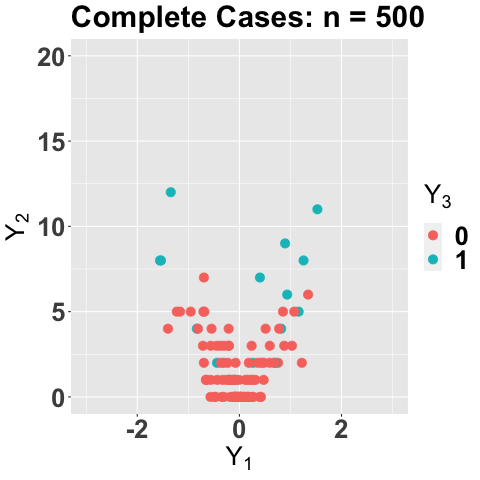}
    \includegraphics[width = .33\textwidth, keepaspectratio]{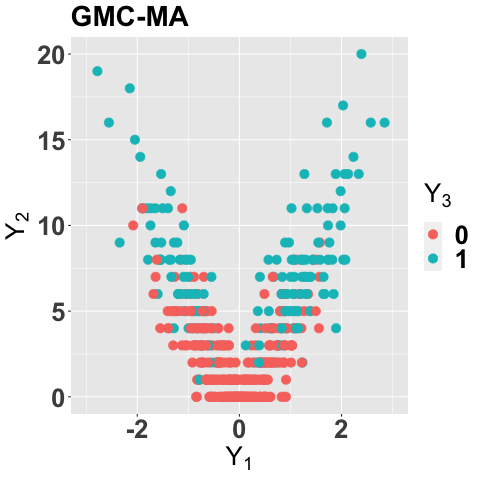}
    \includegraphics[width = .33\textwidth, keepaspectratio]{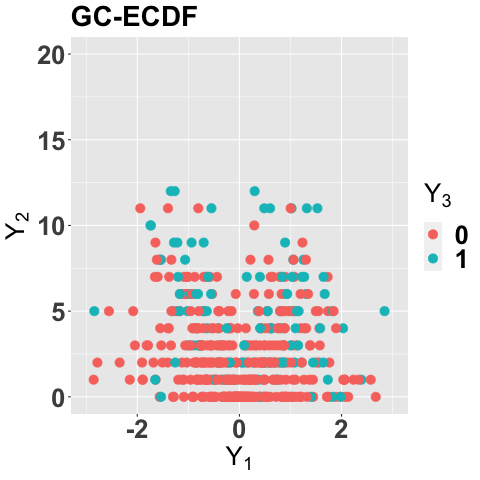}
    \caption{$n = 500, \beta = 1$}
    \label{500_1}
\end{figure}
\begin{figure}[h]
    \centering
    \includegraphics[width = .33\textwidth, keepaspectratio]{Images/GTn1000.png}
    \includegraphics[width = .33\textwidth, keepaspectratio]{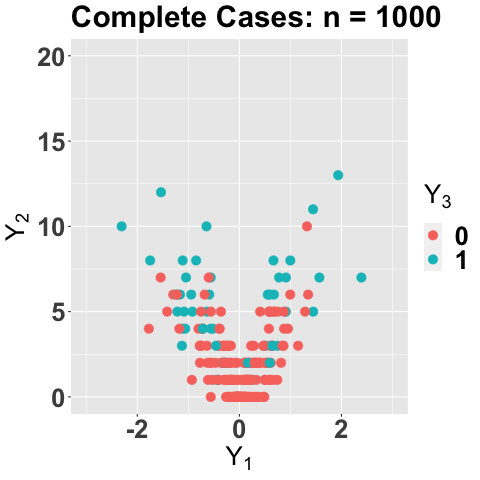}
    \includegraphics[width = .33\textwidth, keepaspectratio]{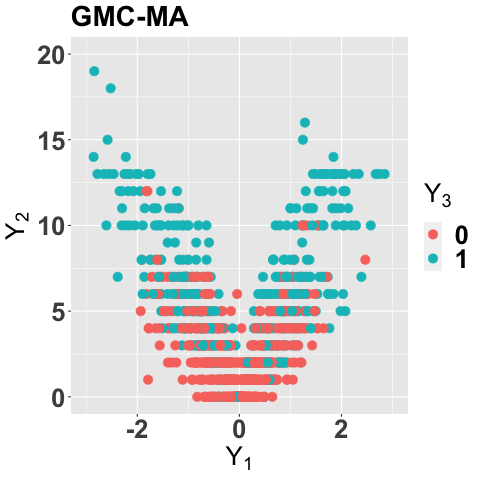}
    \includegraphics[width = .33\textwidth, keepaspectratio]{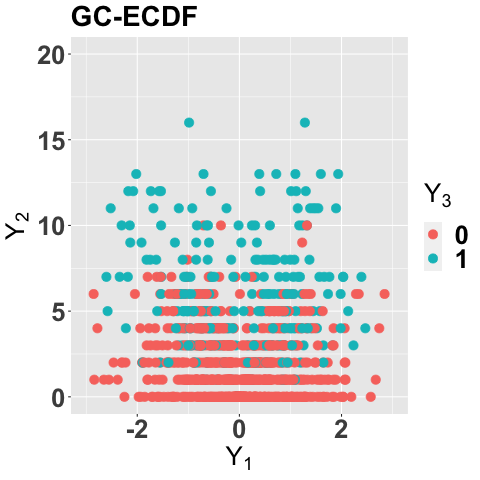}
    \caption{$n = 1000, \beta = 1$}
    \label{1000_1}
\end{figure}
\begin{figure}[h]
    \centering
    \includegraphics[width = .33\textwidth, keepaspectratio]{Images/GTn2000.png}
    \includegraphics[width = .33\textwidth, keepaspectratio]{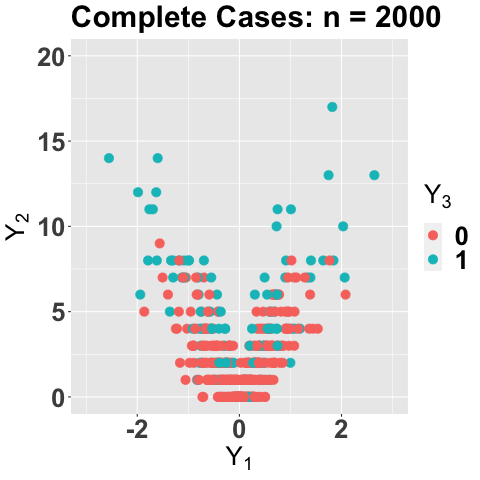}
    \includegraphics[width = .33\textwidth, keepaspectratio]{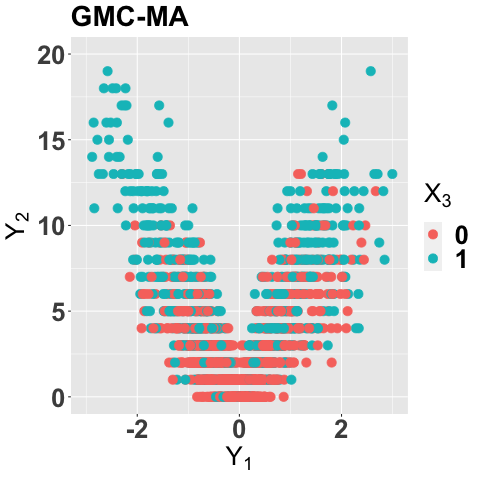}
    \includegraphics[width = .33\textwidth, keepaspectratio]{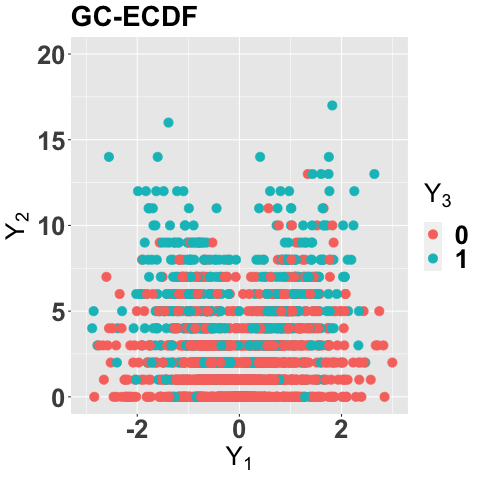}
    \caption{$n = 2000, \beta = 1$}
    \label{2000_1}
\end{figure}

Next, we include analogous plots to Figure~\ref{imp} in the main paper for each additional $(n,\beta)$ combination in Figures~\ref{500_5}-\ref{2000_1}. In each imputation procedure, the proposed approach is able to model non-linearity in the data, whereas the Gaussian copula \citep{hoff2007extending} is ineffective. Notice the consistency with which GMC-MA imputations capture specific features in the data,  from the curvature in the relationship between $Y_{1}$ and $Y_{2}$ to the enhanced probability that $Y_{3} = 1$ for large values of both $Y_{1}$ and $Y_{2}$, regardless of sample size and the amount of missingness. In addition, the margin adjustment relieves reliance on the ECDF for multiple imputation, yielding much more broad support in realized values of missing $Y_{2}$.


\subsection{Imputation for Regression Analysis}

\begin{figure}[h]
    \centering
        \includegraphics[width = .4\textwidth, keepaspectratio]{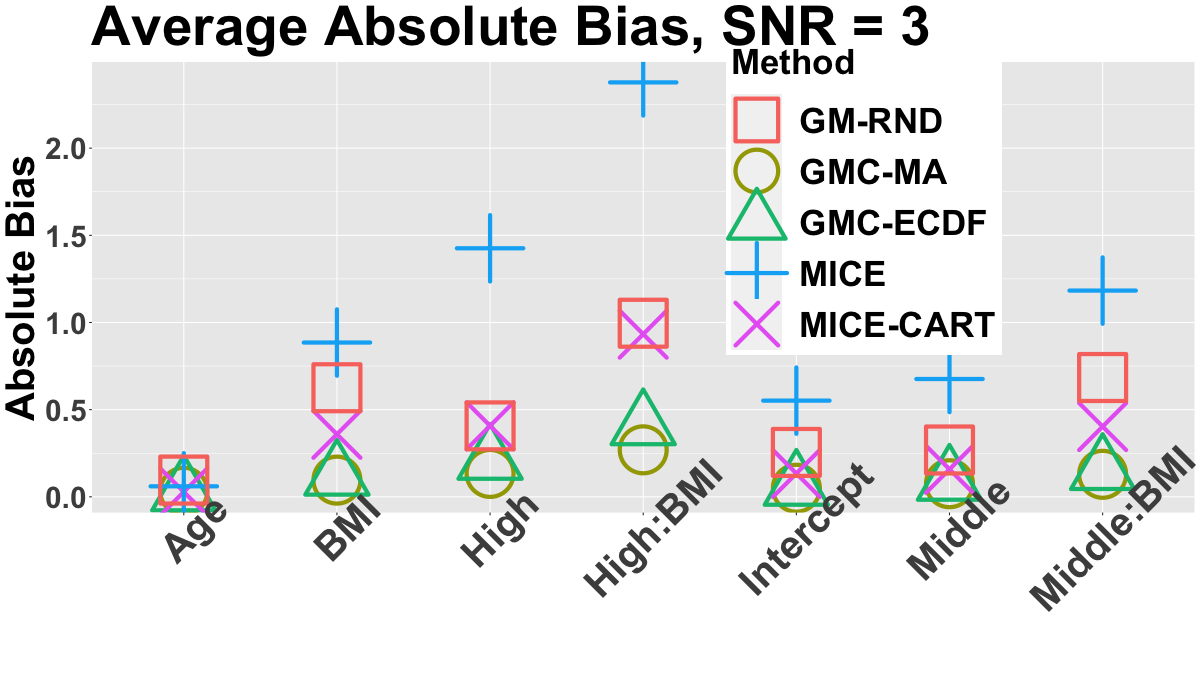}
        \includegraphics[width = .4\textwidth, keepaspectratio]{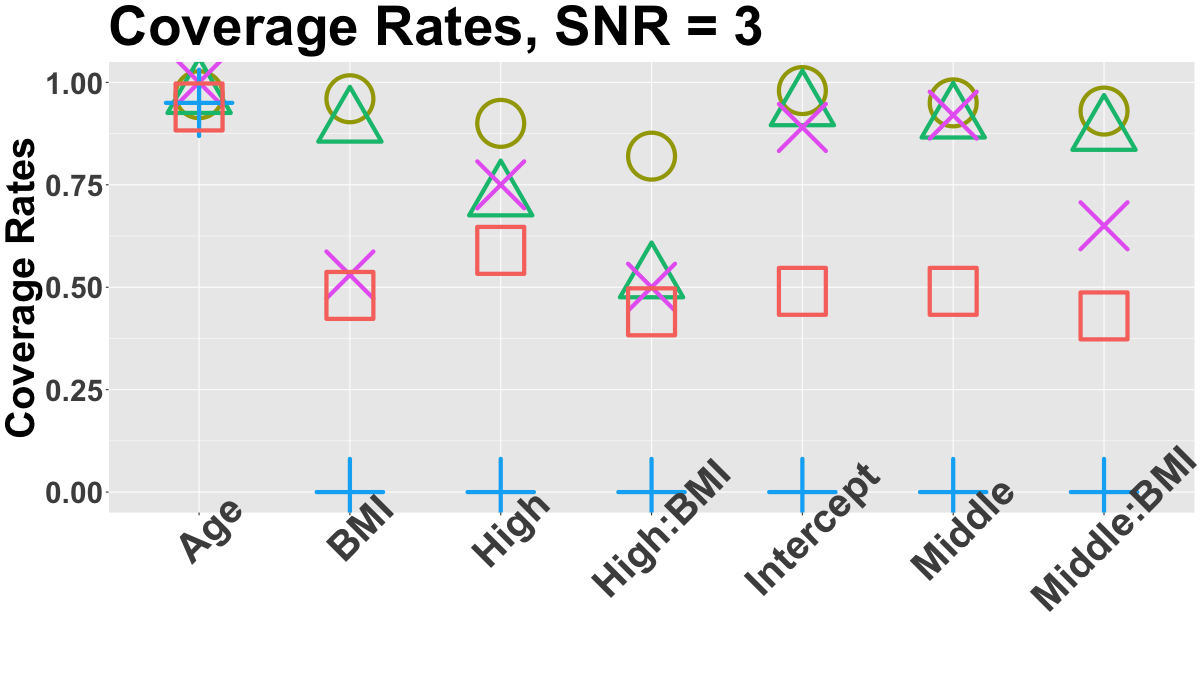}
        \includegraphics[width = .4\textwidth, keepaspectratio]{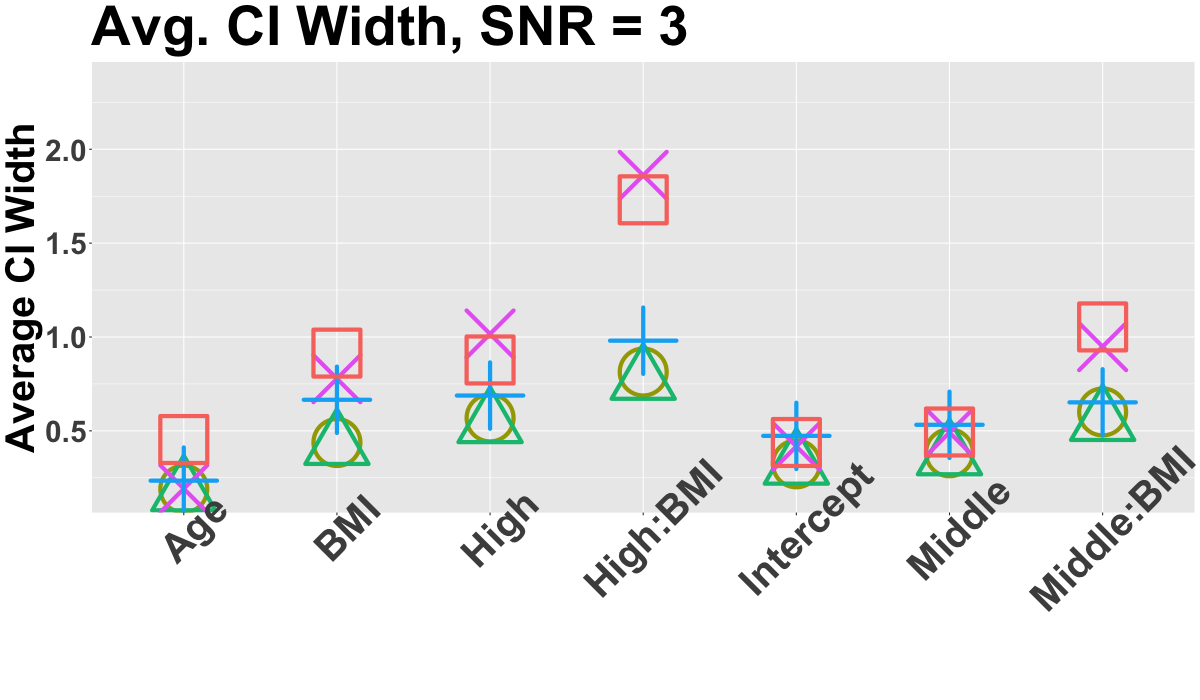}
    \caption{\small $\mbox{SNR} = 3$ absolute bias (left), interval coverage rates (center), and interval widths (right) for point and 99\% interval estimates computed under each imputation method. The GMC-MA approach consistently provides the most accurate point estimates (small absolute bias), the most well-calibrated intervals (large coverage rates), and highly precise inference (small interval widths).} 
    \label{coverage3}
\end{figure}

For completeness, we include in Figure~\ref{coverage3} simulation results for $\mbox{SNR} = 3$, noting similarly exceptional performance of the GMC-MA in terms of point estimation, interval width, and interval calibration. In addition, the shortcomings of the competitors are once again apparent

\begin{figure}[h]
    \centering
        \includegraphics[width = .65\textwidth, keepaspectratio]{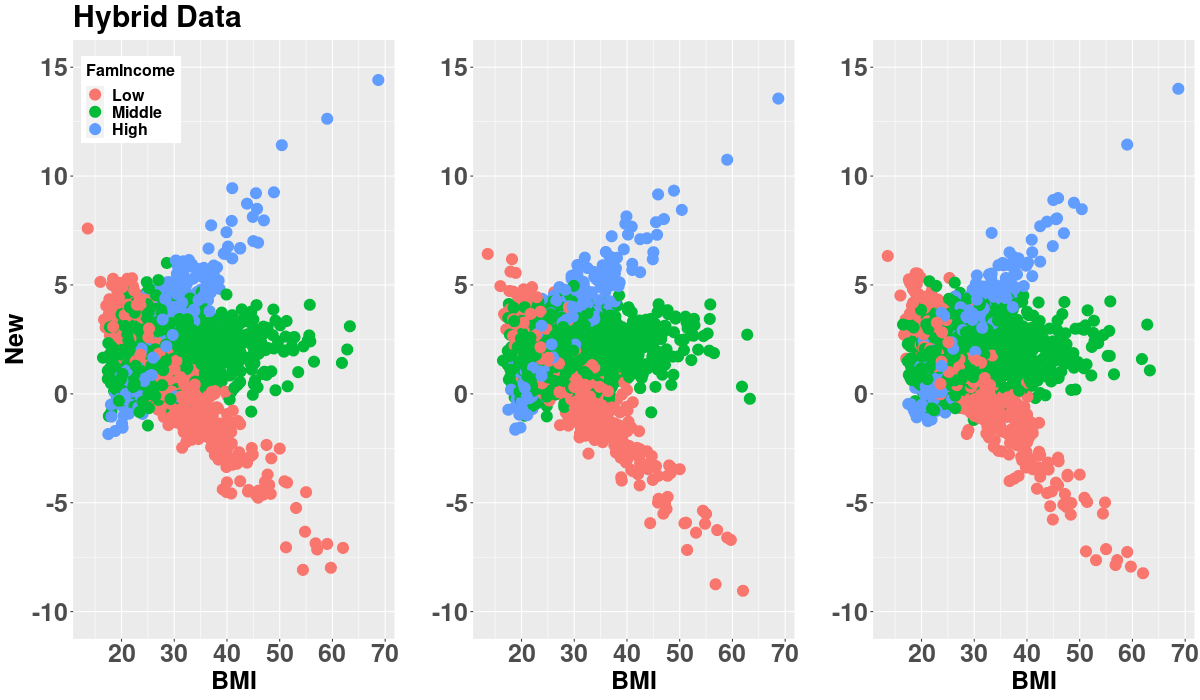}
         \includegraphics[width = .65\textwidth, keepaspectratio]{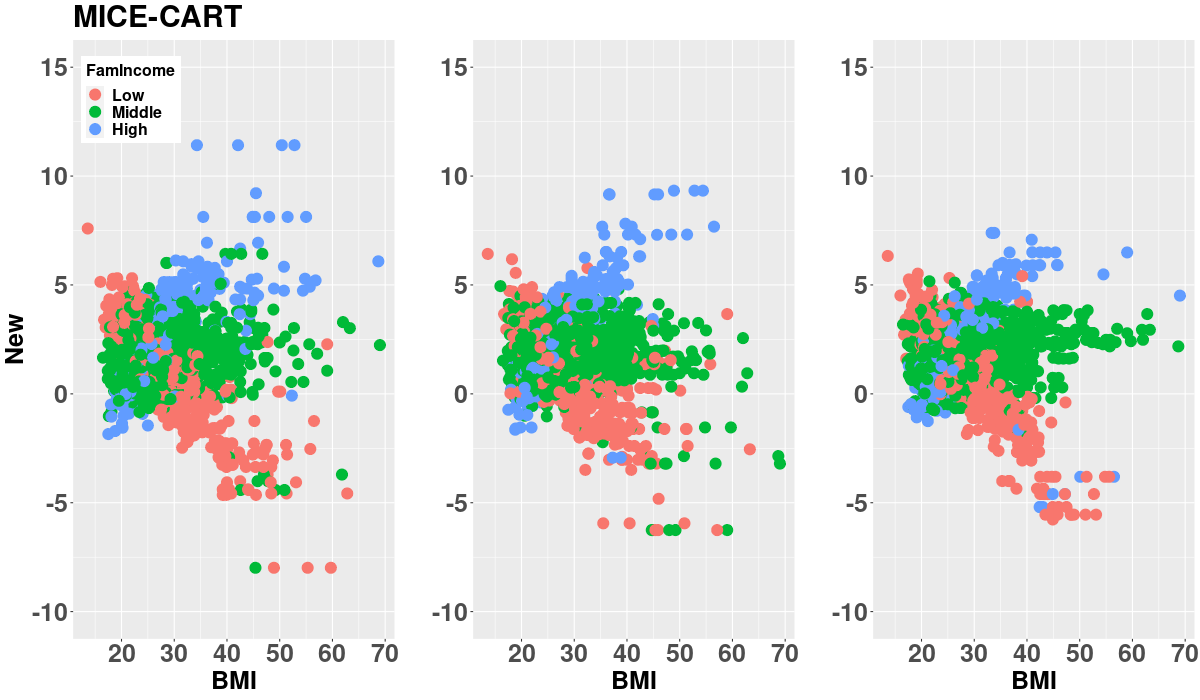}
         \includegraphics[width = 0.65\textwidth]{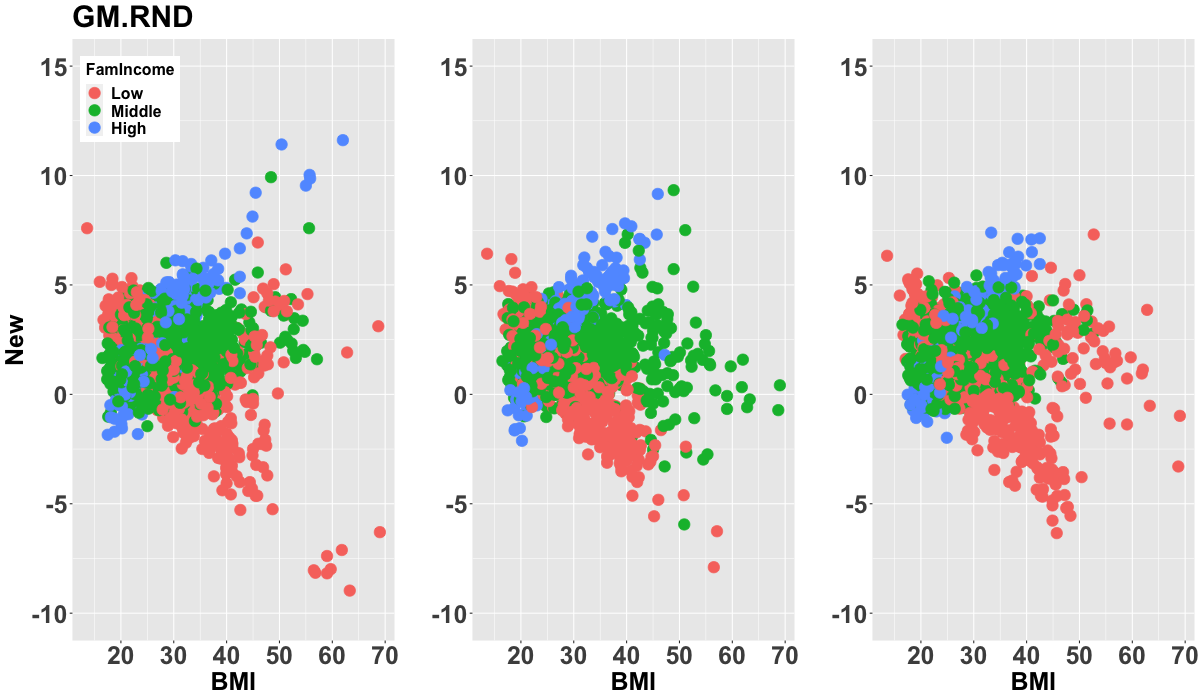}
    \caption{MICE-CART completed data sets compared to ground truth from the 1st (left panel), 50th (middle panel) and 100th (right panel) iterations of the repeated simulation study from Section~\ref{hybrid}. MICE with CART is unable to capture the interactive relationship between \vtt{New}, \vtt{Age}, and \vtt{BMI}, as demonstrated by misclassified \vtt{FI} at the tails of \vtt{New} and \vtt{BMI}. These classifications leverage the regression fit, yielding inaccurate estimates and uncertainty quantification for the regression model of interest. For GM-RND, the continuous relaxation of \vtt{FamIncome} clearly leads to many erroneous imputations, which greatly biases the regression inference.}
    \label{Impcomp2}
\end{figure}

Also highlighted in Section~\ref{hybrid}, MICE-CART yields substantial coverage gaps in estimation of the interaction terms in the regression model of interest. We hypothesize that this is due to model inefficiency of CART when a parametric model is suitable. To visualize this, we plot the interaction between \vtt{BMI}, \vtt{FI} and \vtt{New} using the 10th completed data set under MICE with CART from several iterations of the repeated simulation study in Figure~\ref{Impcomp2}. We include corresponding ground truth data sets without missing values for comparison in the top row of the figure.

The first-order linear interaction model is clear, but MICE-CART is unable to model the differing linear slopes by family income. For instance, in the bottom-left panel, a group of individuals with high \vtt{BMI} and low values for \vtt{New} are classified as having high family income. However, in the ground truth data sets, there is a strong positive association between \vtt{New} and \vtt{BMI} for individuals with high family income. Clearly, MICE-CART is producing implausible imputed values, demonstrating the inadequacy of this method when simpler models suffice. 

Finally, we also present imputations under GM-RND, the Bayesian nonparametric competitor. As mentioned in the main text, the continuous treatment of ordinal variable \vtt{FamIncome} clearly results in poor imputation, which affects the downstream regression analysis.\color{black}
\section{Real Data Application}
\begin{figure}[h]
    \centering
    \includegraphics[width = .24\textwidth,keepaspectratio]{Images/White-Female-No-Comp.png}
    \includegraphics[width = .24\textwidth,keepaspectratio]{Images/Black-Female-No-Comp.png}
    \includegraphics[width = .24\textwidth,keepaspectratio]{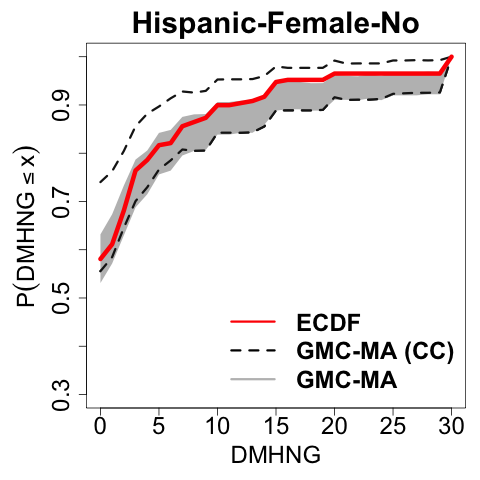}
    \includegraphics[width = .24\textwidth,keepaspectratio]{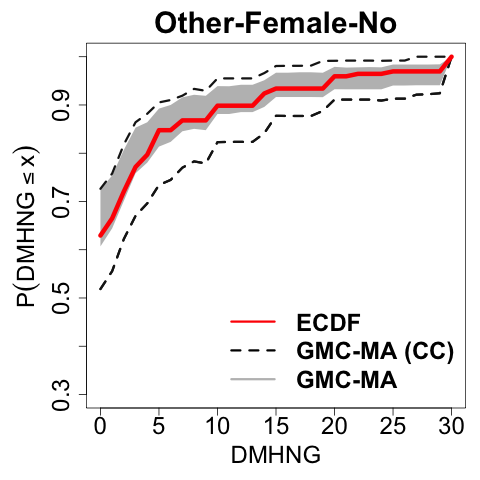}
    \includegraphics[width = .24\textwidth,keepaspectratio]{Images/White-Female-Yes-Comp.png}
    \includegraphics[width = .24\textwidth,keepaspectratio]{Images/Black-Female-Yes-Comp.png}
    \includegraphics[width = .24\textwidth,keepaspectratio]{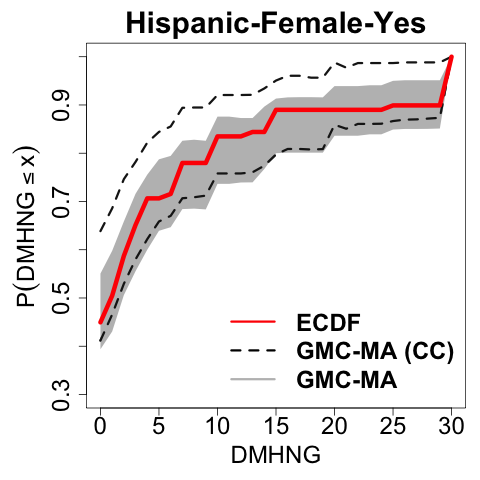}
    \includegraphics[width = .24\textwidth,keepaspectratio]{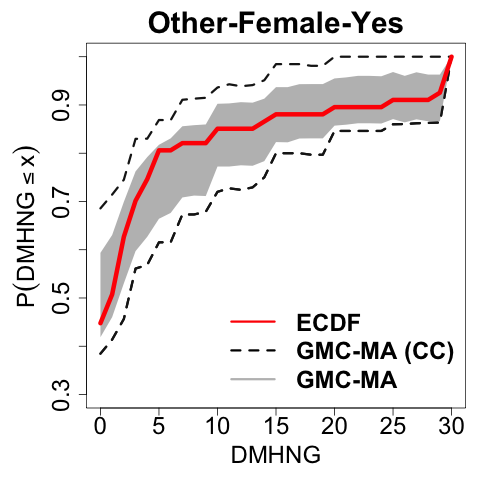}
    \caption{\small Posterior predictive summaries for models fit to the complete case (CC) dataset (GMC-MA (CC)) and the full dataset (GMC-MA). Completing the information presented for females, for each race-gender-marijuana use stratum, we compare the 95\% HPD intervals for the posterior predictive ECDFs, and include the ECDF on the CC data for reference. The GMC-MA (CC) output is well-calibrated to the observed data. By comparison, the GMC-MA fit to the full dataset produces intervals that are narrower and shifted, which suggests that the missingness mechanism is MAR---and that CC analysis is unreliable. These results were consistent across stratum}
    \label{compecdffull}
\end{figure}

\begin{figure}[h]
    \centering
    \includegraphics[width = .24\textwidth,keepaspectratio]{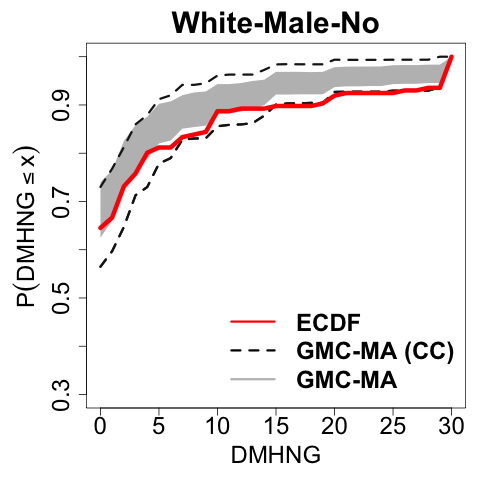}
    \includegraphics[width = .24\textwidth,keepaspectratio]{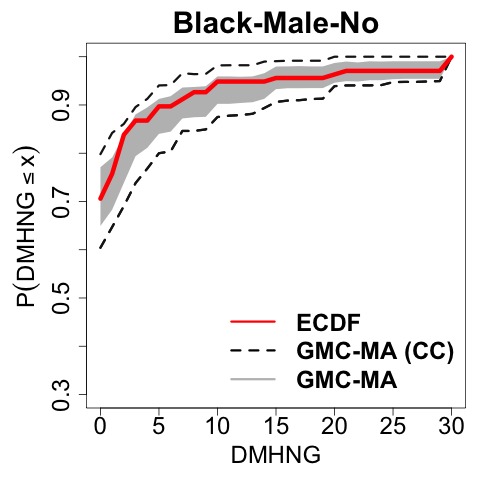}
    \includegraphics[width = .24\textwidth,keepaspectratio]{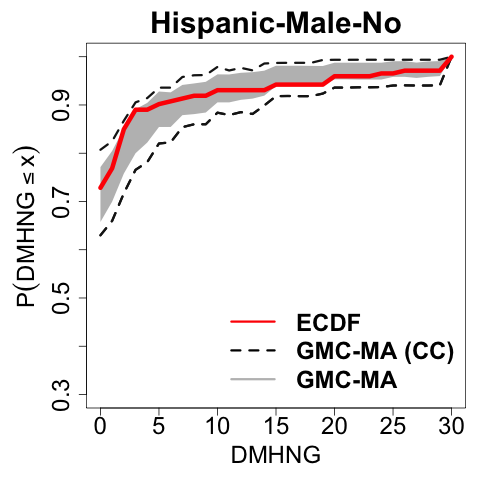}
    \includegraphics[width = .24\textwidth,keepaspectratio]{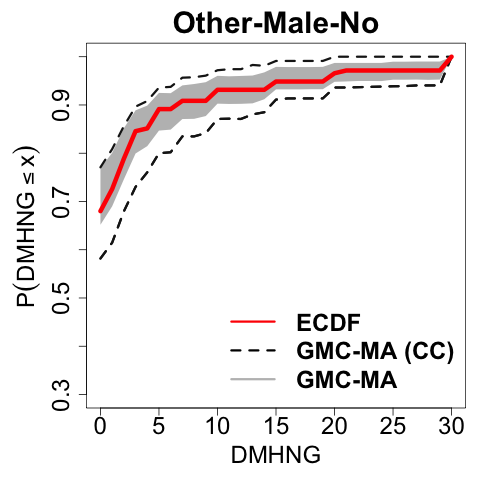}
    \includegraphics[width = .24\textwidth,keepaspectratio]{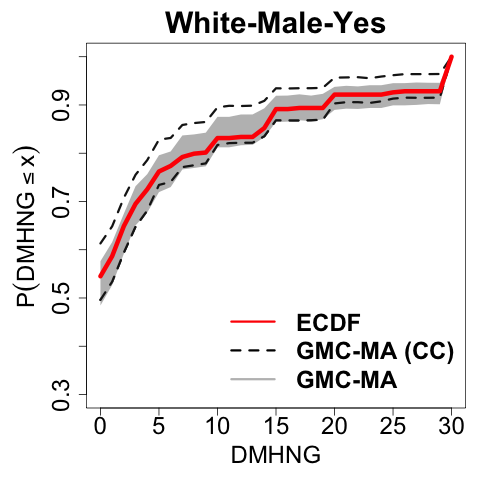}
    \includegraphics[width = .24\textwidth,keepaspectratio]{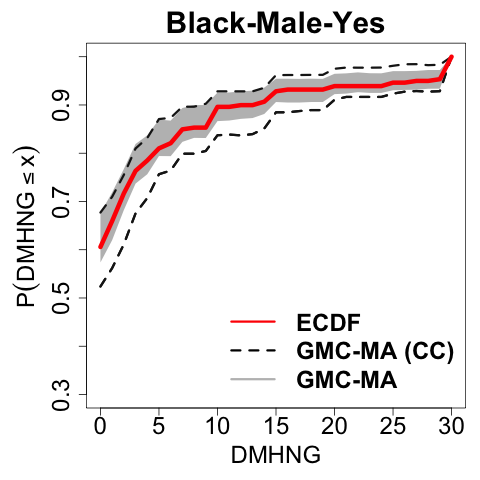}
    \includegraphics[width = .24\textwidth,keepaspectratio]{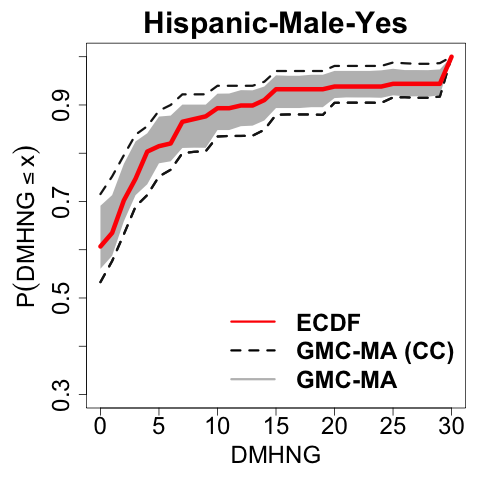}
    \includegraphics[width = .24\textwidth,keepaspectratio]{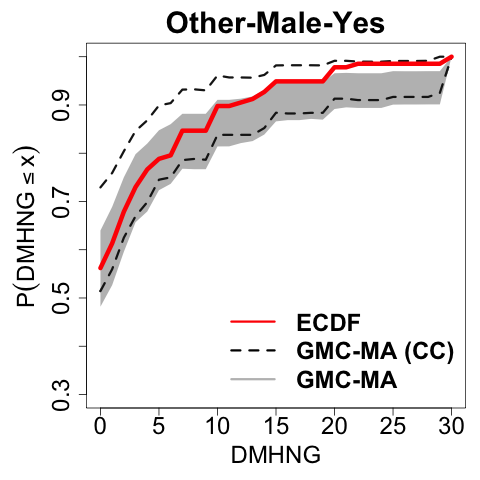}
    \medskip
    \caption{Posterior predictive summaries for models fit to the complete case (CC) dataset (GMC-MA (CC)) and the full dataset (GMC-MA). Among males, for each male race-gender-marijuana use stratum, we compare the 95\% HPD intervals for the posterior predictive ECDFs, and include the ECDF on the CC data for reference. The GMC-MA (CC) output is well-calibrated to the observed data. By comparison, the GMC-MA fit to the full dataset produces intervals that are narrower and shifted, which suggests that the missingness mechanism is MAR---and that CC analysis is unreliable.}
    \label{CCcalibmale}
\end{figure}

In Section~\ref{realdat} of the main paper, we check model calibration of the GMC-MA for several of the race-gender-marijuana stratum among females.  We complete the information presented in Figure~\ref{compecdf} in Figure~\ref{compecdffull} and include the same visual checks highlighted in Figure~\ref{CCcalibmale} for complete case males with similar conclusions. In each stratum, there is substantial overlap with the posterior predictive inference under the GMC-MA (CC) fit and ECDF estimates, which subsequently fails in certain cases for the full data fit. This result is consistent with what is presented in Section \ref{realdat}, which supports the notion that missing data may yield a biased complete case analysis.

\begin{figure}[h]
    \centering
\includegraphics[width = .8\textwidth, keepaspectratio]{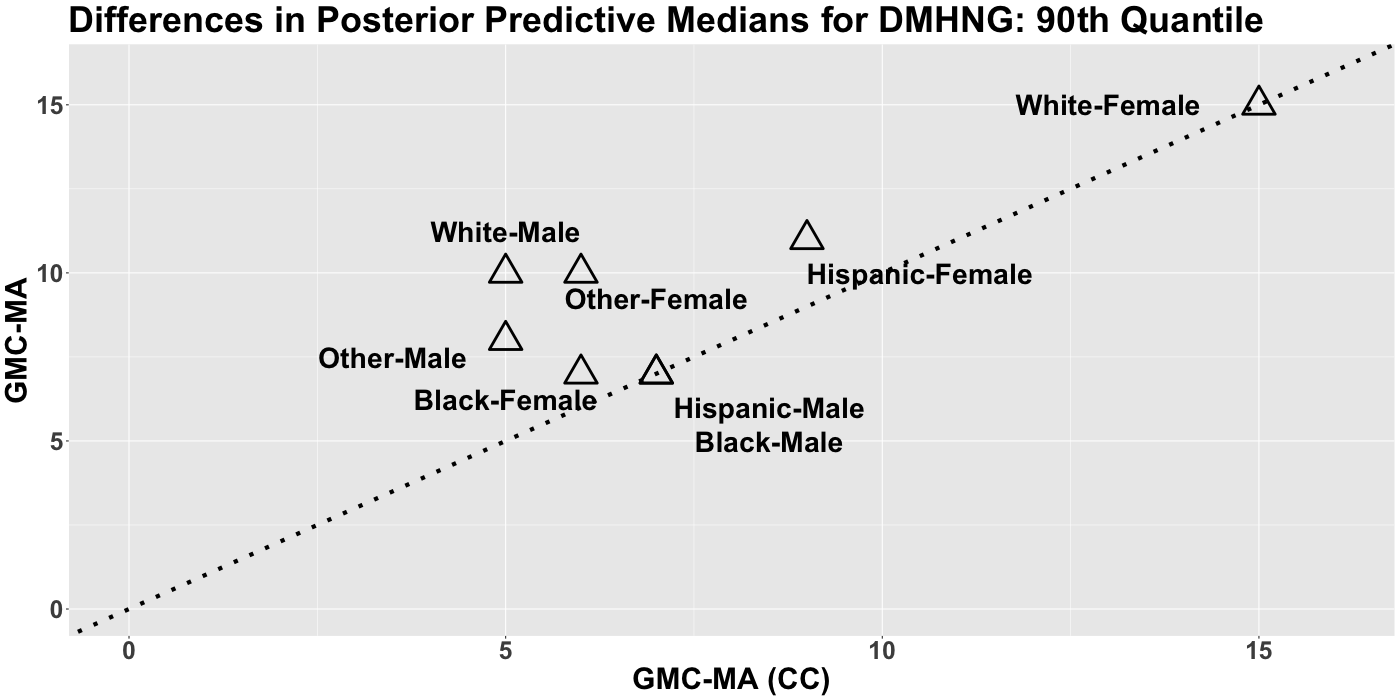}
\caption{\small Difference in posterior predictive medians for the predictive 90th quantiles of \vtt{DMHNG} between marijuana users and non-users and comparing models fit to the complete case (CC) dataset (GMC-MA (CC)) and the full dataset (GMC-MA).  The CC point estimates  attenuate the differences between  marijuana users and non-users across all strata, which dilutes the strong, significant, and adverse effects detected by GMC-MA fit to the full dataset.}
    \label{qcomp-point}
\end{figure}

\begin{figure}[h]
    \centering
        \includegraphics[width = .8\textwidth, keepaspectratio]{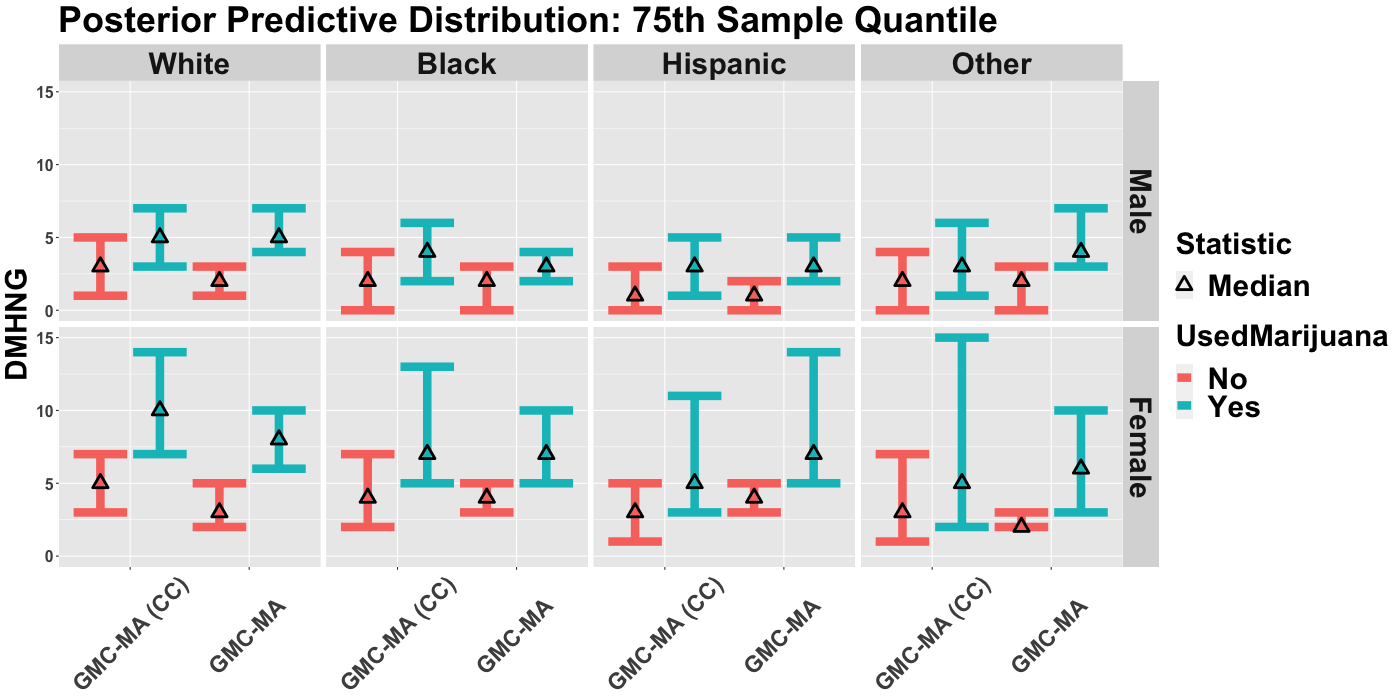}
        \caption{Posterior predictive medians and 95\% HPD intervals for the predictive 75th quantiles of \vtt{DMHNG} by race-gender-marijuana use and comparing models fit to the complete case (CC) dataset (GMC-MA (CC)) and the full dataset (GMC-MA).  The CC analysis produces wider intervals with more overlap between  marijuana users and non-users across all strata, which dilutes the strong, significant, and adverse effects detected by GMC-MA fit to the full dataset.}
        \label{q-int75}
\end{figure}

We include additional visualizations for comparing the estimated effect of marijuana use in each stratum for the 90th sample quantile by model fit. Point estimates are computed by taking the difference between posterior predictive medians for marijuana users and non-marijuana users, and we compare  these point estimates between the GMC-MA and GMC-MA (CC) in Figure~\ref{qcomp-point}. We complete the information in the main paper by comparing point estimates and uncertainty using the posterior predictive distribution of the 75th sample quantile in Figures~\ref{q-int75}-\ref{q-point75}, as was done in Figure~\ref{qcomp-int} in the main paper and Figure~\ref{qcomp-point} in the supplement. As expected, the differences between the full and complete case fits are not as pronounced, owing to fact that for most strata, between 70 and 90\% of individuals have $\leq 10 \ \verb|DMHNG|$. However, several discrepancies still do arise, and some intervals that  substantially overlap in the CC fit are much more clearly separated on the full data set.

\begin{figure}[h]
        \includegraphics[width = .8\textwidth, keepaspectratio]{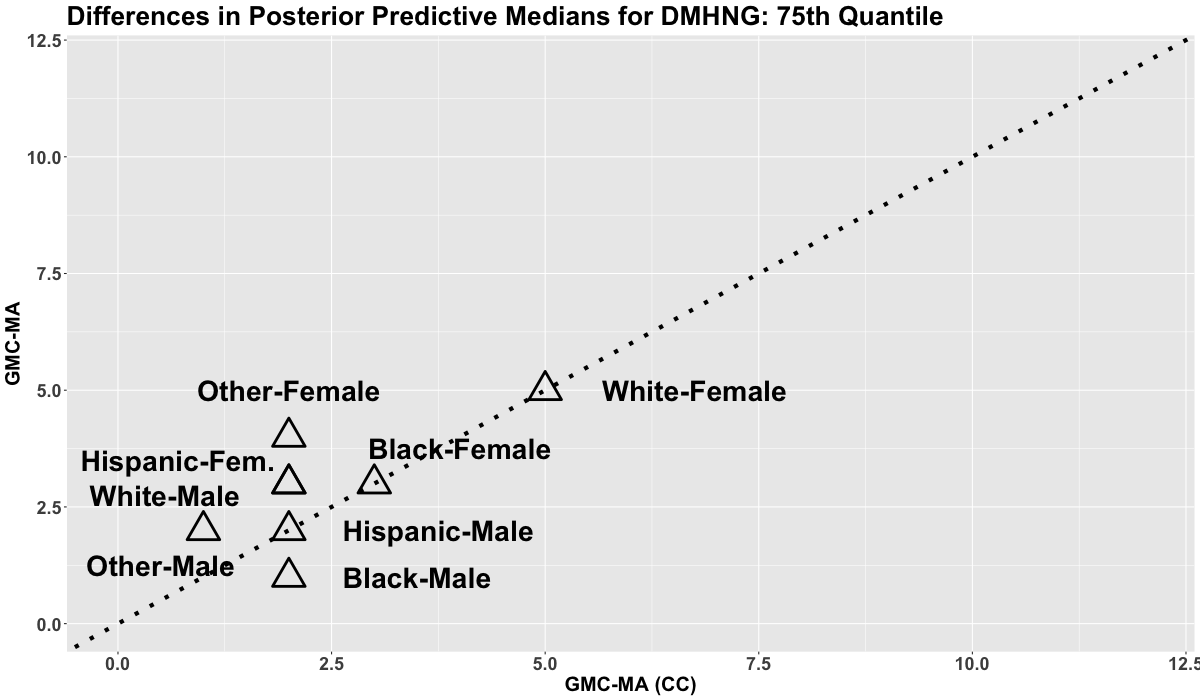}
        \caption{Difference in posterior predictive medians for the predictive 75th quantiles of \vtt{DMHNG} between marijuana users and non-users and comparing models fit to the complete case (CC) dataset (GMC-MA (CC)) and the full dataset (GMC-MA).  The CC point estimates  attenuate the differences between  marijuana users and non-users across nearly all strata, which dilutes the strong, significant, and adverse effects detected by GMC-MA fit to the full dataset.}

    \label{q-point75}
\end{figure}

\end{document}